\documentclass[11pt]{article}

\usepackage[letterpaper,margin=1in]{geometry}

\usepackage{fullpage}
\usepackage[T1]{fontenc}
\usepackage{mathpazo}
\usepackage{eufrak}
\usepackage{tgpagella}

\RequirePackage{graphicx,xcolor}

\usepackage{amsmath,amssymb,amsthm}
\usepackage{mathtools}

\usepackage{graphicx}
\usepackage{booktabs}
\usepackage{subcaption}

\usepackage{color}
\usepackage{amsmath}
\usepackage{todonotes}
\usepackage{comment}
\usepackage{tikz}
\usepackage{amsmath}
\usepackage{filecontents}
\usepackage{graphicx}
\usepackage{amsthm}
\usepackage{amssymb}
\usepackage{thmtools}
\usepackage{color}
\usepackage{enumitem}
\usepackage{bm}

\usepackage{amsmath}  
\usepackage{amssymb}
\usepackage{amsthm}
\usepackage[colorlinks=true,linkcolor=blue,citecolor=blue,allcolors=blue,hypertexnames=false]{hyperref}
\hypersetup{
    colorlinks=true,
    linkcolor=blue,
    filecolor=blue,
    urlcolor=blue,
    allcolors=blue
}
\usepackage[utf8]{inputenc}  %
\DeclareUnicodeCharacter{1EAF}{\'{\u{a}}} %
\usepackage{CJKutf8}         %

\newtheorem{theorem}{Theorem}[section]    %
\newtheorem{lemma}[theorem]{Lemma}        %
  
\newtheorem{corollary}[theorem]{Corollary}  

\newtheorem{claim}[theorem]{Claim}

\newtheorem{definition}[theorem]{Definition}

\usepackage{placeins} %
\usepackage[noend,linesnumbered, lined, boxed]{algorithm2e}
\SetKwInput{KwData}{Input}
\SetKwInput{KwResult}{Output}

\usepackage{cleveref}
\crefname{algorithm}{algorithm}{algorithms}
\Crefname{algorithm}{Algorithm}{Algorithms}
\crefname{claim}{Claim}{Claims}
\crefname{fact}{Fact}{Facts}

\newcommand{\veps}{\varepsilon}
\newcommand{\calA}{\mathcal{A}}
\newcommand{\calB}{\mathcal{B}}
\newcommand{\calD}{\mathcal{D}}

\newcommand{\calG}{\mathcal{G}}
\newcommand{\calH}{\mathcal{H}}
\newcommand{\calM}{\mathcal{M}}
\newcommand{\calN}{\mathcal{N}}
\newcommand{\calP}{\mathcal{P}}
\newcommand{\calQ}{\mathcal{Q}}
\newcommand{\calR}{\mathcal{R}}
\newcommand{\calS}{\mathcal{S}}

\newcommand{\calX}{\mathcal{X}}
\newcommand{\calY}{\mathcal{Y}}

\newcommand{\RR}{\mathrm{RR}}
\newcommand{\Aut}{\mathrm{Aut}}
\newcommand{\SUM}{\mathrm{SUM}}
\newcommand{\E}{\operatorname{\mathbb{E}}}

\newcommand{\bmA}{\mathbf{A}}
\newcommand{\bmB}{\mathbf{B}}
\newcommand{\bmC}{\mathbf{C}}
\newcommand{\bmD}{\mathbf{D}}
\newcommand{\bmE}{\mathbf{E}}
\newcommand{\bmF}{\mathbf{F}}
\newcommand{\bmG}{\mathbf{G}}

\newcommand{\bmI}{\mathbf{I}}
\newcommand{\bmP}{\mathbf{P}}
\newcommand{\bmQ}{\mathbf{Q}}

\newcommand{\bmS}{\mathbf{S}}
\newcommand{\bmW}{\mathbf{W}}
\newcommand{\bmX}{\mathbf{X}}
\newcommand{\bmY}{\mathbf{Y}}
\newcommand{\bmZ}{\mathbf{Z}}
\newcommand{\bma}{\mathbf{a}}

\newcommand{\bme}{\mathbf{e}}
\newcommand{\bmg}{\mathbf{g}}

\newcommand{\bmx}{\mathbf{x}}

\newcommand{\hf}{\widehat{f}}

\newcommand{\nats}{\mathbb{N}}

\newcommand{\reals}{\mathbb{R}}

\newcommand{\Var}{\mathrm{Var}}
\newcommand{\Cov}{\mathrm{Cov}}
\newcommand{\Lap}{\textrm{Lap}}
\newcommand{\Clip}{\textrm{Clip}}

\newcommand{\poly}{\operatorname{poly}}

\allowdisplaybreaks

\begin{document}
\title{The Power of Interaction\\in Locally Edge-Private Subgraph Counting}
\author{Xian Chen\textsuperscript{$\dagger$}\qquad
        Pan Peng\textsuperscript{$\dagger$}}
\date{}
\pagenumbering{roman}
\pagestyle{empty}
\maketitle
\thispagestyle{empty}

\begingroup
\renewcommand{\thefootnote}{\ensuremath{\dagger}}
\footnotetext{University of Science and Technology of China.
Emails: \texttt{xianchen@mail.ustc.edu.cn},
\texttt{ppeng@ustc.edu.cn}.}

\begin{abstract}
We study (non-induced) subgraph counting under interactive local edge differential privacy (LEDP). Given a fixed pattern graph $H$ with $k$ vertices, $\ell\ge1$ edges, and $q$ connected components, our goal is to approximately count its copies in an arbitrary $n$-vertex undirected graph while preserving edge privacy.

We give two complementary accuracy guarantees. First, for every $H$, we give \emph{Linear Noise Cancellation} (LNC), a two-round $\varepsilon$-LEDP algorithm with additive error and root mean square error both $O(n^{k-3/2}/\varepsilon^\ell)$ on every $n$-vertex graph. For fixed sufficiently small $\varepsilon$, this improves the optimal noninteractive additive error for triangles by a factor of $\sqrt n$ and gives the same improvement in worst-case root mean square error for all cliques on at least three vertices. To our knowledge, this is the first separation between interactive and noninteractive LEDP subgraph counting, showing that a single additional round can improve worst-case accuracy polynomially.

Second, we determine how accuracy improves when the graph has maximum degree at most a public bound $d$. Our \emph{Predict and Correct} (PC) framework gives an $(\ell+2)$-round $\varepsilon$-LEDP algorithm with additive error $O(n^{q-1/2}d^{k-q-1}\log^{3\ell-2}n/\varepsilon^\ell)$. For connected patterns, its polynomial factor is $\sqrt n\,d^{k-2}$, giving sublinear error on constant-degree graphs at fixed privacy. We prove matching lower bounds in $n$ and $d$, up to logarithmic factors at fixed privacy, even under arbitrary interaction and approximate privacy with sufficiently small $\delta$. At $d=\Theta(n)$, the lower bound also establishes the optimality of LNC's dependence on $n$.
\end{abstract}

\clearpage
\tableofcontents

\clearpage
\pagestyle{plain}
\pagenumbering{arabic}

\section{Introduction}
\label{sec:introduction}

Subgraph counting turns a large network into a collection of interpretable structural statistics by recording how often fixed small patterns occur. Triangles reflect local clustering~\cite{becchetti2008efficient}, stars capture degree moments~\cite{gonen2011counting}, and paths, cycles, and cliques characterize different aspects of network structure~\cite{gonen2009approximating,fan2021characterizing,palla2005uncovering}. These statistics facilitate comparisons between networks~\cite{prvzulj2007biological} and help reveal their local organization~\cite{milo2002network}.

In applications involving sensitive relational data, the usefulness of these statistics must be balanced against the privacy risks of releasing them. Differential privacy provides a formal framework for releasing such statistics while protecting information about individual relationships or participants~\cite{dwork2006calibrating}. Two modeling choices are fundamental. In the \emph{central} model, a trusted curator collects the graph and releases private statistics. Although the release is protected, participants must still entrust their original data to the curator. This requirement can be undesirable when the graph is distributed across participants who are unwilling to disclose their relationships to a central party. The \emph{local} model removes the need for such trust: participants randomize their reports before sharing them with an untrusted server. Each participant reports information about its own connections, and the server combines these reports to estimate global network statistics~\cite{kasiviswanathan2011can}.

Independently of this trust model, one must decide what information the privacy guarantee protects. \emph{Edge privacy} protects the presence or absence of a single relationship, whereas \emph{node privacy} protects a participant's entire set of relationships~\cite{HayLMJ09}. Node privacy provides a stronger guarantee, but can also entail a greater loss of accuracy~\cite{kasiviswanathan2013graphs}. When individual relationships are the intended unit of protection, edge privacy is a natural choice and may allow more accurate estimates of aggregate network statistics~\cite{HayLMJ09,raskhodnikova2026localnode}. It has also been extensively studied in private graph analysis, including subgraph counting in the local model~\cite{HayLMJ09,karwa2014private,qin2017synthetic,dhulipala2022differential,eden2025triangle}.

Within the local model, interaction introduces another important distinction. In the \emph{noninteractive} setting, each participant sends a randomized report based on its local information, without adapting to other participants' reports. In the \emph{interactive} setting, the server may use earlier reports to determine subsequent queries, and participants may respond over multiple rounds~\cite{kasiviswanathan2011can}. For graph data, the server can send summaries of earlier reports back to the participants, who can combine this information with their own neighbor lists to produce further randomized reports~{\cite{imola2021locally,imola2022communication}}. This capability is particularly relevant to subgraph counting, where the relationships forming a pattern are distributed across several participants. By allowing participants to combine their local information with privatized information from others, interaction may improve the accuracy of subgraph count estimates~\cite{imola2021locally,imola2022communication,eden2025triangle}.

We focus on understanding the power of interaction for subgraph counting under \emph{local edge differential privacy} (LEDP). Given a fixed pattern graph $H$ with $k$ vertices, $\ell$ edges, and $q$ connected components, our goal is to design interactive local algorithms that approximately estimate the number of its non-induced copies in an arbitrary $n$-vertex undirected graph while preserving edge privacy.

Previous work provides noninteractive algorithms for counting pattern $H$ that output an estimate with additive error $O_\veps(n^{k-1})$ and success probability at least $2/3$. The same bound holds for root mean square error (RMSE; see \Cref{eq:rmse}), which is optimal in its dependence on $n$ for cliques on at least three vertices when $\varepsilon$ is a sufficiently small constant~\cite{suppakitpaisarn2025graphlets}. However, whether interaction can improve the worst-case dependence on $n$ has remained unresolved, even for triangle counting. On general $n$-vertex graphs, previous triangle-counting algorithms achieve additive error $O_\veps(n^2)$ both with and without interaction~{\cite{imola2022communication,eden2025triangle,hillebrand2025cycle}}. Although this rate is optimal in the noninteractive setting, the known lower bound for interactive algorithms is only $\Omega_{\varepsilon}(n^{3/2})$~\cite{eden2025triangle}. This leaves a polynomial gap between the upper and lower bounds, and motivates the question of whether interaction can improve the exponent of $n$ in the worst-case error for arbitrary fixed patterns.

We resolve this question with a two-round LEDP algorithm, which we call \emph{Linear Noise Cancellation} (LNC), for every fixed pattern. For triangles and fixed sufficiently small $\veps$, LNC improves the optimal noninteractive additive error by a factor of $\sqrt n$, with success probability at least $2/3$. Its RMSE guarantee yields the same separation in worst-case RMSE for all cliques on at least three vertices. Our lower bound shows that even arbitrarily many rounds cannot improve the resulting exponent of $n$.

These results characterize the optimal worst-case accuracy over all $n$-vertex graphs. A finer question is how the accuracy achievable under interaction depends on structural properties of the input graph beyond its size. Many real-world networks are sparse~\cite{clauset2004finding}. Worst-case upper and lower bounds expressed only in terms of $n$ do not reveal how achievable accuracy varies with the structural properties of the input graph. For example, constant-degree graphs contain only $O(n)$ triangles, so even an optimal unrestricted-graph error bound of order $n^{3/2}$ at fixed privacy can exceed the entire range of possible counts. 

We therefore consider the maximum degree $d$ as a natural additional parameter for refining accuracy guarantees, building on an extensive line of research on degree-dependent bounds for private subgraph counting. This includes algorithms and error analyses for stars and triangles~\cite{imola2021locally,imola2022communication}, four-cycles~\cite{guo2026noisy}, odd cycles~\cite{hillebrand2025cycle}, and acyclic patterns, including paths and trees~\cite{hu2026acyclic}. These results demonstrate the benefits of exploiting graph structure, but do not determine the optimal joint dependence on $n$ and $d$ for arbitrary fixed patterns under unrestricted interaction. Although degree-dependent lower bounds are known for some patterns in the noninteractive setting~\cite{imola2021locally}, they do not resolve this interactive question. We therefore seek matching upper and lower bounds parameterized jointly by $n$ and $d$. For fixed privacy parameters, our second algorithm \emph{Predict and Correct} (PC) achieves the optimal polynomial dependence on both parameters, up to logarithmic factors, while retaining privacy on all input graphs.

\subsection{Our Results}

For a fixed pattern graph $H$, we refer to the problem of estimating the number of non-induced copies of $H$ in an input graph as $H$-counting. We measure privacy using $(\veps,\delta)$-LEDP (see \Cref{def:ledp} for the formal definition), which limits how much the distribution of the complete interaction transcript can change when a single edge is added or removed. The case $\delta=0$ corresponds to pure $\veps$-LEDP. Throughout, these privacy guarantees hold on all input graphs. We first obtain a two-round $\veps$-LEDP algorithm LNC for subgraph counting with the following error guarantee.

\begin{theorem}
\label{thm:interactive-upper-bound-subgraph}
For any fixed pattern graph $H$ with $k$ vertices and $\ell\ge 1$ edges, and any $\veps\in(0,1]$, there is a two-round $\veps$-LEDP algorithm LNC (\Cref{alg:interactive-correction-subgraph}) for $H$-counting with additive error $O(n^{k-3/2}/\veps^\ell)$ with probability at least $2/3$. Its RMSE satisfies the same bound.
\end{theorem}

For comparison, any noninteractive $\veps$-LEDP algorithm for triangle counting with success probability at least $2/3$ must incur worst-case additive error $\Omega(n^2)$ for fixed sufficiently small $\veps$~\cite{eden2025triangle}. LNC therefore improves the optimal noninteractive additive error by a factor of $\sqrt n$. More generally, for $k$-vertex cliques with $k\ge3$, the noninteractive worst-case RMSE is $\Omega(n^{k-1})$ for fixed sufficiently small $\veps$~\cite{suppakitpaisarn2025graphlets}. LNC's additional RMSE guarantee gives the same factor-$\sqrt n$ separation under this measure, where the worst case is over all $n$-vertex input graphs. To our knowledge, this is the first proved separation between interactive and noninteractive LEDP subgraph counting in worst-case accuracy.

\begin{table}[!t]
\begin{minipage}{\linewidth}
\centering
\footnotesize
\setlength{\tabcolsep}{3pt}
\renewcommand{\arraystretch}{1.3}
\begin{tabular}{@{}p{0.14\linewidth}p{0.28\linewidth}p{0.12\linewidth}p{0.16\linewidth}p{0.09\linewidth}@{}}
\toprule
& \multicolumn{2}{c}{Prior Results} & \multicolumn{2}{c}{Our Results} \\
\cmidrule(lr){2-3}\cmidrule(lr){4-5}
Pattern & Additive Error & Rounds & Additive Error & Rounds \\
\midrule
Triangle $C_3$ & $O_{\veps}(\sqrt n\,d^{\frac32})$\cite{imola2021locally} & $2$ & $\widetilde O_{\veps}(\sqrt n\,d)$ & ${3}$ \\
\addlinespace[3pt]
Four-cycle $C_4$ & $\widetilde O_{\veps}(\sqrt n\,d^{\frac52}+nd^{\frac32})$\cite{guo2026noisy} & $2$ & $\widetilde O_{\veps}(\sqrt n\,d^2)$ & ${4}$ \\
\addlinespace[3pt]
Odd cycles $C_k$ & $\widetilde O_{\veps}(n^{\frac{k-2}{2}}d^{\frac{k}{2}})$\cite{hillebrand2025cycle} & $3$ & $\widetilde O_{\veps}(\sqrt n\,d^{k-2})$ & $k+2$ \\
\addlinespace[3pt]
Paths $P_k$ & $\widetilde O_{\veps}(\sqrt n\,d^{k-1})$\cite{hu2026acyclic} & $k-1$ & $\widetilde O_{\veps}(\sqrt n\,d^{k-2})$ & $k+1$ \\
\addlinespace[3pt]
Trees $T_k$ & $\widetilde O_{\veps}(\sqrt n\,d^{k-1})$\cite{hu2026acyclic} & $k+1-L$ & $\widetilde O_{\veps}(\sqrt n\,d^{k-2})$ & $k+1$ \\
\bottomrule
\end{tabular}
\caption{Comparison of additive-error bounds under pure $\veps$-LEDP on $n$-vertex graphs of maximum degree at most $d$, for fixed $H$ and $\veps$, with success probability at least $2/3$. Here $L$ denotes the number of leaves in the tree row. The notation $\widetilde O$ suppresses logarithmic factors. For a uniform comparison, we report general error upper bounds for each pattern family in terms of $n$ and $d$, upper-bounding finer structural parameters where necessary. Our error bounds and round counts follow from \Cref{gh:pure-main-theorem} and \Cref{cor:pc-star-remainder}.%
}
\label{tab:prior-degree-comparison}
\end{minipage}
\end{table}

Our second algorithm PC exploits a degree bound to obtain a more refined accuracy guarantee for every fixed pattern. \Cref{tab:prior-degree-comparison} compares our bounds with prior results. Our algorithms substantially improve the polynomial dependence on $n$ and $d$ while using only a constant number of rounds for each fixed pattern $H$.

\begin{theorem}
\label{gh:pure-main-theorem}
For any fixed pattern graph $H$ with $k$ vertices, $\ell\ge1$ edges and $q$ connected components, and any $\veps\in(0,1]$, there is an $(\ell+2)$-round $\veps$-LEDP algorithm PC. On every $n$-vertex graph of maximum degree at most $d$, it solves $H$-counting with additive error
\begin{align*}
O\left(\frac{n^{q-\frac{1}{2}}d^{k-q-1}\log^{3\ell-2}n}{\veps^\ell}\right)=\widetilde O_{\veps}\left(n^{q-\frac{1}{2}}d^{k-q-1}\right)
\end{align*}
with probability at least $2/3$. Its worst-case running time is $\operatorname{poly}(n,d,1/\veps)$.
\end{theorem}

{For certain patterns, \textsc{PC} uses fewer rounds without changing the error bound. If $H$ is connected and deleting a vertex $y$ leaves a star centered at $x$, it uses at most $|E(H-x)|+2$ rounds. A related improvement applies when $H$ has a vertex adjacent to every other vertex (\Cref{cor:pc-star-remainder}). In particular, triangles require only three rounds and four-cycles only four.}

We complement PC's accuracy guarantee with a lower bound matching its polynomial dependence on $n$ and $d$, even under arbitrary interaction. We state an informal version for fixed $\veps$. \Cref{thm:degree-lower-bound} gives the precise parameter conditions and the full dependence on $\veps$ for any $\veps\in(0,1]$.

\begin{theorem}[Informal]
\label{thm:degree-lower-informal}
Fix a pattern graph $H$ with $k$ vertices, $q$ connected components and at least one edge, and $\veps \in (0, 1]$. Let $C$ be a sufficiently large absolute constant. For all sufficiently large $n$, every integer $d$ satisfying $1\le d\le n-1$ and $d\ge6(k-2)$, and every $\delta\in[0,n^{-C}]$, any $(\veps,\delta)$-LEDP algorithm for $H$-counting with success probability at least $2/3$ must incur worst-case additive error $\Omega(n^{q-\frac{1}{2}}d^{k-q-1}/\veps)$ on $n$-vertex graphs of maximum degree at most $d$, even under arbitrary interaction.
\end{theorem}

Together, \Cref{gh:pure-main-theorem,thm:degree-lower-informal} show that PC achieves optimal polynomial dependence on both $n$ and $d$, up to logarithmic factors at fixed privacy. Taking $d=\Theta(n)$ gives a lower bound of $\Omega(n^{k-3/2}/\veps)$, so LNC (\Cref{thm:interactive-upper-bound-subgraph}) also achieves the optimal dependence on $n$ under arbitrary interaction, using only two rounds and without a logarithmic loss.

\subsection{Technical Overview}
\label{sec:overview}

Our two upper bounds use interaction to control the error in public edge reports. In \Cref{overview:interaction}, we explain how \emph{Linear Noise Cancellation} (LNC) uses a second round to cancel terms linear in the reporting noise, achieving the optimal dependence on $n$. We then turn to \emph{Predict and Correct} (PC), which exploits a degree bound by learning public predictions that reduce the coefficients of the noise. For ease of exposition, \Cref{overview:degree} introduces this idea through triangle counting; \Cref{overview:general} extends it to arbitrary patterns and explains how random hashing and signs are used to estimate the corrections. \Cref{overview:pure} presents a pure-DP replacement for the Gaussian releases used by PC, preserving its accuracy guarantee. Finally, \Cref{overview:lower} outlines the matching lower bounds under arbitrary interaction. Throughout this overview, we consider $H$-counting on an input graph $G$ with vertex set $[n]=\{1,\ldots,n\}$ and its adjacency matrix $\bmA$.

\subsubsection{LNC: Canceling Linear Randomization Error}
\label{overview:interaction}

A natural starting point is to estimate the subgraph count using a noisy adjacency matrix. In the first round, each potential edge is assigned to one of its endpoints, which reports its adjacency bit using standard debiased randomized response (\Cref{lemma:debiased-RR}). The server assembles these unbiased edge estimates into a symmetric public matrix $\bmX$ with zero diagonal. We write $\bmX=\bmA+\bmZ$ to separate the true adjacency information from the randomization noise. This noise has mean zero and is independent across distinct unordered vertex pairs. Substituting these edge estimates into the counting formula gives the polynomial estimator
\begin{align}
 P(\bmX)=\frac{1}{|\Aut(H)|}
 \sum_{\phi:V(H)\hookrightarrow[n]}
 \prod_{e\in E(H)}X_{\phi(e)}.\label{def:P(X)}
\end{align}
Here $\phi$ is an injective vertex map, $\phi(e)$ is the unordered pair corresponding to edge $e$, and $|\Aut(H)|$ is the number of automorphisms of $H$. The polynomial gives the exact $H$-count at $\bmA$ and an unbiased estimate at $\bmX$. We expand $P(\bmX)-P(\bmA)$ into a sum of products of entries of $\bmA$ and $\bmZ$. In the part linear in $\bmZ$, the coefficient of $Z_{\{u,v\}}$ counts the ways to complete $\{u,v\}$ to a copy of $H$ using the other edges of $G$. In a complete graph, these coefficients are of order $n^{k-2}$ for order $\Theta(n^2)$ pairs, so the linear terms alone contribute a variance of order $n^2\cdot (n^{k-2})^2=n^{2k-2}$ at fixed privacy. This produces RMSE of order $n^{k-1}$, exceeding the order $n^{k-3/2}$ we seek.

In the second round, we cancel the terms linear in $\bmZ$ by adding a correction $(A_e-X_e)\partial_e P(\bmX)$ for each potential edge $e\in\binom{[n]}{2}$, where $\partial_e P$ denotes the derivative of $P$ with respect to $X_e$ (see \eqref{eq:partialP}). Since $\partial_e P(\bmX)$ depends only on public reports, the endpoint responsible for $e$ can compute this correction using its own adjacency bit. Each vertex releases the sum of its corrections with standard sensitivity-calibrated Laplace noise (\Cref{lemma:Laplace mechanism}). Each correction term has mean zero, so the estimator 
\[
\widehat{P}(\bmX) =P(\bmX)+\sum_{e\in\binom{[n]}{2}}(A_e-X_e)\partial_e P(\bmX)
\]
obtained by adding the corrections from all vertices to the initial estimate $P(\bmX)$ remains unbiased for the exact $H$-count $P(\bmA)$, even after the final Laplace noise is added.

To see the effect of the correction, expand $\widehat{P}(\bmX) - P(\bmA)$ into products of entries of $\bmA$ and $\bmZ$. A product containing $s\ge 1$ entries of $\bmZ$ appears once in the expansion of $P(\bmX)$ and is subtracted once by the correction associated with each of those $s$ entries. Its original coefficient is therefore multiplied by $1-s$. Consequently, every term containing exactly one entry of $\bmZ$ disappears.

Every remaining error term contains entries of $\bmZ$ indexed by at least two distinct unordered vertex pairs, whose endpoints span $c\ge3$ vertices. Since entries indexed by distinct unordered pairs are independent and have mean zero, the expected product of two terms vanishes unless they contain exactly the same entries of $\bmZ$. For each fixed configuration of these pairs, there are $O(n^c)$ choices of their endpoints and $O(n^{k-c})$ extensions of each choice to an injective vertex map. Thus, for each $c$, the number of pairs of terms that can contribute to the variance is at most
\[
 O\!\left(n^c(n^{k-c})^2\right)=O\left(n^{2k-c}\right),
\]
which is largest at $c=3$. Accounting for the dependence on $\veps$, the corrected estimator has variance $O(n^{2k-3}/\veps^{2\ell})$. The Laplace noise needed to release the local corrections privately contributes at most the same order of variance, so the final estimator retains this bound. Thus LNC has RMSE $O(n^{k-3/2}/\veps^\ell)$ and the same order of additive error, as proved in \Cref{sec:two-round}.

\subsubsection{PC: Exploiting a Degree Bound for Triangles}\label{overview:degree}
LNC achieves the optimal dependence on $n$, raising the natural question of whether its correction mechanism can also exploit a degree bound to achieve optimal dependence on both $n$ and $d$.

For triangles, expanding $\widehat{P}(\bmX)-P(\bmA)$ gives products of entries of $\bmA$ and $\bmZ$, each containing at least two entries of $\bmZ$. A term containing exactly two entries of $\bmZ$ contains only one adjacency entry. This restricts the vertex pair corresponding to the adjacency entry to one of the $O(nd)$ edges in $G$, but leaves $O(n)$ choices for the third vertex, since the noise is present even on nonedges. These terms, therefore, contribute variance $O_\veps(n^2d)$, which gives RMSE of order $O_\veps(n\sqrt{d})$ that exceeds our target $\widetilde O_\veps(\sqrt n\,d)$ when $d$ is small compared with $n$. Thus, directly incorporating the degree bound into LNC's analysis does not yield the desired accuracy. 

This motivates constructing local estimates that make fuller use of the true adjacency information. For a candidate triangle on distinct vertices $i,j,t$, vertex $i$ knows $A_{i,j}$ and $A_{i,t}$ and can form the local estimate $A_{i,j}A_{i,t}X_{j,t}$, where $X_{j,t}=A_{j,t}+Z_{j,t}$ is obtained using standard debiased randomized response. When these local estimates are summed, the coefficients of $Z_{j,t}$ are proportional to the common-neighbor counts $(\bmA^2)_{j,t}$. PC aims to reduce these coefficients by learning public predictions and using them to construct more accurate correction estimates. These predictions are represented by a matrix $\bmP$, called a \emph{public center}, whose entries are fractional predictions of the adjacency bits. To reflect the degree bound, we require $\bmP$ to be symmetric with zero diagonal, entries in $[0,1]$, and row sums at most $d$.

PC replaces one edge factor of the ordered triangle count by $\bmP$ and estimates the difference between the original and predicted counts. To see how the center controls accuracy, fix $\bmP$ and consider estimating this difference using fresh randomized responses for the third edge. Ignoring the additional Laplace noise needed to release the local corrections privately, the additive estimation error is
\begin{equation}
 \widehat{\Delta} - \Delta
 =\sum_{i,j,t}(A_{i,j}-P_{i,j})A_{i,t}Z_{j,t}
 =\sum_{j,t}\bigl((\bmA-\bmP)\bmA\bigr)_{j,t}Z_{j,t}.
 \label{eq:overview-triangle}
\end{equation}
After summing over $i$, the reporting noise is weighted by the entries of $(\bmA-\bmP)\bmA$, giving the variance bound $O(\|(\bmA-\bmP)\bmA\|_F^2/\veps^2)$ conditional on $\bmP$, where $\|\mathbf M\|_F=(\sum_{i,j}M_{i,j}^2)^{1/2}$ is the Frobenius norm. The additional Laplace noise used to release these local estimates contributes total variance $O_\veps(nd^2)$, keeping its error contribution within our target bound. %
This motivates learning a center satisfying $\|(\bmA-\bmP)\bmA\|_F=\widetilde O_{\veps}(\sqrt n\,d)$.

\paragraph{Learning the center from private sketches.}We learn the public center through two rounds of private linear queries. The procedure begins by compressing the adjacency rows using a standard Rademacher sketch~\cite{achlioptas2003database}, represented by a public matrix $\bmS\in\reals^{n\times m}$ with $m=O(d\log n)$ and independent uniform entries $\pm1/\sqrt m$. In two rounds, the vertices release noisy rows of $\bmA\bmS$ and then $\bmA\bmB$, giving
\[
 \bmB=\bmA\bmS+\bmG_1,
 \qquad
 \bmC=\bmA\bmB+\bmG_2.
\]
Here $\bmG_1$ and $\bmG_2$ are independent noise matrices from the standard Gaussian mechanism (\Cref{lemma:Gaussian mechanism}), with privacy budgets allocated to ensure an overall $(\veps,\delta)$-LEDP guarantee. The normalization of $\bmS$ bounds the first query's sensitivity, while public norm checks control the second query's sensitivity. The server then chooses a public center $\bmP$ by approximately minimizing $\|\bmP\bmB-\bmC\|_F^2$ over the centers satisfying the constraints above. The feasible centers form a bounded convex set defined by linear constraints; the public checks ensure that every row of $\bmB$ and $\bmC$ has polynomially bounded Euclidean norm. We can therefore minimize the convex quadratic objective to constant additive error in polynomial time using standard convex optimization tools, such as the ellipsoid method (\Cref{lemma:ellipsoid-method}).

To understand why the fitted center $\bmP$ achieves the required prediction error bound, ignore the privacy noise momentarily: $\bmB$ and $\bmC$ are sketches of $\bmA$ and $\bmA^2$, respectively. In this setting, $\|\bmP\bmB-\bmC\|_F=\|(\bmA-\bmP)\bmA\bmS\|_F$, so minimizing the fitting error amounts to minimizing the norm of a sketch of the desired prediction residual. Reducing each released row to $O(d\log n)$ coordinates keeps the noise cost tied to $d$, up to logarithmic factors. The main difficulty is to show that fitting these noisy sketches controls the prediction error in the original space. 
We establish a comparison that holds with high probability simultaneously for every feasible center $\bmP'$, ensuring that it also applies to the center selected by the algorithm:
\[
 \|(\bmA-\bmP')\bmA\|_F
 \le O\!\left(\|(\bmA-\bmP')\bmB\|_F\right)
 +\widetilde O_\veps(\sqrt n\,d),
\]
where the suppressed logarithmic factors depend on $n/\delta$. Our proof exploits the degree bound: every residual row has $\ell_1$-norm at most $2d$. We first use moment bounds to obtain simultaneous control over a finite family of sums of signed coordinate vectors, then extend the comparison to all feasible residual rows through an averaging argument. Finally, since $\bmC=\bmA\bmB+\bmG_2$ and the true $\bmA$ is an admissible center on the accuracy domain, the fitted center cannot have a substantially larger fitting error than $\bmG_2$. This bounds $\|(\bmA-\bmP)\bmB\|_F$ by $O(\|\bmG_2\|_F+1)=\widetilde O_\veps(\sqrt n\,d)$. Substituting this bound into the uniform comparison yields the desired guarantee for $\|(\bmA-\bmP)\bmA\|_F$.

Combining this two-round center construction with the correction gives a three-round algorithm for triangle counting. The randomized responses needed for the correction can be collected alongside the first center reports, and the parties release their local estimates in round three. The center construction currently provides approximate privacy through Gaussian noise. %
We first extend PC to arbitrary patterns, then explain how to obtain pure privacy in \Cref{overview:pure}.

\subsubsection{PC: Extending to Arbitrary Patterns}
\label{overview:general}

To extend PC to an arbitrary pattern $H$, we first remove isolated vertices, restoring their contribution through a known combinatorial factor at the end. We then use the standard quotient-graph inversion formula (\Cref{lemma:quotient-inversion}) to express the remaining pattern's non-induced copy count as a fixed linear combination of homomorphism counts of quotient graphs. A homomorphism preserves every pattern edge but may map distinct pattern vertices to the same input vertex. This linear combination preserves the asymptotic error bound for fixed $H$, so we focus on homomorphism counting. This change of viewpoint serves two purposes: first, homomorphism counts factor over connected components, so we can estimate each component separately and control the error of their product using bounds on the component counts; second, allowing repeated images lets us combine local computations through interaction without enforcing distinctness across vertex assignments.

Consider a connected component $H'$ of a quotient graph, with $k'\ge2$ vertices, and suppose first that it has no cut vertex, meaning that deleting any one vertex leaves it connected. Such a graph forms a single \emph{block}. %
PC organizes prediction and correction as a sequence of edge replacements. Each pattern edge $e$ contributes a factor to the homomorphism count, represented by a matrix $\bmW_e$. Initially, all these matrices equal $\bmA$; as edges are processed, their factors are replaced by public centers. When replacing the factor on edge $\{v,u\}$ by $\bmP$, we estimate the difference $\Delta$ between the counts before and after replacement. Each vertex assignment contributes a product of matrix entries, using $\bmA-\bmP$ on $\{v,u\}$ and the current factors on all other edges. Summing these products over all vertex assignments gives $\Delta$. Once every edge factor has been replaced, the server can evaluate the final public count; adding the estimated differences gives an estimate of the original homomorphism count.

The main challenge is estimating these differences with the desired degree dependence. For triangles, a vertex knows the two incident adjacency bits, and randomized response suffices to estimate the closing edge. Already for a four-cycle, the vertex opposite the reporting vertex lies outside its local view. Directly substituting randomized responses for the remaining edges leaves sums over images that are not restricted by true local adjacency bits; because randomized-response noise is present even on nonedges, these sums can introduce an extra factor of $n$ in the variance. Learning a center must therefore be accompanied by a correction procedure that controls this remaining computation. Our construction combines interaction along a spanning tree with random hashing and signs.

\paragraph{Estimating corrections with hashing and signs.}For a replacement on $\{v,u\}$, choose a spanning tree of $H'-v$ rooted at $u$, then add the edge $\{v,u\}$ to obtain a spanning tree of $H'$ rooted at $v$. This makes $u$ the only tree child of $v$, allowing us to first aggregate the contributions from $H'-v$ at $u$ and then combine them with the edge factors incident to $v$. {We orient non-tree edges incident to $v$ away from $v$, so that each party evaluates all factors incident to $v$ using its own adjacency row and public information when $v$ is mapped to it. The remaining non-tree edges can be oriented arbitrarily.} We process the tree from leaves toward the root. For each possible image of a pattern vertex, the corresponding party combines its local edge factors with public reports summarizing the computations already performed at its children. %
For each non-tree edge $e$ directed from $x$ to $y$, use independent uniform signs $\xi_{e,j}\in\{-1,1\}$: the computation at $x$ multiplies the factor for a candidate image $j$ of $y$ by $\xi_{e,j}$, and the computation at $y$ uses the same sign at its own image. Averaging over the signs removes contributions in which these two images disagree.

To control the size of the local sums, we use random hashing to distribute the possible images of each nonroot pattern vertex among $d$ buckets, independently across pattern vertices. For each pattern vertex $x$ and candidate image $i$, party $i$ computes the contribution with $x$ mapped to $i$, using its own edge factor rows and public information. For each tree child of $x$, it sums the corresponding edge factor entries weighted by the child's public reports. For each outgoing non-tree edge of $x$, it instead weights these entries by the random signs. Each sum ranges over candidate images of the other endpoint in the corresponding bucket. Party $i$ then multiplies these edge sums and the signs associated with incoming non-tree edges to form its local contribution.

These local values are normalized and clipped to control sensitivity, with Laplace noise added whenever a private release is needed. Each nonroot vertex uses the bucket containing its own image. At the root $v$, each party uses the residual factor $\bmA-\bmP$ on $\{v,u\}$ and sums over all buckets; the server then aggregates these reports over all possible images of $v$. The complete procedure is given in \Cref{gh:correction-algorithm}.

The key to controlling the error in this computation is to avoid paying a factor of $d$ for every edge. Without hashing, normalizing each local edge sum by its worst-case bound $d$ would accumulate such factors, amplifying the privacy noise when these normalizations are undone. Hashing into $d$ buckets instead bounds all bucket sums by $O(\log n)$ with high probability, so the local normalizations incur only logarithmic factors. Clipping keeps the queries bounded on all graphs and is inactive with high probability on bounded-degree graphs. It remains to account for the contributions discarded by hashing. In the noiseless, unclipped expansion, after averaging over the signs, an assignment survives exactly when the $k'-1$ hash values associated with vertices other than $v$ all agree, which occurs with probability $d^{2-k'}$. Recovering the desired correction in expectation therefore requires a single multiplier $d^{k'-2}$. Up to logarithmic factors, the rescaling cost thus depends on the number of vertices rather than the number of edges. Hashing further allows us to run $d$ independent repetitions without a proportional increase in privacy noise: a public hash check ensures that a changed adjacency bit affects only $O(\log n)$ coordinates of each internal report vector, and clipping bounds the change in each coordinate. Averaging these repetitions reduces the variance by a factor of $d$, thereby reducing the estimation error.

\paragraph{Bounding the correction variance.}To analyze the variance of our correction procedure, condition on the centers and extract the edge factors incident to $v$, which do not depend on the fresh hashes, signs, or reporting noise. Summing over the image of $v$ gives a coefficient $b(\phi_0)$ for each assignment $\phi_0$ of images to the neighbors of $v$. For one repetition, write $U_{\phi_0}$ for the remaining random computation, including the hash indicators, signs, and internal reporting noise. Apart from deterministic rescaling and final scalar noise, the unclipped estimator is their weighted sum, whose variance satisfies
\[
 \Var\!\left(\sum_{\phi_0}b(\phi_0)U_{\phi_0}\right)
 \le \left(\sum_{\phi_0}b(\phi_0)^2\right)
 \max_{\phi_0}\sum_{\psi_0}
 |\Cov(U_{\phi_0},U_{\psi_0})|.
\]
This separates the analysis into controlling the squared coefficients through the center and bounding the covariance of the remaining computation.

For the first factor $\sum_{\phi_0}b(\phi_0)^2$, we express the coefficients as the entries of $(\bmA-\bmP)\bmF$, extending the triangle residual $(\bmA-\bmP)\bmA$. The matrix $\bmF$ combines the factors on the edges incident to $v$ other than $\{v,u\}$. Its rows correspond to images $i$ of $v$, and its columns to assignments $\psi$ of images to the other neighbors $u_1,\ldots,u_r$. Each entry $F_{i,\psi}=\prod_{\alpha=1}^r(\bmW_{\{v,u_\alpha\}})_{i,\psi(u_\alpha)}$ records their joint contribution. Multiplying by $\bmA-\bmP$ incorporates the residual edge and sums over the image of $v$, so the first factor is precisely $\|(\bmA-\bmP)\bmF\|_F^2$. The center construction extends accordingly: independently sketch the incident factors, multiply their sketch entries coordinate by coordinate with the appropriate normalization, and apply the same second linear query and convex fit. A similar analysis gives the required bound on the norm (\Cref{gh:centers}).

To bound the remaining covariance term, we expand pairs of computations into auxiliary graphs. Our assumption that $H'$ has no cut vertex ensures that $H'-v$ is connected. This connectivity, together with sign matching and independence, guarantees that every component contributing to the covariance contains an image fixed by $\phi_0$. We can therefore sum over the remaining images along a forest rooted at these fixed images, paying at most $d$ for each edge without introducing a free root with $n$ possible images. Combining this covariance bound with the center guarantee, and accounting for bucket scaling, averaging, and final scalar noise, yields the desired additive error $\widetilde O_{\veps}(\sqrt n\,d^{k'-2})$ for connected patterns without cut vertices (\Cref{gh:correction-analysis}).

\paragraph{Handling cut vertices.}It remains to handle connected patterns with cut vertices. In fact, the preceding correction analysis still applies whenever deleting the chosen endpoint $v$ preserves connectivity. However, both endpoints of an edge may be cut vertices, so neither can serve directly as the correction endpoint. To handle this case, we organize the replacements using the standard block-cut tree construction (\Cref{lem:block-cut-tree}). Root this tree at any block and process it from the leaves upward. A current leaf block meets the rest of the pattern only at its attachment vertex $v_0$. Choosing the correction endpoint $v\ne v_0$ ensures that deleting $v$ preserves connectivity, as required by the correction procedure.

Once a leaf block is public, the server sums over the images of its vertices other than $v_0$, obtaining a contribution that depends only on the image of $v_0$. This contribution becomes a public vertex weight at $v_0$, allowing the block to be removed before processing its parent. Homomorphism counting makes this elimination possible without tracking overlaps between the images of the removed and remaining vertices. We normalize each weight to lie in $[0,1]$ and keep track of the normalization factors separately. The weights then simply multiply the corresponding rows of $\bmF$ and local computation values, preserving the center and correction guarantees (\Cref{gh:replacement-schedule}). Combining the component estimates, applying quotient-graph inversion, and restoring the isolated-vertex factor gives error $\widetilde O_{\veps}(n^{q-1/2}d^{k-q-1})$.

\paragraph{Scheduling the reports.}Although the replacements are sequential, their reporting rounds overlap. The first-round private sketches depend only on the adjacency rows and fresh public randomness, so they can all be collected together. Once the preceding centers are known, the server incorporates the public factors into the sketch for the next center, which is then learned in one further reporting round. Each correction runs alongside later center constructions. The homomorphism counts for the connected components of all quotient graphs are also estimated in parallel. For a component with $s$ edges, the $j$th center is available by round $j+1$, and its correction needs at most $s-j+1$ further rounds. Every correction therefore finishes by round $s+2\le\ell+2$ (\Cref{gh:counting-analysis}). For some patterns, stopping the replacements early and using local reports for the remaining count gives fewer rounds (\Cref{cor:pc-star-remainder}).

\subsubsection{PC: From Approximate to Pure Privacy}%
\label{overview:pure}
PC's correction reports already satisfy pure privacy; the remaining obstacle is the Gaussian noise used to construct its public centers. A natural approach is to replace each Gaussian coordinate by independent Laplace noise. However, calibrating this noise to the sensitivity of the whole report can introduce an additional dependence on the sketch dimension, losing the degree-dependent accuracy established above. We therefore need a pure-private replacement that preserves the accuracy of these Gaussian reports.

Our key idea is to exploit the particular directions in which a center query can change. Each party holds a binary adjacency row $\bma$ and releases a noisy version of $\bma\bmW$ for a public matrix $\bmW\in\reals^{n\times m}$. Changing one bit shifts this query by a signed row $\pm\bmW_j$. Pure privacy requires a uniform likelihood-ratio bound under these shifts. The Gaussian log-density has unbounded slopes in the tails, so rare outputs prevent such a bound. We refer to its negative log-density, up to an additive constant, as its \emph{potential}. We modify this potential by bounding its slopes along all public sensitive directions, while preserving its quadratic form on a region containing almost all of the Gaussian probability mass.

More concretely, normalize the query and matrix by a public bound $\Delta_{\bmW}>0$ on the Euclidean row norms, and write $\bmD=\bmW/\Delta_{\bmW}$. Let $\gamma$ be a fixed fraction of the report's privacy budget. We restrict the slope vector $\mathbf p$ to the set $\Omega$ defined by $\|\mathbf p\|_\infty\le\gamma$ and $|\bmD_j\mathbf p^\top|\le\gamma$ for every $j\in[n]$, and use the potential
\[
h(\mathbf z)=\max_{\mathbf p\in\Omega}
\left(\mathbf z\mathbf p^\top-\frac{\sigma^2}{2}\|\mathbf p\|_2^2\right),
\]
where $\sigma$ is the Gaussian noise standard deviation on this scale. Without the constraint $\mathbf p\in\Omega$, the maximizer is $\mathbf z/\sigma^2$ and the value is the Gaussian potential $\|\mathbf z\|_2^2/(2\sigma^2)$. The two potentials therefore agree whenever this maximizer is feasible. Meanwhile, the directional constraints $|\bmD_j\mathbf p^\top|\le\gamma$ give $|h(\mathbf z\pm\bmD_j)-h(\mathbf z)|\le\gamma$ for every $\mathbf z$.

To construct a pure-private replacement for the Gaussian report, we use $h$ in place of the Gaussian quadratic potential and define a target density proportional to $e^{-h(\mathbf y-\bma\bmD)}$ on a public cube. The cube contains all possible values of $\bma\bmD$ with a noise margin and is shared by all private inputs for fixed $\bmW$; the final report is $\Delta_{\bmW}\mathbf y$. The bound on $|h(\mathbf z\pm\bmD_j)-h(\mathbf z)|\le\gamma$ ensures that, after normalization over the common cube, the density ratio under neighboring inputs is at most $e^{2\gamma}$. To implement sampling efficiently with approximate evaluations of $h$, we use the sampling framework of \cite{bassily2014differentially} and its extension to approximate function evaluations in \cite{lowy2026differentially}. These results give a small \emph{pointwise multiplicative} sampling error, meaning that the ratio between the implemented and target densities is uniformly close to one at every output. We reserve enough privacy margin to accommodate this error, so that the implemented report also satisfies pure privacy (\Cref{gh:pure-query-algorithm}).

We next show that the new reports remain close to their Gaussian counterparts, allowing PC to retain its accuracy guarantee. Choosing $\sigma$ sufficiently large ensures that, for Gaussian noise $\mathbf z$, the unconstrained maximizer $\mathbf z/\sigma^2$ lies in $\Omega$ with high probability. Thus $h$ agrees with the Gaussian quadratic potential on a region containing almost all of the Gaussian probability mass. On this region, the modified and Gaussian unnormalized densities agree. Their normalized densities may still differ because the modified tails change the total mass. We use the standard Gaussian isoperimetric inequality (\Cref{gh:gaussian-enlargement}) to bound this additional tail mass and the resulting change in normalization. Combining this bound with the errors from restriction to the public cube and approximate sampling gives a small total-variation distance between the pure and Gaussian reports.

We formalize the construction and analysis above in \Cref{gh:pure-linear-query}, proving that Gaussian releases of linear queries on binary data admit pure-private replacements with an explicit bound on total variation distance (see \Cref{def:tv-dis}). Applying this result to each Gaussian release in PC converts its approximate-privacy guarantee to pure privacy. Choosing the error of each replacement sufficiently small keeps the final output distribution close to that of the Gaussian version, preserving its accuracy guarantee. The replacements also preserve the reporting schedule and can be implemented in worst-case polynomial time, yielding \Cref{gh:pure-main-theorem}.

\subsubsection{Lower Bounds via Private Gap Testing}
\label{overview:lower}
A natural starting point is the triangle-counting reduction of~\cite{eden2025triangle} from private summation, where each party holds one bit and the goal is to estimate their sum under local differential privacy. Their construction encodes each private bit as the presence of an edge in a matching and connects all matching endpoints to a common set of public neighbors. Every present private edge creates the same number of triangles, and no triangle contains more than one private edge. The triangle count is therefore a known multiple of the bit sum, so an additive approximation to the triangle count can be rescaled into an additive approximation to the sum.%

To adapt this construction to a degree bound, we use $\Theta(n/d)$ connected components of $\Theta(d)$ vertices each. Within each connected component, a public clique is joined to a disjoint set of vertices partitioned into pairs. Each private bit determines whether the edge between its corresponding pair is present. The sizes are chosen so that the maximum degree is at most $d$, while the construction encodes $N=\Theta(n)$ private bits. The public cliques allow us to extend each present private edge to copies of any fixed pattern $H$. The single-edge pattern is a special case, for which a private matching suffices.

For general $H$, however, the count need not be a function of the bit sum. A copy may contain several private edges, and the total number of copies can depend on how the present private edges are distributed among the connected components of the constructed graph.
Thus, the count need not have a fixed linear relation to the bit sum, so the rescaling used for triangles does not directly extend to general patterns.

We instead reduce from \emph{gap testing}: distinguish the all-zero input from inputs containing at least $r$ ones. This requires only a lower bound on the increase over the count in the public graph obtained when all private edges are absent. Fix one edge of $H$ and map it to a present private edge. The other vertices of its component in $H$ can be mapped into the public clique in the connected component containing that private edge, while each remaining component of $H$ can be mapped into any public clique, with all images chosen to be distinct. This gives $\Omega(n^{q-1}d^{k-q-1})$ new copies using exactly that one private edge. These copies are distinct for different private edges, so $f_H(G_x)-f_H(G_0)\ge\Lambda\sum_i x_i$ for some $\Lambda=\Theta(n^{q-1}d^{k-q-1})$, where $G_0$ is the public graph obtained when all bits are zero. Additional copies involving several private edges only increase this gap.

We prove that this gap-testing problem requires $r=\Omega(\sqrt N/\veps)$ at fixed privacy and sufficiently small $\delta$, even under arbitrary interaction (\Cref{lem:bit-sum-testing}). Our proof compares the all-zero input with a random input consisting of independent Bernoulli bits whose expected sum is a sufficiently large constant multiple of $r$. Since the random input contains at least $r$ ones with high probability, a gap-testing algorithm would also distinguish these two input distributions. Applying the local hypothesis-testing lower bound of~\cite{joseph2019role} then yields the claimed bound on $r$.

Finally, we reduce gap testing to $H$-counting. Suppose an LEDP algorithm estimates the $H$-count with additive error $\alpha$. Each party simulates the two vertices corresponding to its bit, and the server simulates the remaining public vertices. Since changing one bit changes exactly one edge, the reduction preserves privacy under arbitrary interaction. Comparing the counter's output with $f_H(G_0)+\alpha$ distinguishes the two testing cases whenever $\Lambda r>2\alpha$. The testing lower bound therefore yields $\alpha=\Omega(\Lambda\sqrt N/\veps)=\Omega(n^{q-1/2}d^{k-q-1}/\veps)$, matching PC's polynomial dependence on $n$ and $d$. 

\subsection{Related Work}
\paragraph{LEDP subgraph counting.}%
For stars, \cite{imola2021locally} exploit the fact that each vertex can determine the number of copies centered at it from its own adjacency vector. Summing noisy versions of these local counts gives a one-round algorithm with RMSE $O(\sqrt n\,d^{r-1}/\veps)$ for an $r$-leaf star under a public degree bound $d$. For triangles, their two-round algorithm combines true incident edges with randomized reports of the closing edge, achieving RMSE $O_{\veps}(\sqrt n\,d^{3/2})$. The follow-up work~\cite{imola2022communication} reduces communication by sampling edge reports and selectively downloading those needed for local computations. For triangle counting, \cite{eden2025triangle} prove an $\Omega_\veps(n^2)$ noninteractive lower bound and an $\Omega_\veps(n^{3/2})$ lower bound under arbitrary interaction, for fixed sufficiently small $\veps$. They also give tight variance bounds for the debiased randomized response estimator. For a fixed pair of vertices on the same side of a bipartite graph, \cite{he2024common} give a two-round unbiased estimator of their common-neighbor count with RMSE $O_{\veps}(\sqrt d)$ when the maximum degree is at most $d$.

Subsequent work improves accuracy by exploiting graph structure and extends counting to more general patterns. For triangles and odd cycles, \cite{hillebrand2025cycle} order vertices by their noisy reported degrees, obtaining triangle RMSE $\widetilde O_{\veps}(\sqrt n(a^{3/2}+\sqrt{ad}+1))$, where $a$ is the degeneracy, the largest minimum degree among nonempty subgraphs. Further refinements of triangle estimation use more entries of the noisy adjacency matrix and tighter global sensitivity bounds~\cite{he2025robust}, compress edge reports through hashing~\cite{hillebrand2025hash}, or combine randomized edge reports with a private vertex ordering that limits the number of later neighbors, giving error bounds in terms of degeneracy~\cite{mundra2025practical}. For arbitrary fixed patterns, \cite{suppakitpaisarn2025graphlets} sum products of debiased edge reports, achieving noninteractive RMSE $O_{\veps}(n^{k-1})$, with a matching noninteractive lower bound for cliques on at least three vertices and interactive lower bounds for certain cycles. To count trees, \cite{hu2026acyclic} combine counts of smaller patterns over multiple rounds and assign each vertex of $G$ a random position in the pattern to enforce distinct images, obtaining additive error $\widetilde O_{H,\veps}(\sqrt n\,d^{k-1})$ for trees on $k$ vertices. Their length-three walk estimator, combined with corrections for repeated vertices, also yields $\widetilde O_{\veps}(\sqrt n\,d^2)$ for the four-vertex path.

\paragraph{Interaction in local privacy.}
The study of~\cite{joseph2019role} distinguishes local algorithms that contact each party once from those that may return to the same party, and proves hypothesis-testing lower bounds under arbitrary interaction. For pointer-chasing problems, in which each answer determines the next location to query, \cite{joseph2020exponential} show that one additional round of interaction can exponentially reduce the sample complexity. \cite{dagan2020interaction} show that interaction can be necessary to achieve polynomial sample complexity in learning and stochastic convex optimization, and can improve estimation of a mean with few nonzero coordinates~\cite{acharya2022role}.

\paragraph{Other privacy models.}
Central algorithms for counting triangles~\cite{nissim2007smooth} and other small patterns~\cite{karwa2014private} calibrate noise to sensitivity bounds that adapt to the input graph. \cite{chen2026range} study subgraph counts in induced subgraphs selected by ranges of public vertex attributes, giving central algorithms and nearly matching lower bounds for general fixed patterns. Under node privacy, which protects a vertex together with all its incident edges, \cite{blocki2013restricted,kasiviswanathan2013graphs} control sensitivity through restrictions such as bounded degree, while \cite{chen2013recursive} give algorithms for general subgraph counting. Other decentralized models allow parties to observe edges beyond those incident to themselves or use a shuffler to hide the association between reports and their senders. The algorithms of~\cite{sun2019extended} count subgraphs from extended local views, in which a party can observe edges beyond those incident to itself. In the shuffle model, a trusted shuffler hides the association between randomized reports and their senders. Using such reports about paths of length two, \cite{imola2022shuffle} estimate triangles and four-cycles; Fang and Yi~\cite{fang2025shuffle} improve these estimates and also count paths of length three. More recently, \cite{raskhodnikova2026localnode} introduce local node privacy and give algorithms for degree statistics, and \cite{raskhodnikova2026subgraphnode} give algorithms and lower bounds for subgraph counting in the noninteractive setting.

\paragraph{From approximate to pure privacy.}
The conversion of~\cite{bun2018heavy} requires privacy under arbitrary input replacements, which would increase the noise scale and worsen our accuracy bounds. The more general purification method of~\cite{lin2026purifying} uses randomized post-processing on a bounded output domain, but its dependence on the output range and $\delta^{1/m}$ for $m$-dimensional outputs do not directly preserve our accuracy guarantees as the sketch dimension grows. The Huber mechanism~\cite{gowtham2022introducing} modifies Gaussian tails to obtain pure privacy, but its calibration to $\ell_1$-sensitivity can also introduce additional dimension dependence.

\subsection{Organization}

\Cref{sec:preliminaries} introduces the notation and background used throughout the paper. \Cref{sec:two-round} presents and analyzes our two-round $\veps$-LEDP algorithm LNC for counting arbitrary fixed patterns on general graphs. \Cref{gh:main} develops our algorithm PC for counting arbitrary fixed patterns by learning public predictions and correcting their errors, and establishes its degree-dependent accuracy guarantee. Finally, \Cref{sec:degree-lower} proves a lower bound that holds under arbitrary interaction.

\section{Preliminaries}\label{sec:preliminaries}

For any $x,\ell',r'\in\reals$, we use $\Clip_{[\ell',r']}(x)$ to denote the clipping of $x$ to the interval $[\ell',r']$, i.e., $\Clip_{[\ell',r']}(x)=\min(r',\max(\ell',x))$. Let $[x]_+ := \max\{x,0\}$ denotes the positive part of $x$.

\subsection{Graph-Theoretic Notation and Tools}
\label{ssec:graph-notation}

We only consider undirected simple graphs. Let $\calG_{n,d}$ denote the family of all graphs on vertex set $[n]$ with maximum degree at most $d$. For a graph $K$, write $N_K(v)=\{u\in V(K):\{u,v\}\in E(K)\}$ and $\Delta(K)=\max_{v\in V(K)}|N_K(v)|$.

\paragraph{Subgraph Counting.}We retain $G=([n],E)$ for the input graph and $\bmA=(A_{i,j})_{i,j\in[n]}\in\{0,1\}^{n\times n}$ for its adjacency matrix. For each $i\in[n]$, let $\bmA_i\in\{0,1\}^n$ denote the $i$-th row of $\bmA$, which we refer to as the adjacency vector of vertex $i$. Let $d_i:=\sum_{j=1}^n A_{i,j}$ denote the degree of vertex $i$. Let $H$ be a fixed graph with $q$ connected components, $k:=|V(H)|$ vertices and $\ell:=|E(H)|$ edges. Let $\Aut(H)$ denote the automorphism group of $H$, i.e., the set of all permutations $\sigma$ of $V(H)$ such that, for every $u,v\in V(H)$, $\{u,v\}\in E(H)$ if and only if $\{\sigma(u),\sigma(v)\}\in E(H)$. We denote by $f_H(G)$ the number of non-induced copies of $H$ contained in $G$. Equivalently,
$$
f_H(G):=\frac{1}{|\Aut(H)|}
\sum_{\phi:V(H)\hookrightarrow[n]}
\prod_{\{u,v\}\in E(H)}
A_{\phi(u),\phi(v)},
$$
where the sum ranges over all injections from $V(H)$ to $[n]$.

Given an arbitrary matrix $\bmX=(X_{i,j})_{i,j\in[n]}$, we use the following shorthand notation. For an undirected edge $e=\{i,j\}$, let $X_e:=X_{\min(i,j),\max(i,j)}$ so that $X_e$ always refers to the corresponding upper-triangular entry of $\bmX$. For an injection $\phi:V(H)\hookrightarrow[n]$ and an edge $e=\{u,v\}\in E(H)$, let $\phi(e):=\{\phi(u),\phi(v)\}$. For a set $S\subseteq E(H)$, let $\phi(S):=\{\phi(e)\mid e\in S\}$. For an edge set $E'$, let $V(E')$ denote the set of vertices incident to the edges in $E'$. The operation $\operatorname{vec}$ lists all entries of a matrix or tensor in lexicographic order as a row vector. 

\paragraph{Quotient Graphs and M\"obius Inversion.}
For any graph $F$, let $\hom(F,G)$ denote the number of homomorphisms from $F$ to $G$, i.e.,
$$
\hom(F,G):=
\sum_{\phi:V(F)\to[n]}
\prod_{\{u,v\}\in E(F)}
A_{\phi(u),\phi(v)},
$$
where the maps $\phi$ need not be injective. 
Let $\Pi(H)$ denote the set of all partitions $\pi$ of $V(H)$ whose parts are independent sets in $H$.
For each $\pi\in\Pi(H)$, let $H/\pi$ denote the quotient graph of $H$ under $\pi$, whose vertices are the parts of $\pi$, with two distinct parts adjacent if $H$ contains an edge between them. 
\begin{lemma}[Quotient Graph Inversion {\cite{lovasz2012large}}]
\label{lemma:quotient-inversion}
Let $H$ and $G$ be two graphs. For each $\pi\in\Pi(H)$, let $\mathcal H_\pi$ denote the set of connected components of $H/\pi$. Then
\begin{align*}
f_H(G)
&=\frac{1}{|\Aut(H)|}\sum_{\pi\in\Pi(H)}\mu(\pi)\hom(H/\pi,G)\\
&=\frac{1}{|\Aut(H)|}\sum_{\pi\in\Pi(H)}\mu(\pi)\prod_{H'\in\mathcal H_\pi}\hom(H',G),
\end{align*}
where $\mu(\pi) := \prod_{V'\in\pi}(-1)^{|V'|-1}(|V'|-1)!$.
\end{lemma}

The second equation holds because homomorphisms on distinct connected components can be chosen independently; their images in $G$ need not be disjoint.

\paragraph{Block-Cut Tree.} A vertex $v$ of a graph $G = (V, E)$ is called a cut vertex if removing $v$ increases the number of connected components. A block of $G$ is a maximal connected subgraph with no cut vertex. The block-cut tree of a connected graph $G$ is a tree whose vertices correspond to the blocks and cut vertices of $G$, with a block adjacent to a cut vertex if and only if the block contains that vertex. The following lemma states that the block-cut tree can be constructed in linear time.

\begin{lemma}[Construction of a Block-Cut Tree, \cite{tarjan1972depth}]\label{lem:block-cut-tree}
The block-cut tree of a connected graph $G=(V,E)$ can be constructed in $O(|V|+|E|)$ time.
\end{lemma}

\subsection{Linear Algebra}
\label{ssec:linear-algebra}

Let $\bmI_t$ be the $t\times t$ identity matrix and, for $a\in[t]$, let $\bme_a\in\reals^t$ be the row vector with a one in coordinate $a$ and zeros elsewhere. For a row vector $\bmx=(x_1,\ldots,x_t)\in\reals^t$ and $1\le p<\infty$, its $\ell_p$ norm is $\|\bmx\|_p:=\left(\sum_{i=1}^t|x_i|^p\right)^{1/p}$, and $\|\bmx\|_\infty:=\max_{i\in[t]}|x_i|$. For a matrix $\bmW\in\reals^{s\times t}$, its Frobenius norm is $\|\bmW\|_F:=\sqrt{\sum_{i=1}^s\sum_{j=1}^{t}W_{i,j}^2}$. Writing $\bmW_i$ for its $i$th row, we have $\|\bmW\|_F^2=\sum_{i=1}^s\|\bmW_i\|_2^2$.  The Frobenius norm is also submultiplicative: for any two matrices $\bmW$ and $\bmW'$ of compatible dimensions, we have $\|\bmW\bmW'\|_F\le\|\bmW\|_F\|\bmW'\|_F$. This follows by applying the Cauchy--Schwarz inequality to each pair consisting of a row of $\bmW$ and a column of $\bmW'$, and then summing over all entries.

\subsection{Probabilistic and Optimization Tools}
For a randomized estimate $\widehat{f}_H(G)$ of $f_H(G)$ on a fixed input graph $G$, its \emph{root mean square error (RMSE)} is
\begin{equation}
\label{eq:rmse}
\operatorname{RMSE}(\widehat{f}_H;G)
:=\sqrt{\E\!\left[(\widehat{f}_H(G)-f_H(G))^2\right]},
\end{equation}
where the expectation is over the randomness of the estimator.

For two probability distributions $\calP$ and $\calQ$ over a common domain $\Omega$ with probability density functions $p$ and $q$, respectively, their total variation distance is defined as 
\begin{align}
d_{\mathrm{TV}}(\calP,\calQ)=\frac12\int_{\Omega}|p(x)-q(x)|\,\mathrm{d}x=\int_{\Omega}[p(x)-q(x)]_+\,\mathrm{d}x. \label{def:tv-dis}
\end{align}
We will use the following probabilistic and optimization tools.

\begin{lemma}[Gaussian Norm Tail Bound {\cite[Example~2.28]{wainwright2019high}}]
\label{lemma:gaussian-norm-tail}
Let $\bmg\sim\calN(0,\sigma^2\bmI_D)$, where $\sigma>0$. Then, for every $t>0$,
\[
\Pr\left[\|\bmg\|_2\ge(1+t)\sigma\sqrt D\right]\le\exp\left(-\frac{Dt^2}{2}\right).
\]
\end{lemma}

\begin{lemma}[Gaussian Isoperimetric Inequality {\cite[Theorem~3.1, Eq.~(3.1)]{borell1975brunn}}]
\label{gh:gaussian-enlargement}
Let $\bmg\sim\calN(0,\bmI_m)$ and let $K\subseteq\reals^m$ be a nonempty closed set. Write $\Phi$ for the standard Gaussian distribution function and $\operatorname{dist}(\bmx,K):=\inf_{\mathbf y\in K}\|\bmx-\mathbf y\|_2$. If $\Pr[\bmg\in K]\ge\Phi(b)$ for some $b\in\reals$, then, for every $t\ge0$,
\[
\Pr[\operatorname{dist}(\bmg,K)>t]\le1-\Phi(b+t).
\]
\end{lemma}

\begin{lemma}[Ellipsoid Method, Theorem 5.2.1 in \cite{ben2021lectures}]
\label{lemma:ellipsoid-method}
Let $K\subseteq\reals^N$, with $N\ge2$, be a closed convex set contained in the ball $E_0 = \{\bmx:\|\bmx\|_2\le R\}$ and containing a Euclidean ball of radius $r>0$. Let $f:\reals^N\to\reals$ be convex and differentiable. Suppose that we have access to a separation oracle that tests membership in $K$ and returns a separating hyperplane for any point outside $K$, and an oracle that evaluates $f$ and its gradient at any point. Then, for every $\eta>0$, the ellipsoid method returns a point $\widehat{\bmx}$ satisfying $\widehat{\bmx}\in K$ and $f(\widehat{\bmx})-\min_{\bmx\in K}f(\bmx)\le\eta$ after at most
\[
\left\lceil
2N^2\left(
\ln\!\left(\frac{R}{r}\right)
+\ln\!\left(
\frac{\eta+\max_{\bmx \in E_0}f(\bmx)-\min_{\bmx \in E_0}f(\bmx)}{\eta}
\right)
\right)
\right\rceil+1
\]
iterations. Each iteration uses at most one call to each oracle and $O(N^2)$ additional arithmetic operations.
\end{lemma}

\subsection{Differential Privacy}

We begin with the standard definition of differential privacy. 

\begin{definition}[Differential Privacy \cite{dwork2006calibrating}]
Let $\veps > 0$ and $\delta \in [0,1)$. A randomized algorithm $\calM$ is $(\veps,\delta)$-\emph{differentially private (DP)} if, for every pair of neighboring datasets $D \sim D'$ and every measurable event $S$ in the output space of $\calM$,
$$
\Pr[\calM(D) \in S] \le e^\veps \Pr[\calM(D') \in S] + \delta.
$$
When $\delta = 0$, we say that $\calM$ is $\veps$-\emph{differentially private}.
\end{definition}

We will make use of the following properties of differential privacy.

\begin{lemma}[Basic Composition Theorem \cite{dwork2006calibrating}]\label{lemma:basic composition}
If $\calM_i$ is an $(\veps_i,\delta_i)$-DP algorithm for each $i\in[t]$, then their composition is $(\sum_{i=1}^t \veps_i,\sum_{i=1}^t \delta_i)$-DP.
\end{lemma} 

\begin{lemma}[Post-processing Property \cite{dwork2006calibrating}]\label{lemma:post-processing}
    Let $\calM$: $\reals^{d_1} \rightarrow \reals^{d_2}$ be an $(\veps,\delta)$-DP algorithm and let $f$: $\reals^{d_2} \rightarrow \reals^{d_3}$ be an arbitrary function. Then, the function $f \circ \calM$: $\reals^{d_1} \rightarrow \reals^{d_3}$ is also $(\veps,\delta)$-DP.
\end{lemma}

A common way to release a function’s value under differential privacy is to use noise scaled to its sensitivity. For $b>0$, let $\Lap(b)$ denote the Laplace distribution with density $p(z)=\exp(-|z|/b)/(2b)$, mean $0$, and variance $2b^2$. Let $\calN(\mu,\sigma^2)$ denote the Gaussian distribution with mean $\mu$ and variance $\sigma^2$, and let $\calN(\mu,\sigma^2 \bmI_m)$ denote the distribution of a row vector with $m$ independent coordinates drawn from $\calN(\mu,\sigma^2)$.

\begin{lemma}[Laplace Mechanism \cite{dwork2006calibrating}]
\label{lemma:Laplace mechanism}
Let $\veps>0$ and $\Delta>0$, and suppose that $f:\calX\rightarrow\reals^t$ has $\ell_1$-sensitivity at most $\Delta$, that is, $\|f(x)-f(x')\|_1\le\Delta$ for every neighboring pair $x\sim x'$ in $\calX$. The mechanism that outputs $M(x)=f(x)+(Y_1,\dots,Y_t)$, where $Y_1,\dots,Y_t$ are independent samples from $\Lap(\Delta/\veps)$, is $\veps$-DP.
\end{lemma}

\begin{lemma}[Gaussian Mechanism \cite{dwork2006our}]
\label{lemma:Gaussian mechanism}
Let $\veps,\delta\in(0,1)$ and $\Delta>0$, and suppose that $f:\calX\rightarrow\reals^t$ has $\ell_2$-sensitivity at most $\Delta$, that is, $\|f(x)-f(x')\|_2\le\Delta$ for every neighboring pair $x\sim x'$ in $\calX$. For any $\sigma\ge(\Delta/\veps)\sqrt{2\ln(2/\delta)}$, the mechanism that outputs $M(x)=f(x)+(Y_1,\dots,Y_t)$, where $Y_1,\dots,Y_t$ are independent samples from $\calN(0,\sigma^2)$, is $(\veps,\delta)$-DP.
\end{lemma}

For graph datasets, we use the edge-neighboring relation. An algorithm satisfying differential privacy for every edge-neighboring graph pair is called \emph{edge DP}.

\begin{definition}[Edge-Neighboring Graphs]
Two graphs $G$ and $G'$ are \emph{edge-neighboring}, denoted by $G \sim G'$, if they differ in exactly one edge.
\end{definition}

\paragraph{Local Differential Privacy}
We first recall the local model \cite{joseph2019role}. Consider a distributed input $\bmx=(x_1,\ldots,x_n)$, where party $i\in[n]$ holds $x_i$ as its private input. A local algorithm interacts with the parties only through \emph{local randomizers}, each of which is a DP algorithm and acts on the private input of a single party.

An interactive local algorithm proceeds in rounds. Let $\Pi_t$ denote the transcript after round $t$, with $\Pi_0$ being the empty transcript. At round $t$, based on $\Pi_{t-1}$, the algorithm may select a collection of parties, assign local randomizers to the selected parties, and specify the privacy parameters of these randomizers. Each selected party applies its assigned randomizer to its private input and sends the resulting message to the algorithm. The choices made at round $t$ may be arbitrary randomized functions of $\Pi_{t-1}$. The transcript records all information generated during the interaction, including the selected parties, the assigned local randomizers and privacy parameters, and the messages returned by the parties. We write $\Pi(\bmx)$ for the complete transcript produced on input $\bmx$.

\begin{definition}[Local Differential Privacy]
Let $\veps>0$ and $\delta\in[0,1)$. A randomized local algorithm $\mathcal{A}$ is $(\veps,\delta)$-\emph{locally differentially private (LDP)} if the randomized algorithm that outputs its complete transcript $\Pi(\bmx)$ is $(\veps,\delta)$-DP. When $\delta=0$, we say that $\mathcal{A}$ is $\veps$-LDP.
\end{definition}

We consider the local model for graph data \cite{eden2025triangle}. For an $n$-vertex graph $G$, each vertex $i\in[n]$ is treated as a party and holds its adjacency vector $\bmA_i\in\{0,1\}^n$ as its private input. In this setting, we use the edge-neighboring relation on distributed graph inputs.

\begin{definition}[Local Edge Differential Privacy]\label{def:ledp}
Let $\veps>0$ and $\delta\in[0,1)$. A randomized local algorithm $\mathcal{A}$ is $(\veps,\delta)$-\emph{locally edge differentially private (LEDP)} if the randomized algorithm that outputs its complete transcript $\Pi(G)$ is $(\veps,\delta)$-DP with respect to the edge-neighboring relation. When $\delta=0$, we say that $\mathcal{A}$ is $\veps$-LEDP.
\end{definition}

In the graph setting, a local randomizer used by a party acts on its adjacency vector.

\begin{definition}[Local Randomizer]
Let $\veps>0$ and $\delta\in[0,1)$. An $(\veps,\delta)$-\emph{local randomizer} is a randomized mapping $\calR:{0,1}^n\to\calY$ such that, for every pair $\bma,\bma'\in\{0,1\}^n$ differing in exactly one coordinate and every measurable set $S\subseteq\calY$,
$$
\Pr[\calR(\bma)\in S]
\le
e^\veps\Pr[\calR(\bma')\in S]+\delta.
$$
When $\delta=0$, we simply call $\calR$ an $\veps$-local randomizer.
\end{definition}

An LEDP algorithm is called \emph{noninteractive} if it consists of a single round in which the local randomizers are fixed independently of any previous messages. Otherwise, it is called \emph{interactive}. Unless stated otherwise, the LEDP algorithms considered in this paper may be fully interactive and may use an arbitrary number of rounds.

Many DP algorithms in the local model are built from the randomized response primitive of \cite{warner1965randomized,kasiviswanathan2011can}.

\begin{definition}[Randomized Response]
Given a privacy parameter $\veps > 0$ and a $t$-bit vector $\bma$, the algorithm $\RR_{\veps}(\bma)$ outputs a  $t$-bit vector, where each bit is randomized independently: for each $i \in [t]$, bit $i$ is $a_i$ with probability $e^{\veps}/(e^{\veps}+1)$ and $1-a_i$ otherwise.
\end{definition}

\begin{lemma}\label{lemma:RR}
Randomized Response is an $\veps$-local randomizer.
\end{lemma}

The standard method for debiasing randomized response is analyzed, e.g., in \cite{eden2025triangle}.

\begin{lemma}\label{lemma:debiased-RR}
Given output $X = \RR_{\veps}(b)$, the debiased estimate $Y = (X(e^{\veps}+1)-1)/(e^{\veps}-1)$ is $\veps$-local randomizer and has expectation $\mathbb{E}[Y] = b$. Moreover, $\operatorname{Var}[Y] = e^{\veps}/(e^{\veps}-1)^2 = \Theta(1/\veps^2)$ for any $\veps \in (0, 1]$.
\end{lemma}

\section{LNC: The General Upper Bound}
\label{sec:two-round}

In this section, we present the algorithm LNC for counting an arbitrary fixed subgraph $H$ in \Cref{alg:interactive-correction-subgraph}. In the pseudocode, $[s]$ marks operations performed by the server, and $[v_i]$ marks operations performed by the party corresponding to vertex $i$. Recall the counting polynomial $P$ defined in \Cref{def:P(X)}, for which $P(\bmA)=f_H(G)$. For $i<j$, its partial derivative is
\begin{align}\label{eq:partialP}
    \frac{\partial P(\bmX)}{\partial X_{i,j}} = \frac{1}{|\Aut(H)|}\sum_{e \in E(H)}
        \sum_{\substack{
        \phi:V(H)\hookrightarrow[n]\\
        \phi(e)=\{i,j\}
        }}
        \prod_{e'\in E(H)\setminus \{e\}}X_{\phi(e')}.
\end{align}

\setlength{\algomargin}{5mm}
\begin{algorithm}[t]
    \SetAlgoLined

    \KwData{
        Graph $G=([n],E)$ represented by adjacency row vectors
        $\bmA_1,\ldots,\bmA_n\in\{0,1\}^n$, pattern graph $H$ with $k$ vertices and $\ell$ edges, and privacy budget $\veps>0$.
    }

    \KwResult{
        A private estimate $\hf_H(G)$ of $f_H(G)$.
    }

    \tcc{First round: randomized response}

    $[s]\;$Let $\veps_0 \leftarrow \veps/2$ and $\kappa \leftarrow e^{\veps_0}/(e^{\veps_0} - 1)$\;
    
    \For{$i\gets 1$ \KwTo $n$}{
        \For{$j\gets i+1$ \KwTo $n$}{
            $[v_i]\;Y_{i,j}\leftarrow \RR_{\veps_0}(A_{i,j})$\;
        }
        $[v_i]\;$Upload $(Y_{i,i+1},\ldots,Y_{i,n})$ to the server\;
    }

    \For{$i\gets 1$ \KwTo $n$}{
        \For{$j\gets i+1$ \KwTo $n$}{
        $[s]\;X_{i,j}\leftarrow ((e^{\veps_0} + 1)Y_{i,j}-1)/(e^{\veps_0} - 1)$\;
    }
    }
    \tcc{Second round: cancellation of the linear RR error}

    \For{$i\gets 1$ \KwTo $n$}{
        $[v_i]\;Q_i \leftarrow \sum_{j = i + 1}^{n}(A_{i,j} - X_{i,j})(\partial P(\bmX))/(\partial X_{i,j})$\;
        $[v_i]\;M_i \leftarrow Q_i + \Lap\left(2\ell (n-2)!\kappa^{\ell-1}/(|\Aut(H)|(n-k)!\veps_0)\right)$\;

        [$v_i$] Upload $M_i$ to the server\;
    }

    $[s]\;\hf_H(G)\leftarrow P(\bmX) + \sum_{i=1}^n M_i$\;

    \KwRet{$\hf_H(G)$.}

    \caption{
        Linear Noise Cancellation (LNC) for subgraph counting.
    }
    \label{alg:interactive-correction-subgraph}
\end{algorithm}

\begin{proof}[Proof of \Cref{thm:interactive-upper-bound-subgraph}]
We prove LNC's privacy and accuracy guarantees separately.
\paragraph{Privacy.}
By \Cref{lemma:RR,lemma:post-processing}, the first round is $\veps/2$-LEDP.

Now condition on an arbitrary fixed first-round transcript $X$. Note that $|X_{i,j}| \le \kappa = O(1/\veps)$ and $|X_{i,j} - A_{i,j}| \le \kappa = O(1/\veps)$ for any $1 \le i < j \le n$. Then the sensitivity of $\bmQ(G) := (Q_1, Q_2, \dots, Q_n)$ satisfies 
\begin{align*} 
    \Delta_Q := \max_{G\sim G'}\|\bmQ(G) - \bmQ(G')\|_1 = \max_{1\le i< j\le n}\left|\frac{\partial P(\bmX)}{\partial X_{i,j}}\right|,
\end{align*}
Then we can bound $\Delta_Q$ by bounding $|(\partial P(\bmX))/(\partial X_{i,j})|$. Fix an edge $e=\{u,v\}\in E(H)$. There are two possible ways to map its endpoints to $i$ and $j$. Once these two images are fixed, the remaining $k-2$ vertices of $H$ can be mapped injectively to the remaining $n-2$ vertices in $(n-2)!/(n-k)!$ ways. Summing over the $\ell$ choices of $e$, the derivative $(\partial P(\bmX))/(\partial X_{i,j})$ contains at most $2\ell(n-2)!/(n-k)!$ monomials. Moreover, each monomial contains $\ell-1$ factors, each of absolute value at most $\kappa$, and hence has absolute value at most $\kappa^{\ell-1}$. Therefore,
$$
\Delta_Q = \max_{1\le i< j\le n}\left|\frac{\partial P(\bmX)}{\partial X_{i,j}}\right| \le \frac{2\ell (n - 2)!\kappa^{\ell-1}}{|\Aut(H)|(n - k)!}.
$$
By \Cref{lemma:Laplace mechanism}, the second round is $(\veps/2)$-LEDP. Basic composition therefore gives $\veps$-LEDP for LNC (\Cref{alg:interactive-correction-subgraph}).

\paragraph{Accuracy.} We first show that LNC outputs an unbiased estimator of $f_H(G)$. 

\begin{lemma}\label{claim:unbiased-H}
    $\E[\hat{f}_H(G)] = f_H(G)$. 
\end{lemma}

\begin{proof}
    Note that $\E[X_{i,j}] = A_{i,j}$. Then we have
    \begin{align*}
    \E[{P(\bmX)}] =& \frac{1}{|\Aut(H)|}\sum_{\phi:V(H)\hookrightarrow [n]}\prod_{e\in E(H)}\E\left[X_{\phi(e)}\right] \\
    =& \frac{1}{|\Aut(H)|}\sum_{\phi:V(H)\hookrightarrow [n]}\prod_{e\in E(H)}A_{\phi(e)} = f_H(G).
    \end{align*}
    Therefore, 
    \begin{align*}
    \E[\hat{f}_H(G)] =& \E[P(\bmX)]+\sum_{i = 1}^{n} \left(\E[Q_i] + \E\left[\Lap\left(\dfrac{2\ell (n-2)!\kappa^{\ell-1}}{|\Aut(H)|(n-k)!\veps_0}\right)\right]\right)  \\
    =& \E[P(\bmX)] + \sum_{i = 1}^{n}\left(\sum_{j = i + 1}^{n}(A_{i,j} - \E[X_{i,j}])\E\left[\frac{\partial P(\bmX)}{\partial X_{i,j}}\right] + 0\right) \\
    =& \E[P(\bmX)] = f_H(G),
    \end{align*}
    where the second inequality follows from the fact that the Laplace distribution is unbiased and $(\partial P(\bmX))/(\partial X_{i,j})$ is independent of $X_{i,j}$. 
\end{proof}

Then we bound the variance of $\hf_H(G)$.

\begin{lemma}\label{claim:variance-H}
    $\Var[\hf_H(G)] = O\left(n^{2k- 3}/\veps^{2\ell}\right)$.
\end{lemma}

\begin{proof}
    For every pair $(i, j) \in [n]^2$, define $Z_{i,j} = X_{i,j} - A_{i,j}$. Then we have $\E[Z_{i,j}] = 0$ and $V:= \Var[Z_{i,j}] = \Var[X_{i,j}] = e^{\veps_0}/(e^{\veps_0} - 1)^2 = O\left(1/\veps^2\right)$ by \Cref{lemma:debiased-RR}.
    
    Let $\beta:= P(\bmX) + \sum_{i = 1}^{n}Q_i - f_H(G)$ and $\alpha:= \sum_{i = 1}^{n}\Lap(2\ell(n-2)!\kappa^{\ell-1}/(|\Aut(H)|(n-k)!\veps_0))$. Then $\Var\left[\hf_H(G)\right] = \Var[\beta] + \Var[\alpha]$.

    We first consider $\Var[\beta]$. By \Cref{claim:unbiased-H} and $\E[\alpha]=0$, we have $\E[\beta]=0$, and hence $\Var[\beta]=\E[\beta^2]$. Fix an injection $\phi: V(H) \hookrightarrow [n]$. Its contribution to $\beta$ is
    \begin{align*}
      &\frac{1}{|\Aut(H)|}\left(\prod_{e\in E(H)}\left(A_{\phi(e)} + Z_{\phi(e)}\right) - \sum_{e\in E(H)}Z_{\phi(e)}\prod_{e' \in E(H) \setminus \{e\}}\left(A_{\phi(e')} + Z_{\phi(e')}\right)-\prod_{e\in E(H)}A_{\phi(e)}\right) \\
      =& \frac{1}{|\Aut(H)|}\left(\sum_{\varnothing \subsetneq S\subseteq E(H)}\prod_{e\in S}Z_{\phi(e)}\prod_{e'\in E(H)\setminus S}A_{\phi(e')} - \sum_{e\in E(H)}\sum_{\{e\}\subseteq S\subseteq E(H)}\prod_{e'\in S}Z_{\phi(e')}\prod_{e''\in E(H)\setminus S}A_{\phi(e'')}\right) \\
      =& \frac{1}{|\Aut(H)|}\sum_{\varnothing \subsetneq S\subseteq E(H)}(1 - |S|)\prod_{e\in S}Z_{\phi(e)}\prod_{e'\in E(H)\setminus S}A_{\phi(e')}.
    \end{align*}
    Then 
    \begin{align*}
    &\Var[\beta] = \E[\beta^2] \\
    =& \frac{1}{|\Aut(H)|^2}\E\left[\left(\sum_{\phi:V(H)\hookrightarrow [n]}\sum_{\varnothing \subsetneq S\subseteq E(H)}(1 - |S|)\prod_{e\in S}Z_{\phi(e)}\prod_{e'\in E(H)\setminus S}A_{\phi(e')}\right)^2\right] \\
    \le& \frac{1}{|\Aut(H)|^2}\sum_{\phi,\phi':V(H)\hookrightarrow [n]}\sum_{\varnothing \subsetneq S,T\subseteq E(H)}(1 - |S|)(1 - |T|)\prod_{e\in \phi(S)\triangle \phi'(T)}\E\left[Z_{e}\right]\prod_{e'\in \phi(S)\cap \phi'(T)}\E\left[Z^2_{e'}\right]\\
    =& \frac{1}{|\Aut(H)|^2}\sum_{\phi,\phi':V(H)\hookrightarrow [n]}\sum_{\substack{\varnothing \subsetneq S,T\subseteq E(H)\\ \phi(S) = \phi'(T)}}(1 - |S|)^2\prod_{e\in \phi(S)}\E\left[Z^2_{e}\right],
    \end{align*}
    where the inequality follows from the fact $A_{i,j} \in \{0,1\}$ and the final equation follows from the fact that only the terms satisfying $\phi(S) = \phi'(T)$ contribute to the sum due to $\E[Z_e] = 0$. 
    
    We reorganize the summation according to the common image edge set $C:=\phi(S)=\phi'(T)$. Since $\phi$ and $\phi'$ induce injective maps on $E(H)$, for fixed $C$, $\phi$, and $\phi'$, the sets $S$ and $T$ are uniquely determined by $S=\phi^{-1}(C)$ and $T=(\phi')^{-1}(C)$. Hence,
    \begin{align*}
    \Var[\beta] \le & \frac{1}{|\Aut(H)|^2}\sum_{\substack{C\subseteq E(K_n)\\2 \le |C| \le \ell, |V(C)| \le k}}(1 - |C|)^2\sum_{\substack{\phi,\phi':V(H)\hookrightarrow [n]\\C\subseteq \phi(E(H))\cap\phi'(E(H))}}\prod_{e\in C}\E\left[Z_{e}^2\right]\\
    \le & \frac{(\ell - 1)^2\cdot \max(V,1)^\ell}{|\Aut(H)|^2}
    \left(\sum_{\substack{C\subseteq E(K_n)\\2 \le |C| \le \ell, |V(C)|\le k}}\left(\sum_{\substack{\phi,\phi':V(H)\hookrightarrow [n]\\C\subseteq \phi(E(H))\cap\phi'(E(H))}}1\right)\right)\\
    = & \frac{(\ell - 1)^2\cdot \max(V,1)^\ell}{|\Aut(H)|^2}\sum_{c=3}^{k}\left(\sum_{\substack{C\subseteq E(K_n)\\2 \le |C| \le \ell, |V(C)| = c}}\left(\sum_{\substack{\phi,\phi':V(H)\hookrightarrow [n]\\C\subseteq \phi(E(H))\cap\phi'(E(H))}}1\right)\right)\\
    := & \frac{(\ell - 1)^2\cdot \max(V,1)^\ell}{|\Aut(H)|^2} \sum_{c=3}^{k}g_{H}(n,c),
    \end{align*}
    where the last inequality follows from the fact $\E[Z_e^2] = \Var[Z_e] = V$. The range of $c$ in the last equation follows from the fact that 2 edges must contain at least 3 vertices. 

    It remains to bound $g_H(n,c)$. For each fixed set $C$, the inner sum is at most $((k!/(k-c)!)n^{k-c})^2$. Indeed, for each of the injections $\phi$ and $\phi'$, there are at most $k!/(k-c)!$ ways to select and order the $c$ vertices of $H$ mapped to $V(C)$, while the images of the remaining $k-c$ vertices can be chosen in at most $n^{k-c}$ ways. Moreover, the number of possible choices for $C$ is at most $2^{\binom{c}{2}}n^c$. To see this, we first choose the $c$ vertices in $V(C)$ in at most $n^c$ ways and then choose the edges of $C$ from among the $\binom{c}{2}$ possible edges on these vertices. Consequently,
    \begin{align*}
        g_H(n, c) \le & 2^{\binom{c}{2}}n^c \cdot \left(\frac{k!}{(k-c)!}n^{k-c}\right)^2 = 2^{\binom{c}{2}}
\left(\frac{k!}{(k-c)!}\right)^2 n^{2k-c},\\
        \Var[\beta] \le & \frac{(\ell - 1)^2\cdot \max(V,1)^\ell}{|\Aut(H)|^2}\sum_{c=3}^{k}2^{\binom{c}{2}}
\left(\frac{k!}{(k-c)!}\right)^2 n^{2k-c} = O\left(\frac{n^{2k-3}}{\veps^{2\ell}}\right).
    \end{align*}

    Moreover, we can bound the term $\Var[\alpha]$ directly as 
    \begin{align*}
    \Var[\alpha] =& n \Var\left[\Lap\left(\frac{2\ell(n-2)!\kappa^{\ell-1}}{|\Aut(H)|(n-k)!\veps_0}\right)\right] \\
    =& 2n\left(\frac{2\ell (n-2)!\kappa^{\ell-1}}{|\Aut(H)|(n-k)!\veps_0}\right)^2 = O\left(\frac{n^{2k - 3}}{\veps^{2\ell}}\right).
    \end{align*}
    
    In conclusion, we have 
    \begin{align*}
    \Var[\hf_H(G)] = \Var[\beta] + \Var[\alpha] = O\left(\frac{n^{2k- 3}}{\veps^{2\ell}}\right). \end{align*}
\end{proof}
By \Cref{claim:unbiased-H,claim:variance-H}, for every $n$-vertex graph $G$,
\[
\sqrt{\E\left[\left(\hf_H(G)-f_H(G)\right)^2\right]}
=\sqrt{\Var[\hf_H(G)]}
=O\left(\frac{n^{k-3/2}}{\veps^\ell}\right),
\]
which proves the RMSE guarantee uniformly over all input graphs.
By applying Chebyshev's inequality and \Cref{claim:unbiased-H}, we have 
    \begin{align*}
        \Pr\left[\left|\hf_H(G) - f_H(G)\right| \le \sqrt{3\Var\left[\hf_H(G)\right]}\right] \ge \frac{2}{3}, 
    \end{align*}
    which, together with \Cref{claim:variance-H}, establishes the claimed additive error bound.
\end{proof}

\section{PC: The Degree-Dependent Upper Bound} 
\label{gh:main}

We now give the formal construction and analysis of PC. We first establish its $(\veps,\delta)$-LEDP guarantee using Gaussian releases, then derive the pure-privacy guarantee in \Cref{gh:pure-main-theorem}. 

\subsection{The Counting Algorithm}
\label{gh:counting}
\label{gh:connected}
\label{gh:injective}
PC estimates the subgraph count by progressively replacing private edge factors with public centers and estimating the change at each replacement. \Cref{gh:full-algorithm} describes the complete Gaussian version of PC, in which the public centers are learned from linear queries released with Gaussian noise. The following theorem gives its accuracy guarantee. \Crefrange{gh:counting-state}{gh:replacement-schedule} explain the details of \Cref{gh:full-algorithm}, while \Cref{gh:counting-analysis} proves the theorem using the center and correction guarantees in \Cref{gh:center,gh:stage}.

\begin{theorem}\label{gh:main-theorem}
For any fixed pattern graph $H$ with $k$ vertices, $\ell$ edges and $q$ connected components, and any $\veps\in(0,1]$ and $\delta\in(0,1)$, on every $n$-vertex graph with maximum degree at most $d$, the Gaussian version of PC (\Cref{gh:full-algorithm}) is an $(\ell+2)$-round $(\veps,\delta)$-LEDP algorithm for $H$-counting with additive error
\begin{align*}
|\hf_H(G)-f_H(G)|
 \le O\left(\frac{\log^{3\ell - 2}\left(\frac{n}{\delta}\right)n^{q-\frac{1}{2}}d^{k-q-1}}{\veps^{\ell}}\right)=\widetilde{O}_{\veps,\delta}\left(d^{k-q-1}n^{q - \frac{1}{2}}\right), 
\end{align*}
with probability at least $2/3$. The algorithm runs in $\operatorname{poly}(n,d,1/\veps,\log(1/\delta))$ time. 
\end{theorem}

\begin{algorithm}[!htbp]
\small
\SetAlgoLined
\caption{Predict and Correct (PC) with Gaussian center reports.}
\label{gh:full-algorithm}
\KwData{Graph $G=([n],E)$, represented by adjacency row vectors  $\bmA_1,\ldots,\bmA_n\in\{0,1\}^n$, a target integer degree parameter  $1\le d\le n-1$, pattern graph $H$ with $k$ vertices and $\ell$ edges, and privacy parameters  $\veps\in(0,1]$ and $\delta\in(0,1)$.}
\KwResult{An estimate $\hf_H(G)$.}
\tcc{Public parameters and isolated vertices}
$[s]\;$\If{$n<k$}{\KwRet{$0$}\;}
$[s]\;$Let $H^\circ$ be $H$ with its isolated vertices deleted\;
$[s]\;$\If{$\ell=0$}{\KwRet{$\binom{n}{k}$}\;}
$[s]\;$Construct the set $\Pi$ of all partitions $\pi$ of $V(H^\circ)$ whose parts are independent sets in $H^\circ$\;
$[s]\;$Let $N\gets2|\Pi|\ell$, $\veps_0\gets\veps/N$, $\delta_0\gets\delta/N$, and $\zeta\gets1/(12N)$\;
\tcc{Construct centers and start their corrections}
\ForEach{$\pi\in\Pi$ in parallel}{
 $[s]\;$Form $H_\pi$ with one vertex per part of $\pi$ and one edge between two parts whenever an edge of $H^\circ$ connects them; let $\mathcal H_\pi$ be the connected components of $H_\pi$\;
 \ForEach{$H'\in\mathcal H_\pi$ in parallel}{
  $[s]\;$Let $\widehat t_{\pi,H'}\leftarrow 0$,$\lambda\gets1$, $K$ be a copy of $H'$\;
  $[s]\;$Let $w_z(i)\gets1$ for all $z\in V(K)$ and $i\in[n]$\;
  $[s]\;$Let $\bmW_e\gets\bmA$ (marked private) for all $e\in E(K)$\;
  $[s]\;$Root the block-cut tree of $H'$ at any block\;
  \ForEach{block $K'$ in postorder}{
   $[s]\;$\If{$K'$ is not the root block}{
    Let $v_0$ be the cut vertex corresponding to its parent in the block-cut tree\;
   }
   \ForEach{$e\in E(K')$}{
    $[s]\;$\eIf{$K'$ is the root block}{
     Let $v$ be either endpoint of $e$\;
    }{
     Let $v$ be an endpoint of $e$ different from $v_0$\;
    }
    $[s]\;$Let $u$ be the other endpoint of $e$\;
    Run \Cref{gh:center-algorithm} with $(G,d,(w_v(i))_{i\in[n]}, (\bmW_{\{v,z\}})_{z\in N_K(v)\setminus\{u\}},(\veps_0,\delta_0),\zeta)$ to obtain $\bmP$, allowing the first rounds of different calls to run in parallel\;
    $[s]\;$Let $\calS$ be a copy of $(K,v,u,\bmW,w,\bmP)$, and let $\lambda'$ be a copy of $\lambda$\;
    $[s]\;$Let $\bmW_e\gets\bmP$, and mark $e$ as public\;
    Run \Cref{gh:correction-algorithm} with $(G, \calS, d, \veps_0, \zeta)$ in parallel with subsequent calls to \Cref{gh:center-algorithm}\;
    $[s]\;$When \Cref{gh:correction-algorithm} returns $\widehat\Delta$, let $\widehat t_{\pi,H'}\gets\widehat t_{\pi,H'}+\lambda'\widehat\Delta$, where $\lambda'$ is the value saved for this call\;
   }
   $[s]\;$\If{$K'$ is not the root block}{
    Compute $T_{K',v_0}(i)$ for all $i\in[n]$ by \Cref{gh:block-contribution}\;
    Let $w_{v_0}(i)\gets w_{v_0}(i)T_{K',v_0}(i)/d^{|V(K')|-1}$ for all $i\in[n]$\;
    Let $\lambda\gets\lambda d^{|V(K')|-1}$ and delete $V(K')\setminus\{v_0\}$ from $K$\;
   }
  }
  $[s]\;$Wait until all calls to \Cref{gh:correction-algorithm} return and their contributions are added to $\widehat t_{\pi,H'}$\;
  $[s]\;$Let $\widehat t_{\pi,H'}\gets\widehat t_{\pi,H'}+\lambda T_K(\bmW,w)$, where $T_K(\bmW,w)$ is calculated by \Cref{gh:weighted-count}\;
 }
}
$[s]\;$Let $\hf_H(G)\gets(\binom{n-|V(H^\circ)|}{k-|V(H^\circ)|}/|\Aut(H^\circ)|)\sum_{\pi\in\Pi}(\prod_{V'\in\pi}(-1)^{|V'|-1}(|V'|-1)!)(\prod_{H'\in\mathcal H_\pi}\widehat t_{\pi,H'})$\;
\KwRet{$\hf_H(G)$.}
\end{algorithm}

\subsubsection{Homomorphisms and Weighted Counts}
\label{gh:counting-state}
\label{gh:partition-combination}

Let $H^\circ$ be $H$ with its isolated vertices deleted. We use the standard quotient-graph inversion formula (\Cref{lemma:quotient-inversion}), with $\Pi=\Pi(H^\circ)$ and $H_\pi=H^\circ/\pi$, to reduce $H^\circ$-counting to homomorphism counting. Let $\mathcal H_\pi$ be the connected components of $H_\pi$. Since homomorphism counts factor over connected components, we estimate $\hom(H',\bmA)$ separately for each $H'\in\mathcal H_\pi$, obtaining an estimate $\widehat t_{\pi,H'}$. We then multiply these estimates within each quotient graph, combine the products with coefficients $\mu(\pi)$, and divide by $|\Aut(H^\circ)|$.

Each copy of $H^\circ$ extends to a copy of $H$ by choosing any $k-|V(H^\circ)|$ of the remaining vertices. There are $\binom{n-|V(H^\circ)|}{k-|V(H^\circ)|}$ such choices, so we multiply the estimate of $f_{H^\circ}(G)$ by this factor.

Fix one such connected component $H'$, and let $K$, initially equal to $H'$, be the current pattern. As in \Cref{overview:general}, PC progressively replaces the private edge factors by public centers and estimates the change at each replacement. Each public center belongs to the following convex set of fractional adjacency matrices:
\begin{align*}
\calP_d:=
\left\{\bmP'=(\bmP')^\top:
 P'_{i,i}=0,\quad 0\le P'_{i,j}\le1,\quad
 \sum_{j=1}^nP'_{i,j}\le d\text{ for every }i\in[n]\right\}.
\end{align*}

Each edge $e$ of $K$ contributes a factor matrix $\bmW_e$, initially $\bmA$ and subsequently the public center in $\calP_d$ used to replace it. Public vertex weights $w_v:[n]\to[0,1]$ record contributions from parts of the pattern that have already been processed. Initially all weights are one. At any stage of the procedure, the current weighted count is
\begin{align}
 T_K(\bmW,w):=\sum_{\phi:V(K)\to[n]}
 \prod_{v\in V(K)}w_v(\phi(v))
 \prod_{e=\{u,v\}\in E(K)}(\bmW_e)_{\phi(u),\phi(v)}. \label{gh:weighted-count}
\end{align}

Initially this expression is $\hom(K,\bmA)$. A public multiplier $\lambda$, initialized to one, records the normalization factors introduced when these weights are updated.

Before making any replacements, we root the block-cut tree of $H'$ at any block (\Cref{lem:block-cut-tree}) and fix a postorder traversal of its blocks. For a nonroot block, let $v_0$ be the cut vertex corresponding to its parent in this tree. Completed nonroot blocks are removed from $K$, with their contributions absorbed into the weight at $v_0$.

\subsubsection{Guarantees for Edge Replacements}
\label{gh:report-guarantees}
Within each block $K'$, we replace one private edge factor at a time. Under the postorder traversal, a nonroot block meets the rest of $K$ only at $v_0$. Since $K'$ has no cut vertex, choosing an endpoint $v\ne v_0$ of the selected edge $\{u,v\}$ ensures that $K-v$ remains connected; when the root block is processed, either endpoint is valid. Each edge is processed exactly once because it belongs to a unique block. 

Write $K_-:=K-v$ for the graph obtained by deleting $v$ and its incident edges. The center procedure takes the public weight $w_v$, the other factors incident to $v$, and the private adjacency rows, and returns a public $\bmP\in\calP_d$. Replace the factor on $\{u,v\}$ by $\bmP$ and save the current factors, weights, and normalization multiplier from immediately before this replacement. Saving a private factor means recording its position as an $\bmA$ factor, not publishing any private entries.

Let $\Delta$ be the count before replacement minus the count after replacement. Thus $\Delta$ is the same weighted sum with the factor on $\{u,v\}$ replaced by $\bmA-\bmP$.

\paragraph{The center guarantee.}
Let the edges incident to $v$ other than $\{v,u\}$ be $\{v,u_1\},\ldots,\{v,u_{r}\}$, and write $\bmW_\alpha$ for the edge factor on $\{v,u_\alpha\}$. Suppose $s$ of these factors are still private. Number them first, so $\bmW_\alpha=\bmA$ for $\alpha \in [s]$, whereas the other $\bmW_\alpha\in\calP_d$ are public. For each $i\in[n]$, write $w_i=w_v(i)\in[0,1]$ for the saved public weight.

To express what the center predicts, collect all assignments of the other neighbors of $v$ in one matrix. Define a matrix $\bmF\in\reals^{n\times n^{r}}$ whose row $i$ corresponds to mapping $v$ to $i$ and whose columns are indexed by maps $\phi:\{u_1,\ldots,u_{r}\}\to[n]$. Its entry is the product of the weight and these edge contributions:
\begin{equation}
F_{i,\phi}:=w_i\prod_{\alpha=1}^{r}
 (\bmW_\alpha)_{i,\phi(u_\alpha)}.
\label{gh:feature}
\end{equation}
This matrix is used only to state and prove the guarantee; the center procedure need not construct it explicitly. The following lemma establishes privacy and bounds the prediction error $\|(\bmA-\bmP)\bmF\|_F$, with the center construction and proof given in \Cref{gh:centers}.

\begin{lemma}\label{gh:center}
For any $\veps_0\in(0,1]$, $\delta_0\in(0,1)$, constant $\zeta\in (0,1)$, public weights $w_1,\ldots,w_n\in[0,1]$, and incident edge factors $\bmW_1,\ldots,\bmW_r$ satisfying the conditions above, \Cref{gh:center-algorithm} is a two-round $(\veps_0,\delta_0)$-LEDP algorithm, allowing the first rounds of different calls to run in parallel. On every $n$-vertex graph $G$ with maximum degree at most $d$, put $\bmE:=\bmA-\bmP$, where $\bmA$ is the adjacency matrix of $G$ and $\bmP \in \calP_d$ is the output of the algorithm, and we have
\begin{align*}
\|\bmE\bmF\|_F
 \le O\left(\frac{\sqrt n\,d^{\frac{r+1}{2}}\log^{\frac{r+2s+2}{2}}(\frac{n}{\delta_0})}{\veps_0^{s+1}}\right)
\end{align*}
with probability at least $1-\zeta$. The algorithm runs in $\operatorname{poly}(n,d,1/\veps_0,\log(1/\delta_0))$ time. 
\end{lemma}

\paragraph{The correction guarantee.}
Given a center satisfying \Cref{gh:center}, the following lemma bounds the additive error of the correction estimate $\widehat\Delta$; the correction procedure and proof are given in \Cref{gh:correction-analysis}.

\begin{lemma}[One-edge correction]\label{gh:stage}
Fix the saved replacement $(K,v,u,\bmW,w,\bmP)$ with $K-v$ connected and $\{v,u\}$ marked private. For any $\veps_0\in(0,1]$, $\delta_0\in(0,1)$, constant $\zeta\in (0,1)$, \Cref{gh:correction-algorithm} is an $\veps_0$-LEDP algorithm. On every $n$-vertex graph $G$ with maximum degree at most $d$, if the saved center $\bmP$ satisfies the error bound in \Cref{gh:center}, then
\begin{align*}
 |\widehat\Delta-\Delta|
 \le O\left(\frac{\sqrt n\,d^{k_K-2}\log^{3\ell_K - 2}\left(\frac{n}{\delta_0}\right)}{\veps_0^{\ell_K}}\right)
\end{align*}
with probability at least $1-\zeta$, where $k_K := |V(K)|$ and $\ell_K := |E(K)|$. The number of rounds used by \Cref{gh:correction-algorithm} is at most one plus the number of private edge factors in $K-v$. The algorithm runs in $\poly(n,d)$ time.
\end{lemma}

\subsubsection{Block Elimination and Count Preservation}
\label{gh:replacement-schedule}

\paragraph{Eliminating a completed public block.} Once all edges of a nonroot $K'$ are public, its contribution can be absorbed into the weight at $v_0$. Using the current public factors $\bmW_e$, compute
\begin{equation}
 T_{K',v_0}(i):=\sum_{\substack{\phi:V(K')\to[n]\\\phi(v_0)=i}}
 \prod_{v\in V(K')\setminus\{v_0\}}w_v(\phi(v))
 \prod_{e=\{u,v\}\in E(K')}(\bmW_e)_{\phi(u),\phi(v)}.
 \label{gh:block-contribution}
\end{equation}
The factor $w_{v_0}(i)$ is excluded from $T_{K',v_0}(i)$ because it remains attached to $v_0$ in the rest of the pattern. For any fixed assignment of the remaining vertices with $\phi(v_0)=i$, summing over the images of the removed vertices contributes exactly $T_{K',v_0}(i)$. Thus multiplying $w_{v_0}(i)$ by $T_{K',v_0}(i)$ for each $i\in[n]$ preserves the weighted count, with every vertex weight and edge factor included exactly once. All factors in this computation are public.

A spanning tree rooted at $v_0$ gives $0\le T_{K',v_0}(i)\le d^{|V(K')|-1}$. Indeed, all vertex weights and edge entries lie in $[0,1]$, so omitting the weights and the non-tree edge factors can only increase the sum. With the image of the root fixed to $i$, sum over the images of the other vertices from the leaves upwards. Each leaf sum is at most $d$, because the corresponding public matrix belongs to $\calP_d$ and thus has row sums at most $d$. Removing the $|V(K')|-1$ nonroot vertices in this way gives the bound.

Delete $V(K')\setminus\{v_0\}$, replace $w_{v_0}(i)$ with $w_{v_0}(i)T_{K',v_0}(i)/d^{|V(K')|-1}$ for every $i\in[n]$, and multiply the accumulated multiplier $\lambda$ by $d^{|V(K')|-1}$. The product of $\lambda$ and the remaining weighted count is unchanged, and all weights stay in $[0,1]$. Thus the next center and correction calls still have the required inputs. The final block is evaluated entirely from public factors.

\paragraph{The count preserved throughout the procedure.}
For each component $H'$, \Cref{gh:full-algorithm} builds $\widehat t_{\pi,H'}$ by adding back the estimated difference from each edge replacement. Each estimate $\widehat\Delta$ is multiplied by the saved multiplier $\lambda'$ to account for the normalization introduced by earlier block eliminations. Once all factors in the remaining pattern are public, the algorithm computes its weighted count exactly and adds $\lambda T_K(\bmW,w)$ to the accumulated corrections. Together, these terms account for the original homomorphism count: replacing each $\widehat\Delta$ by the exact difference $\Delta$ makes their sum equal to $\hom(H',\bmA)$.

\subsubsection{Analysis of the Counting Algorithm}
\label{gh:counting-analysis}
Now we give the formal proof of \Cref{gh:main-theorem}.
\begin{proof}[Proof of \Cref{gh:main-theorem}]
We prove these guarantees for the Gaussian version of PC.
\paragraph{Privacy.}
By \Cref{gh:center,gh:stage}, \Cref{gh:center-algorithm,gh:correction-algorithm} are $\veps_0$-LEDP and $(\veps_0,\delta_0)$-LEDP, respectively. Each subroutine is invoked at most $|\Pi|\ell$ times. Since $N=2|\Pi|\ell$, $\veps_0=\veps/N$ and $\delta_0=\delta/N$, their composition is $(\veps,\delta)$-LEDP by \Cref{lemma:basic composition}. All remaining computations are postprocessing, so this Gaussian version of PC (\Cref{gh:full-algorithm}) is $(\veps,\delta)$-LEDP by \Cref{lemma:post-processing}.

\paragraph{Accuracy.}
Fix $G \in \calG_{n,d}$. The cases $n<k$ and $\ell=0$ return exact answers, so assume $n\ge k$ and $\ell\ge1$. We first consider $H$ without isolated vertices. 

By \Cref{gh:center}, each center bound fails with probability at most $\zeta$. Conditional on all center bounds holding, each refined correction bound also fails with probability at most $\zeta$ by \Cref{gh:stage}. Since each procedure is invoked at most $|\Pi|\ell$ times, a union bound shows that all these bounds hold simultaneously with probability at least $1-N\zeta=11/12>2/3$. We bound the estimation error on this event below.

Fix $\pi\in\Pi$ and a component $H'\in\mathcal H_\pi$, and write $\ell_{H'}=|E(H')|$. For a correction saved when the current pattern is $K$, its multiplier is $\lambda'=d^{|V(H')|-k_K}$, since each previously eliminated vertex contributes one factor of $d$. The endpoint choice in \Cref{gh:report-guarantees} ensures that $K-v$ is connected. The refined correction bound therefore gives
\begin{align*}
\lambda'|\widehat\Delta-\Delta|
&\le d^{|V(H')|-k_K}O\left(\frac{\sqrt n\,d^{k_K-2}\log^{3\ell_K - 2}\left(\frac{n}{\delta_0}\right)}{\veps_0^{\ell_K}}\right)\le O\left(\frac{\sqrt n\,d^{|V(H')|-2}\log^{3\ell_{H'} - 2}\left(\frac{n}{\delta}\right)}{\veps^{\ell_{H'}}}\right),
\end{align*}
where we used $\ell_K\le\ell_{H'}$, $\veps_0=\veps/N$, $\delta_0=\delta/N$ and the fact that $N$ depends only on the fixed pattern $H$.

By \Cref{gh:replacement-schedule}, replacing each $\widehat\Delta$ by its exact value $\Delta$ recovers $\hom(H',\bmA)$. Summing the scaled errors over at most $\ell_{H'}$ replacements therefore gives the same bound for $|\widehat t_{\pi,H'}-\hom(H',\bmA)|$. A spanning tree of $H'$ also gives $0\le\hom(H',\bmA)\le nd^{|V(H')|-1}$: its root has $n$ possible images and each subsequent vertex has at most $d$ choices.

We next combine the component estimates. Let $k_\pi=|V(H_\pi)|$ and $q_\pi=|\mathcal H_\pi|$. For $\varnothing\ne S\subseteq\mathcal H_\pi$, let $\ell_S=\sum_{H'\in S}\ell_{H'}\le|E(H_\pi)|\le\ell$. Writing each estimate as its true count plus its error and expanding the product gives
\begin{align*}
&\left|\prod_{H'\in\mathcal H_\pi}\widehat t_{\pi,H'}-\hom(H_\pi,\bmA)\right| \\ 
&= \left|\prod_{H'\in\mathcal H_\pi}\left(\widehat t_{\pi,H'} - \hom(H',\bmA) + \hom(H', \bmA)\right)-\prod_{H'\in \calH_\pi}\hom(H',\bmA)\right|\\
&= \left|\sum_{\varnothing\ne S\subseteq\mathcal H_\pi}
\prod_{H'\in S}\left(\widehat t_{\pi,H'}-\hom(H',\bmA)\right)
\prod_{H'\in\mathcal H_\pi\setminus S}\hom(H', \bmA)\right|\\
&\le\sum_{\varnothing\ne S\subseteq\mathcal H_\pi}
\prod_{H'\in S}|\widehat t_{\pi,H'}-\hom(H',\bmA)|
\prod_{H'\in\mathcal H_\pi\setminus S}nd^{|V(H')|-1}\\
&\le\sum_{\varnothing\ne S\subseteq\mathcal H_\pi}
O\left(\frac{n^{q_\pi-\frac{|S|}{2}}d^{k_\pi-q_\pi-|S|}\log^{3\ell_S - 2}\left(\frac{n}{\delta}\right)}{\veps^{\ell_S}}\right)=O\left(\frac{n^{q_\pi-\frac{1}{2}}d^{k_\pi-q_\pi-1}\log^{3\ell - 2}\left(\frac{n}{\delta}\right)}{\veps^{\ell}}\right).
\end{align*}
The last equation uses $|S|\ge 1$, $\ell_S\le\ell$, $0<\veps\le1$, and the fact that the number of summands depends only on $H$. Identifying vertices cannot increase the number of vertices or connected components, so $k_\pi\le k$ and $q_\pi\le q$. Since $1\le d\le n-1$, we have
\begin{align*}
n^{q_\pi-\frac{1}{2}}d^{k_\pi-q_\pi-1}
&=n^{q-\frac{1}{2}}d^{k-q-1}\left(\frac{d}{n}\right)^{q-q_\pi}d^{k_\pi-k}\le n^{q-\frac{1}{2}}d^{k-q-1}.
\end{align*}
Applying \Cref{lemma:quotient-inversion} now gives the claimed error bound for $\hf_H(G)$, since $|\Pi|$, the coefficients $\mu(\pi)$ and $|\Aut(H)|$ depend only on $H$.

If $H$ has isolated vertices, \Cref{gh:full-algorithm} multiplies the estimate for $H^\circ$ by
\[
 \binom{n-|V(H^\circ)|}{k-|V(H^\circ)|}
 =\Theta\!\left(n^{k-|V(H^\circ)|}\right).
\]
Removing each isolated vertex decreases both the vertex count and the component count by one. Thus this multiplier gives the same stated error bound in terms of $k$ and $q$.

\paragraph{Rounds.}

Fix $\pi\in\Pi$ and $H'\in\mathcal H_\pi$. By \Cref{gh:center}, different calls of \Cref{gh:center-algorithm} run their first rounds in parallel. Number its calls to \Cref{gh:center-algorithm} in the edge-replacement order. The center returned by the $j$th call is available by round $j+1$. In the corresponding saved record, $K-v$ contains at most $|E(H')|-j$ private edge factors. By \Cref{gh:stage}, the correction needs at most $|E(H')|-j+1$ further rounds and hence finishes by round
\begin{align*}
(j+1)+\bigl(|E(H')|-j+1\bigr)=|E(H')|+2\le\ell+2.
\end{align*}
Since the iterations over $\pi$ and $H'$ run in parallel and the final public computations require no additional reporting rounds, \Cref{gh:full-algorithm} uses at most $\ell+2$ rounds.

\paragraph{Runtime.} There are at most $N=2|\Pi|\ell$ calls, where $N$ depends only on $H$. By \Cref{gh:center,gh:stage} and the choices $\veps_0=\veps/N$ and $\delta_0=\delta/N$, their total running time is $\operatorname{poly}(n,d,1/\veps,\log(1/\delta))$. The sums in \Cref{gh:block-contribution,gh:weighted-count} can each be evaluated by enumerating at most $n^k$ maps $\phi$. Since $H$ is fixed, these computations also take polynomial time, which proves the stated running time.
\end{proof}

\subsection{Center and Correction Procedures}
\label{gh:reports-analysis}

We now implement the local correction computation and the public-center construction described in \Cref{overview:general}. 

\subsubsection{Estimating a Correction}
\label{gh:stage-section}
\label{gh:kernel-section}
\label{gh:correction-analysis}

Fix a replacement record $(K,v,u,\bmW,w,\bmP)$ saved by PC, as specified in \Cref{gh:stage}. The correction is the weighted count with factor $\bmA-\bmP$ on $\{v,u\}$. The complete procedure is given in \Cref{gh:correction-algorithm}, and we describe its steps below.

\paragraph{The computation tree.}

Choose a spanning tree of $K-v$ rooted at $u$, attach $v$ through $\{v,u\}$, and root the full tree at $v$. The vertex $u$ is the only tree child of $v$. Direct tree edges from parents to children. Direct non-tree edges incident to $v$ away from $v$, and the others arbitrarily. Evaluate each edge at its initial endpoint.

Here $x,y$ are pattern vertices and $i,j\in[n]$ are input vertices. For each $x$, let $E_{\mathrm{out}}(x)$ and $E_{\mathrm{in}}(x)$ be the sets of non-tree edges directed out of and into $x$, respectively, and let $C(x)$ be its children in the chosen spanning tree.

\paragraph{Hashes, signs, and child reports.}
The algorithm performs $d$ repetitions. Each repetition uses fresh public randomness: independent uniform hash functions $h_x:[n]\to[d]$ for $x\ne v$ and independent uniform signs $\xi_{e,j}\in\{-1,1\}$ for every non-tree edge $e$ and input vertex $j\in[n]$. 
Before computing the reports, the server performs the public hash check; if the check fails, the procedure returns zero.

For $x\ne v$, let $M_{x,i}$ denote the report computed with $\phi(x)=i$ fixed. Reports are computed in postorder, so the child reports are already public when needed. For a tree edge from $x$ to $y$, the local computation at $x$ uses $M_{y,j}$ for each candidate image $j$ of $y$. For a non-tree edge $e$ directed from $x$ to $y$, it uses $\xi_{e,j}$ instead, while the computation at $y$ includes $\xi_{e,\phi(y)}$. Equivalently, each incoming non-tree edge $e\in E_{\mathrm{in}}(x)$ contributes the factor $\xi_{e,i}$ at $x$.

At $x\ne v$, retain a candidate image $j$ of an outgoing neighbor $y$ only when $h_y(j)=h_x(i)$; the root $v$ sums over all buckets. Each party uses only its own factor rows and public inputs. If all outgoing factors are public, the server computes the report.

\paragraph{Local queries.}
Each local query multiplies the vertex weight and incoming signs by one normalized sum for each outgoing edge. The threshold $\tau$, specified in \Cref{gh:correction-algorithm}, bounds each factor row's bucket sum with high probability.

For one repetition, suppress its index. For any $x\ne v$, $i\in [n]$, and $z\in[d]$, compute
\begin{align}
Q_{x,i}(z)
=w_x(i)\prod_{e\in E_{\mathrm{in}}(x)}\xi_{e,i}
\prod_{y\in C(x)}
\frac{\sum\limits_{j\in[n]:h_y(j)=z}(\bmW_{\{x,y\}})_{i,j}M_{y,j}}{\tau}
\prod_{e=\{x,y\}\in E_{\mathrm{out}}(x)}
\frac{\sum\limits_{j\in[n]:h_y(j)=z}(\bmW_e)_{i,j}\xi_{e,j}}{\tau}.
\label{gh:local-query}
\end{align}
At the root $v$, we use $\bmA-\bmP$ on $\{v,u\}$. For any $i\in [n]$ and $z\in[d]$, compute
\begin{align}
Q_{v,i}(z)
=\frac{w_v(i)\sum\limits_{j\in[n]:h_u(j)=z}(\bmA-\bmP)_{i,j}M_{u,j}}{\tau}
\prod_{e=\{v,y\}\in E_{\mathrm{out}}(v)}
\frac{\sum\limits_{j\in[n]:h_y(j)=z}(\bmW_e)_{i,j}\xi_{e,j}}{\tau}.
\label{gh:local-query-v}
\end{align}

\paragraph{Normalization and aggregation.}
In each repetition, for $x\ne v$, evaluate $Q_{x,i}(h_x(i))$ and clip it to $[-1,1]$. If $x$ has a private outgoing edge, party $i$ adds independent Laplace noise of scale $\sigma$ and divides the result by $1+\tau\sigma$ to obtain $M_{x,i}$. Otherwise, the server uses the clipped value directly as $M_{x,i}$.

Let $p$ be the number of vertices in $K-v$ with at least one private outgoing edge. Set $\beta=\tau^{|E(K)|}(1+\tau\sigma)^p$ to account for the normalizations in the edge sums and private reports. At the root $v$, each party $i$ clips $Q_{v,i}(z)$ to $[-2,2]$ and sums these values over all buckets and $d$ independent repetitions. Then it multiplies this sum by $\beta d^{|V(K)|-3}$ and releases the result with the final Laplace noise specified in \Cref{gh:correction-algorithm}. The server sums these final reports to obtain $\widehat\Delta$.

\begin{algorithm}[!htb]
\small
\SetAlgoLined
\caption{PC's correction subroutine for one saved edge replacement.}
\label{gh:correction-algorithm}
\KwData{Graph $G=([n],E)$, represented by adjacency row vectors $\bmA_1,\ldots,\bmA_n\in\{0,1\}^n$, a saved replacement record $(K,v,u,\bmW,w,\bmP)$ with $K-v$ connected and $\{v,u\}$ marked private, a target integer degree parameter $1\le d\le n-1$, privacy parameter $\veps_0$ and failure probability $\zeta$.}
\KwResult{An estimate $\widehat\Delta$ of the count $\Delta$ before replacement minus the count after replacement.}

\tcc{Public setup}
$[s]\;$Let $K_-\gets K-v$, $k_K\gets|V(K)|$, and $\ell_K\gets|E(K)|$\;
$[s]\;$Choose a spanning tree of $K_-$ rooted at $u$, attach $v$ through $\{v,u\}$, and root the full tree at $v$\;
$[s]\;$Direct tree edges from parents to children, non-tree edges incident to $v$ away from $v$, and the remaining edges arbitrarily\;
$[s]\;$Let $p$ count the vertices of $K_-$ with at least one private outgoing edge\;
$[s]\;$Let $\veps_1\gets\veps_0/(2(p+1))$, $\tau\gets\left\lceil\ln\left(128(\ell_K+1)n^3/\zeta\right)\right\rceil$, $\sigma\gets2\ell_K\tau/\veps_1$, and $\beta\gets \tau^{\ell_K}(1+\tau\sigma)^p$\;
$[s]\;$For all $a\in[d],x\in V(K_-)$, draw independent uniform hash functions $h_x^{(a)}:[n]\to[d]$\;
$[s]\;$For all $a\in[d]$, non-tree edge $e$ of $K$, and $j\in[n]$, draw independent uniform signs $\xi_{e,j}^{(a)}\in\{-1,1\}$\;
$[s]\;$\If{$|\{a\in [d]:h_x^{(a)}(i)=h_y^{(a)}(j)\}|>\tau$ for some  $\{x,y\} \in E(K_-)$ and $i,j\in[n]$}{\KwRet{$0$}\;}

\tcc{Remaining-pattern reports: process children before parents}
\ForEach{$x\in V(K_-)$ in postorder}{
 \eIf{$x$ has at least one private outgoing edge}{
 \For{$i\gets 1$ \KwTo $n$}{
  $[v_i]\;$For all $a\in[d]$, let $M_{x,i}^{(a)}\gets(\Clip_{[-1,1]}(Q_{x,i}^{(a)}(h_x^{(a)}(i)))+\Lap(\sigma))/(1+\tau\sigma)$ using Equation (\ref{gh:local-query})\;
  $[v_i]\;$Upload $(M_{x,i}^{(a)})_{a\in[d]}$ to the server\;
 }}{
  $[s]\;$For all $i\in[n], a\in[d]$, let $M_{x,i}^{(a)}\gets\Clip_{[-1,1]}(Q_{x,i}^{(a)}(h_x^{(a)}(i)))$ using Equation (\ref{gh:local-query})\;
 }
}

\tcc{One final round: apply the same product at $v$}
\For{$i\gets1$ \KwTo $n$}{
 $[v_i]\;$Let $Q_i\gets\beta d^{k_K-3}\sum_{a=1}^d\sum_{z=1}^d\Clip_{[-2,2]}(Q_{v,i}^{(a)}(z))$ using Equation (\ref{gh:local-query-v})\;
 $[v_i]\;$Upload $M_i\gets Q_i+\Lap(4\ell_K\beta d^{k_K-2}/\veps_1)$ to the server\;
}
$[s]\;$\KwRet{$\widehat\Delta=\sum_{i=1}^nM_i$}\;
\end{algorithm}

Now we give the formal proof of \Cref{gh:stage}.

\begin{proof}[Proof of \Cref{gh:stage}]
We prove the privacy, accuracy, rounds, and runtime guarantees separately.

\paragraph{Privacy.}
The hash check depends only on public randomness, so returning zero when it fails incurs no privacy loss. If it passes, fix the public hashes, signs, and preceding reports. Clipping and the hash check give the following sensitivity bounds on all graphs. At a private reporting vertex $x\ne v$, party $i$ computes $Q_{x,i}^{(a)}(h_x^{(a)}(i))$ and clips it to $[-1,1]$ for each repetition $a\in[d]$. These $d$ clipped values form the vector $(\Clip_{[-1,1]}(Q_{x,i}^{(a)}(h_x^{(a)}(i))))_{a=1}^{d}$. We bound the $\ell_1$-sensitivity of this entire vector when one raw entry $A_{i,t}$ changes. For each outgoing private edge $(x,y)$, this entry appears in the corresponding edge sum only when $h_y^{(a)}(t)=h_x^{(a)}(i)$. The public hash check guarantees that this occurs in at most $\tau$ repetitions for each edge. Since there are at most $\ell_K$ outgoing private edges, at most $\ell_K\tau$ coordinates of the vector can change. Clipping it to $[-1,1]$ bounds the change in that coordinate by two. Before the division, each clipped query changes by at most two, so the vector sensitivity is at most $2\ell_K\tau$. By \Cref{lemma:Laplace mechanism}, this release is $\veps_1$-DP.

At $v$, party $i$ computes $Q_{v,i}^{(a)}(z)$ for every $a,z\in[d]$ and clips each value to $[-2,2]$. For each outgoing edge $(v,y)$, a change to $A_{i,t}$ can affect its edge sum only in the bucket $z=h_y^{(a)}(t)$. Since there are at most $\ell_K$ outgoing edges, at most $\ell_K$ clipped values can change per repetition, each by at most four. Summing over the $d$ repetitions and multiplying by $\beta d^{k_K-3}$, the sensitivity of $Q_i$ is therefore at most $4\ell_K\beta d^{k_K-2}$. By \Cref{lemma:Laplace mechanism}, the final scalar release is $\veps_1$-DP.

At either changed endpoint, there are $p$ private internal releases and one final scalar release. By \Cref{lemma:basic composition}, the total privacy loss is at most $2(p+1)\veps_1=\veps_0$, so the entire algorithm is $\veps_0$-LEDP.

\paragraph{Accuracy.} 
Fix $G\in\calG_{n,d}$ and the center transcript, and assume that the saved center $\bmP$ satisfies the bound in \Cref{gh:center}. %

Delete all clips and ignore the hash-check return, using the same hashes, signs and internal Laplace noises. Let $\widetilde M_{x,i}^{(a)}$ be the resulting internal reports, and let $\widetilde Q_{x,i}^{(a)}(z)$ be the queries obtained by substituting these reports into Equations (\ref{gh:local-query}) and (\ref{gh:local-query-v}). Define
\[
\widetilde \Delta:=\beta d^{k_K-3}\sum_{a=1}^d\sum_{i=1}^n\sum_{z=1}^d\widetilde Q_{v,i}^{(a)}(z)+\sum_{i=1}^nL_i,
\]
where $L_i\sim\Lap(4\ell_K\beta d^{k_K-2}/\veps_1)$ are the independent final noises used by the algorithm. These variables are defined even when the algorithm returns early. We first show that removing the checks and clips leaves the reports and output unchanged with high probability, then bound the expectation and variance of $\widetilde\Delta$ without conditioning on this agreement.

\begin{lemma}\label{gh:direct-correction-agreement}
With probability at least $1-\zeta/4$, the hash check passes and no clip changes its input. That is, on this event, we have $\widetilde\Delta=\widehat\Delta$ and $\widetilde M_{x,i}^{(a)}=M_{x,i}^{(a)}$ for all $x\in V(K_-), i\in[n], a \in [d]$.
\end{lemma}
\begin{proof}
We can define this event by the following conditions. First, the public hash check passes. Second, for every $i\in[n]$, $a,z\in[d]$, and $e=\{x,y\}$ directed from $x$ to $y$ other than $\{v,u\}$, require the bucket bound $\sum_{j\in[n]:h_y^{(a)}(j)=z}(\bmW_e)_{i,j}\le\tau$. For the residual edge $\{v,u\}$, require the corresponding bucket bounds $\sum_{j\in[n]:h_u^{(a)}(j)=z}A_{i,j}\le\tau$ and $\sum_{j\in[n]:h_u^{(a)}(j)=z}P_{i,j}\le\tau$.  Finally, every Laplace noise added to an internal report $M_{x,i}^{(a)}$, with $x\ne v$, must have a magnitude of at most $\tau\sigma$. This condition does not concern the noises added to the final scalar reports $M_i$. On this event, the bucket and noise bounds, together with \Cref{gh:local-query,gh:local-query-v}, the report formulas in \Cref{gh:correction-algorithm}, and the fact that all vertex weights and signs have absolute value at most one, imply by induction from the leaves towards the root that $|Q_{x,i}^{(a)}(z)|\le1$ and $|M_{x,i}^{(a)}|\le1$ for all $x\in V(K)\setminus\{v\}$, $i\in[n]$, and $a,z\in[d]$. At the root $v$, the separate bucket bounds for $\bmA$ and $\bmP$ give $|Q_{v,i}^{(a)}(z)|\le2$ for all $i\in[n]$ and $a,z\in[d]$. Hence no clipping operation changes its input. 

We first bound the probability that a hash or bucket condition fails. For a fixed edge $\{x,y\}\in E(K_-)$ and fixed $i,j\in[n]$, the hash-check quantity $\sum_{a=1}^{d}\mathbf1_{h_x^{(a)}(i)=h_y^{(a)}(j)}$ is a sum of independent indicators, each with expectation $1/d$, so its expectation is one. Similarly, each bucket sum consists of independent random variables in $[0,1]$. Its expectation is at most one because each index enters the bucket with probability $1/d$ and the nonnegative entries of the factor row sum to at most $d$. This applies to every saved edge factor $\bmW_e$ and separately to $\bmA$ and $\bmP$.

For any such sum $U$, the Chernoff bound for independent $[0,1]$-valued random variables with total expectation at most one gives $\Pr[U>\tau]\le\exp(e-1-\tau)\le e^{2-\tau}$. There are at most $\ell_Kn^2$ hash-check quantities. For the bucket conditions, there are $\ell_K-1$ ordinary edge factors and two matrices, $\bmA$ and $\bmP$, for the residual edge. Each contributes $nd^2$ sums, indexed by the input row, repetition, and bucket. Since $d\le n-1$, the total number of hash and bucket conditions is at most $\ell_Kn^2+(\ell_K+1)nd^2\le(\ell_K+1)n^3$. By the union bound and the choice of $\tau$, the probability that any of these conditions fails is at most $(\ell_K+1)n^3e^{2-\tau}\le e^2\zeta/128$.

There are $pnd\le(k_K-1)nd$ internal Laplace noises. For each $Y\sim\Lap(\sigma)$, we have $\Pr[|Y|>\tau\sigma]=e^{-\tau}$. Since $k_K-1\le\ell_K$, the probability that at least one internal Laplace noise has magnitude greater than $\tau\sigma$ is at most $(k_K-1)nd\,e^{-\tau}\le\zeta/128$.

Consequently, the probability that the desired event does not occur is at most $(e^2+1)\zeta/128<\zeta/4$.
\end{proof}

\begin{lemma}\label{gh:direct-correction-mean}
$\E[\widetilde\Delta]=\Delta$.
\end{lemma}
\begin{proof}
Consider one repetition and suppress its index. Let $Y:=\beta\sum_{i=1}^n\sum_{z=1}^d\widetilde Q_{v,i}(z)$ and fix the hashes $h$ and signs $\xi$. The non-tree edge sums and their signs are fixed under this conditioning. Repeatedly substitute the report formulas into $Y$. Choosing the noise summand at $x$ removes all child reports below $x$; choosing its query summand multiplies reports from distinct child subtrees. Thus each expanded product contains each internal noise at most once. Since each internal noise appears at most once in a term and the noises are independent with mean zero, only the terms with no internal noise remain in $\E[Y\mid h,\xi]$. These terms contain one divisor $\tau$ per edge of $K$ and one divisor $1+\tau\sigma$ per private reporting vertex, which together cancel the factor $\beta$ exactly.

Let $\phi:V(K)\to[n]$ record the images selected by the root and tree-edge sums. The root edge requires $h_u(\phi(u))=z$, and each tree edge from $x$ to its child $y$ requires $h_y(\phi(y))=h_x(\phi(x))$. Since the tree spans $K_-$, these conditions give
\[
\mathbf1_{h_u(\phi(u))=z}
\prod_{x\in V(K_-)}\prod_{y\in C(x)}
\mathbf1_{h_y(\phi(y))=h_x(\phi(x))}
=\prod_{x\in V(K_-)}\mathbf1_{h_x(\phi(x))=z}.
\]
Expanding the root and tree-edge sums therefore yields
\begin{align*}
\E[Y\mid h,\xi]
=&\sum_{z=1}^d\sum_{\phi:V(K)\to[n]}
\mathbf1_{\forall x\in V(K_-),h_x(\phi(x))=z}
E_{\phi(v),\phi(u)}
\prod_{x\in V(K)}w_x(\phi(x))\\
&\cdot
\prod_{x\in V(K_-)}\prod_{y\in C(x)}
(\bmW_{\{x,y\}})_{\phi(x),\phi(y)}
\prod_{x\in V(K)}
\prod_{e=\{x,y\}\in E_{\mathrm{out}}(x)}
\sum_{j\in[n]:h_y(j)=z}
(\bmW_e)_{\phi(x),j}\xi_{e,j}\xi_{e,\phi(y)}.
\end{align*}
For each non-tree edge $e=\{x,y\}$ directed from $x$ to $y$, we have
\begin{align*}
&\E_\xi\left[
\sum_{j\in[n]:h_y(j)=z}(\bmW_e)_{\phi(x),j}\xi_{e,j}\xi_{e,\phi(y)}
\,\middle|\,h\right]\\
=&\sum_{j\in[n]:h_y(j)=z}(\bmW_e)_{\phi(x),j}
\E_\xi[\xi_{e,j}\xi_{e,\phi(y)}]=\sum_{j\in[n]:h_y(j)=z}(\bmW_e)_{\phi(x),j}\mathbf1_{j=\phi(y)}=(\bmW_e)_{\phi(x),\phi(y)}\mathbf1_{h_y(\phi(y))=z}.
\end{align*}
The sign families for distinct edges are independent. Since $y\in V(K_-)$, the condition $h_x(\phi(x))=z$ for all $x\in V(K_-)$ already implies $h_y(\phi(y))=z$, so the extra indicator can be omitted.

Taking conditional expectations gives
\begin{align*}
\E[Y\mid h]
={}&\sum_{\phi:V(K)\to[n]}
E_{\phi(v),\phi(u)}
\prod_{x\in V(K)}w_x(\phi(x))
\prod_{e=\{x,y\}\in E(K)\setminus\{\{v,u\}\}}
(\bmW_e)_{\phi(x),\phi(y)}\sum_{z=1}^d\prod_{x\in V(K_-)}
\mathbf1_{h_x(\phi(x))=z}.
\end{align*}
For each fixed $\phi$, the $k_K-1$ variables $h_x(\phi(x))$ are independent and uniform on $[d]$. Hence
\begin{align*}
\E\left[\sum_{z=1}^d\prod_{x\in V(K_-)}
\mathbf1_{h_x(\phi(x))=z}\right]=\sum_{z=1}^d\prod_{x\in V(K_-)}\Pr[h_x(\phi(x))=z]=d\left(\frac1d\right)^{k_K-1}=d^{2-k_K}.
\end{align*}
By the tower property of conditional expectation again and the definition of the correction $\Delta$, it follows that
\begin{align*}
\E[Y]
&=d^{2-k_K}\sum_{\phi:V(K)\to[n]}
E_{\phi(v),\phi(u)}
\prod_{x\in V(K)}w_x(\phi(x))
\prod_{e=\{x,y\}\in E(K)\setminus\{\{v,u\}\}}
(\bmW_e)_{\phi(x),\phi(y)}=d^{2-k_K}\Delta.
\end{align*}
Finally, the independent final Laplace noises have mean zero, so
\begin{align*}
\E[\widetilde\Delta]
&=d^{k_K-3}\sum_{a=1}^d\E[Y^{(a)}]+\sum_{i=1}^n\E[L_i]=d^{k_K-3}\,d\,d^{2-k_K}\Delta+0
=\Delta.
\end{align*}
\end{proof}
\begin{lemma}\label{gh:direct-correction-variance}
$\Var[\widetilde\Delta]=O(nd^{2k_K-4}\log^{6\ell_K - 4}(n/\delta_0)/\veps_0^{2\ell_K})$.
\end{lemma}
\begin{proof}
We first bound the variance of $Y:=\beta\sum_{i=1}^n\sum_{z=1}^d\widetilde Q_{v,i}(z)$ for one repetition, with its index suppressed. We then use $\widetilde\Delta=d^{k_K-3}\sum_{a=1}^dY^{(a)}+\sum_{i=1}^nL_i$ and independence to combine the repetitions and final noises. Let $B:=N_K(v)$, $r:=|B|-1$, and $\ell_{K_-}:=|E(K_-)| = \ell_K - r-1$, and let $s$ be the number of private factors on the edges incident to $v$ other than $\{v,u\}$. For each $\phi_0:B\to[n]$, by the symmetry of $\bmE$ and the definition of $\bmF$ in \Cref{gh:feature}, define
\begin{align}
b(\phi_0)
&:=\sum_{i=1}^n w_v(i)E_{i,\phi_0(u)}
\prod_{x\in B\setminus\{u\}}(\bmW_{\{v,x\}})_{i,\phi_0(x)}\notag=(\bmE\bmF)_{\phi_0(u),\,\phi_0|_{B\setminus\{u\}}}.\notag
\end{align}
By the assumed guarantee of $\bmP$ in \Cref{gh:center}, we have
\begin{align}
\sum_{\phi_0:B\to[n]}b(\phi_0)^2
&=\|\bmE\bmF\|_F^2
\le O\left(\frac{nd^{r+1}\log^{r+2s+2}(\frac{n}{\delta_0})}
{\veps_0^{2(s+1)}}\right).
\label{gh:stage-center-bound}
\end{align}
We express $Y$ as a sum with coefficients $b(\phi_0)$, so that the preceding squared-norm bound can be applied.
Since $u$ is the only tree child of $v$, Equation (\ref{gh:local-query-v}) gives
\begin{align*}
Y
={}&\beta\sum_{i=1}^n\sum_{z=1}^d 
\frac{w_v(i)\sum_{j\in[n]:h_u(j)=z}E_{i,j}\widetilde M_{u,j}}{\tau}
\prod_{x\in B\setminus\{u\}}
\frac{\sum_{j\in[n]:h_x(j)=z}(\bmW_{\{v,x\}})_{i,j}\xi_{\{v,x\},j}}{\tau}.
\end{align*}
Put $\beta_-:=\beta/\tau^{r+1}$. There are $|B|=r+1$ divisions by $\tau$ in this expression, so their combination with $\beta$ gives $\beta_-$. Expanding the product of sums amounts to choosing an input vertex $\phi_0(x)$ for each $x\in B$. All these choices must lie in bucket $z$. Hence
\begin{align*}
Y=&\beta_-\sum_{i=1}^n\sum_{z=1}^d
\sum_{\substack{\phi_0:B\to[n]\\
\forall x\in B,h_x(\phi_0(x))=z}}
w_v(i)E_{i,\phi_0(u)}\widetilde M_{u,\phi_0(u)}
\prod_{x\in B\setminus\{u\}}
(\bmW_{\{v,x\}})_{i,\phi_0(x)}
\xi_{\{v,x\},\phi_0(x)}.\\
=&\beta_-\sum_{\phi_0:B\to[n]}
b(\phi_0)\widetilde M_{u,\phi_0(u)}
\prod_{x\in B\setminus\{u\}}\xi_{\{v,x\},\phi_0(x)}\sum_{z=1}^d
\mathbf1_{\forall x\in B,h_x(\phi_0(x))=z}.
\end{align*}
For a fixed $\phi_0$, the last sum equals one if all the displayed hash values are equal, and zero otherwise. Indeed, when they agree, the unique admissible bucket is $h_u(\phi_0(u))$. Consequently, for each boundary map $\phi_0:B\to[n]$, let
\[
U_{\phi_0}:=
\mathbf1_{\forall x\in B,h_x(\phi_0(x))=h_u(\phi_0(u))}
\widetilde M_{u,\phi_0(u)}
\prod_{x\in B\setminus\{u\}}\xi_{\{v,x\},\phi_0(x)}.
\]
Then we have $Y=\beta_-\sum_{\phi_0}b(\phi_0)U_{\phi_0}$. Note that $b(\phi_0)$ does not depend on the hashes or signs drawn for this correction, then
\begin{align*}
\Var[Y]
&=\beta_-^2\sum_{\phi_0,\psi_0:B\to [n]}
b(\phi_0)b(\psi_0)\Cov(U_{\phi_0},U_{\psi_0}).
\end{align*}
The key step is to establish the following bound on the covariance row sums.
\begin{claim}\label{gh:correction-covariance-row-sum}
$\max_{\phi_0:B\to [n]}\sum_{\psi_0:B\to [n]}|\Cov(U_{\phi_0},U_{\psi_0})| = O\left(d^{-r}\right)$.
\end{claim}
Then we can bound $\Var[Y]$. By the inequality $2|ab| \le a^2 + b^2$ and \Cref{gh:correction-covariance-row-sum}, 
\begin{align*}
\Var[Y]
&\le\beta_-^2\sum_{\phi_0,\psi_0:B\to [n]}
|b(\phi_0)b(\psi_0)||\Cov(U_{\phi_0},U_{\psi_0})|\\
&\le\frac{\beta_-^2}{2}
\sum_{\phi_0,\psi_0:B\to [n]}
\bigl(b(\phi_0)^2+b(\psi_0)^2\bigr)|\Cov(U_{\phi_0},U_{\psi_0})|\\
&=\frac{\beta_-^2}{2}
\left(\sum_{\phi_0:B\to [n]}b(\phi_0)^2\sum_{\psi_0:B\to [n]}|\Cov(U_{\phi_0},U_{\psi_0})|
+\sum_{\psi_0:B\to [n]}b(\psi_0)^2\sum_{\phi_0:B\to [n]}|\Cov(U_{\phi_0},U_{\psi_0})|\right)\\
&\le O\left(\beta_-^2d^{-r}\sum_{\phi_0}b(\phi_0)^2\right) = O\left(\beta_-^2d^{-r}\|\bmE\bmF\|_F^2\right).
\end{align*}
Each of the $p$ private reporting vertices accounts for a distinct private edge, so $p\le\ell_{K_-} =\ell_{K} - r - 1$. Since $\tau=\Theta(\log n)$, $\veps_1=\Theta(\veps_0)$ and $\sigma = \Theta(\log n/\veps_0)$, we have $\beta = \tau^{\ell_K}(1+\tau\sigma)^p = O(\log^{3\ell_K - 2r - 2} n/\veps_0^{\ell_K - r - 1})$. Then by $s \le r$, $\beta_{-} = \beta/\tau^{r + 1}$, and Inequality (\ref{gh:stage-center-bound}), we have
\[
 \Var[Y] \le O\left(\frac{\log^{6\ell_K - 6r - 6} n}{\veps_0^{2\ell_K - 2r - 2}}\cdot\frac{nd\log^{r + 2s + 2}(\frac{n}{\delta_0})}{\veps_0^{2(s+1)}}\right) = O\left(\frac{nd\log^{6\ell_K - 4}(\frac{n}{\delta_0})}{\veps_0^{2\ell_K}}\right).
\]
Consequently,
\begin{align*}
\Var[\widetilde\Delta]
&=d^{2k_K-6}\sum_{a=1}^d\Var[Y^{(a)}]+\sum_{i=1}^n\Var[L_i]\\
&=d^{2k_K-5}\Var[Y]+2n\left(\frac{4\ell_K\beta d^{k_K-2}}{\veps_1}\right)^2\le O\left(\frac{nd^{2k_K-4}\log^{6\ell_K -4}(\frac{n}{\delta_0})}{\veps_0^{2\ell_K}}\right).
\end{align*}

Now we only need to prove \Cref{gh:correction-covariance-row-sum}.

\begin{proof}[Proof of \Cref{gh:correction-covariance-row-sum}]
Fix $\phi_0:B\to[n]$. 
Write each internal Laplace noise as $\sigma\eta_{x,i}$, where the $\eta_{x,i}$ are independent samples from $\Lap(1)$. At a private reporting vertex $x$, the corresponding unclipped report has the form 
\begin{align}
\widetilde M_{x,i}
=\frac{\widetilde Q_{x,i}(h_x(i))}{1+\tau\sigma}
+\frac{\sigma}{1+\tau\sigma}\eta_{x,i}.
\label{gh:Q-expansion}
\end{align}
Let $\mathcal S$ be the family of all subsets $S$ of the private reporting vertices in $K_-$ such that no vertex in $S$ is an ancestor of another in the chosen tree. Repeated substitution of \Cref{gh:Q-expansion} and the product formula for $\widetilde Q_{x,i}$ gives
\begin{equation}
U_{\phi_0}
=c_{\phi_0,\varnothing}(h,\xi)
+\sum_{S\in\mathcal S\setminus\{\varnothing\}}\sum_{\theta:S\to[n]}
c_{\phi_0,S,\theta}(h,\xi)
\prod_{x\in S}\eta_{x,\theta(x)}.
\label{gh:correction-noise-expansion}
\end{equation}
Here $\theta$ specifies the image indexing each noise term in the product, and $c_{\phi_0,S,\theta}(h,\xi)$ is its coefficient. For $S=\varnothing$, there is a single coefficient, denoted by $c_{\phi_0,\varnothing}(h,\xi)$.

Since $\E[\eta_{x,i}]=0$ and $\E[\eta_{x,i}^2]=2$, independence gives
\[
\E_\eta\left[
\prod_{x\in S}\eta_{x,\theta(x)}
\prod_{y\in S'}\eta_{y,\theta'(y)}
\right]
=
\begin{cases}
2^{|S|},&S=S'\ \text{and }\theta=\theta',\\
0,&\text{otherwise}.
\end{cases}
\]
Thus Equation (\ref{gh:correction-noise-expansion}) yields the exact identities
\begin{align}
\E[U_{\phi_0}]
=&\E_{h,\xi}[c_{\phi_0,\varnothing}(h,\xi)],\notag\\
\E[U_{\phi_0}U_{\psi_0}]
=&\E_{h,\xi}[c_{\phi_0,\varnothing}(h,\xi)c_{\psi_0,\varnothing}(h,\xi)]+\sum_{S\in\mathcal S\setminus\{\varnothing\}}2^{|S|}\sum_{\theta:S\to[n]}
\E_{h,\xi}[c_{\phi_0,S,\theta}(h,\xi)c_{\psi_0,S,\theta}(h,\xi)],\notag\\
\Cov(U_{\phi_0},U_{\psi_0})=&\E[U_{\phi_0}U_{\psi_0}] - \E[U_{\phi_0}]\E[U_{\psi_0}]\notag\\
=&\Cov(c_{\phi_0,\varnothing}(h,\xi),c_{\psi_0,\varnothing}(h,\xi))+\sum_{S\in\mathcal S\setminus\{\varnothing\}}2^{|S|}
\sum_{\theta:S\to[n]}
\E_{h,\xi}[c_{\phi_0,S,\theta}(h,\xi)c_{\psi_0,S,\theta}(h,\xi)].\notag
\end{align}
Thus, we can write $\Cov(U_{\phi_0},U_{\psi_0})=\sum_{S\in\mathcal S}2^{|S|}c_{\phi_0,\psi_0}(S)$, where
\begin{align}
c_{\phi_0,\psi_0}(S) = \begin{cases}\Cov(c_{\phi_0,\varnothing}(h,\xi),c_{\psi_0,\varnothing}(h,\xi)),&S=\varnothing,\\\sum_{\theta:S\to[n]}
\E_{h,\xi}[c_{\phi_0,S,\theta}(h,\xi)c_{\psi_0,S,\theta}(h,\xi)],&\text{otherwise}.\end{cases}
\label{eq:def-c(S)}
\end{align}
Then we can bound $\sum_{\psi_0:B\to[n]}|\Cov(U_{\phi_0},U_{\psi_0})|$ by bounding $|c_{\phi_0,\psi_0}(S)|$. In bounding the absolute value of $c_{\phi_0,\psi_0}(S)$, we may replace its normalization factors by one, since $1/\tau\le1$, $1/(1+\tau\sigma)\le1$, and $\sigma/(1+\tau\sigma)\le1$. This does not worsen the worst-case powers of $\log n$ and $\veps_0^{-1}$ in the total variance bound, since the final Laplace noises already attain these powers.

Each contribution to $c_{\phi_0,\psi_0}(S)$ pairs an expanded term from $U_{\phi_0}$ with one from $U_{\psi_0}$. To bound $|c_{\phi_0,\psi_0}(S)|$, we represent each such pair by an auxiliary graph $G_{\phi_0, \psi_0}(S)$. 

For a term from $U_{\phi_0}$, start with a copy of the chosen spanning tree of $K_-$ rooted at $u$, and delete the proper descendants of vertices in $S$. Write $\phi(x)$ for the image assigned to the retained vertex $x$, with $\phi(u)=\phi_0(u)$; a retained tree edge $\{x,y\}$ represents $(\bmW_{\{x,y\}})_{\phi(x),\phi(y)}$.

For each non-tree edge $e=\{x,y\}$ directed as $(x,y)$ with $x$ retained and $x\notin S$, the edge sum introduces a separate candidate image $j$, which need not equal the image assigned to $y$ even if $y$ is retained. We represent $j$ by a new leaf and connect it to $x$ with factor $(\bmW_e)_{\phi(x),j}$. This leaf is associated with pattern vertex $y$ and uses the same hash variable $h_y(j)$. Also add an isolated vertex carrying $\phi_0(x)$ for each $x\in B\setminus\{u\}$. These vertices and the root give $|B|$ designated vertices with images fixed by $\phi_0$.

Add the vertices and edges for the term from $U_{\psi_0}$ in the same way, using its own candidate images and the prescribed images $\psi_0(x)$. To obtain $G_{\phi_0, \psi_0}(S)$, merge all auxiliary vertices associated with the same pattern vertex and image, equivalently the same variable $h_x(i)$. %

We show that every connected component of $G_{\phi_0, \psi_0}(S)$ with a nonzero contribution to $c_{\phi_0,\psi_0}(S)$ contains a designated vertex with image $\phi_0(x)$ for some $x\in B$. This allows us to choose a spanning forest rooted at these designated vertices and sum over each remaining image at a cost of at most $d$, without an extra factor of $n$ from an unfixed root image. Avoiding this extra factor of $n$ is necessary to obtain the desired bound $O(d^{-r})$.

We prove the property by contradiction. Every auxiliary vertex introduced for the term from $U_{\phi_0}$ lies in a connected component containing a designated vertex with image $\phi_0(x)$ for some $x\in B$. Hence a component without one contains only vertices from the graph for $U_{\psi_0}$.

If this component contains no vertex of the auxiliary tree for $U_{\psi_0}$, it contains an originally isolated vertex carrying $\psi_0(x)$. The isolated vertex carrying $\psi_0(x)$ has no matching occurrence: every other occurrence for this edge lies in one of the two trees or at the designated vertex carrying $\phi_0(x)$, and an equal index would merge that vertex into the component. Thus the corresponding sign $\xi_{\{v,x\},\psi_0(x)}$ occurs only once, giving zero product expectation and hence zero contribution for $S\ne\varnothing$. For $S=\varnothing$, $c_{\phi_0,\psi_0}(\varnothing)=\E[c_{\phi_0,\varnothing}(h,\xi)c_{\psi_0,\varnothing}(h,\xi)]-\E[c_{\phi_0,\varnothing}(h,\xi)]\E[c_{\psi_0,\varnothing}(h,\xi)]$ by \Cref{eq:def-c(S)}; at least one individual expectation of the paired terms is also zero, so their contribution vanishes.

Otherwise, the component contains the auxiliary tree for $U_{\psi_0}$. If $S\ne\varnothing$, by \Cref{eq:def-c(S)}, any $x\in S$ remains in both trees with the same image $\theta(x)$ indexing the noise, so merging these vertices joins the two trees, a contradiction. Therefore $S=\varnothing$. For each $x\in B\setminus\{u\}$, let $i'_x$ be the image assigned to the retained vertex $x$ in the tree for $U_{\psi_0}$. No auxiliary vertex introduced for $U_{\phi_0}$ can have the same pattern vertex $x$ and image $i'_x$, since it would then be merged with the retained vertex $x$ in the auxiliary tree for $U_{\psi_0}$. For the contribution to be nonzero, the corresponding sign $\xi_{\{v,x\},i'_x}$ must match $\xi_{\{v,x\},\psi_0(x)}$, forcing $i'_x=\psi_0(x)$ and merging the isolated vertex carrying $\psi_0(x)$ with the retained vertex $x$. All the originally isolated vertices for $U_{\psi_0}$ therefore merge into its tree. The entire graph for $U_{\psi_0}$ is disjoint from the graph for $U_{\phi_0}$ after merging, so the two terms use disjoint sets of independent hashes and signs. Their covariance is zero, so their contribution to $c_{\phi_0,\psi_0}(\varnothing)$ in \Cref{eq:def-c(S)} is zero, a contradiction. This proves the required component property.

For fixed $S$, group assignments according to which occurrences of the same pattern vertex have equal images. There are $O(1)$ possible identifications, since $H$ is fixed and the auxiliary graphs have $O(1)$ vertices. Fix one such identification contributing a nonzero term, and let $q$ be the number of vertices after merging, or equivalently the number of distinct hash variables. The hash constraints in the two terms require these $q$ independent uniform variables on $[d]$ to have the same value, with probability $d^{1-q}$. Each absolute contribution is bounded by $O(d^{1-q})$ times its vertex-weight and matrix factors.

Choose a spanning forest rooted at the $|B|$ designated vertices, with exactly one root in each tree; the component property ensures that such a forest exists. These roots remain distinct because they are associated with different pattern vertices. The forest has $q-|B|$ edges. Discard all vertex-weight factors and all matrix factors on edges outside the forest; they lie in $[0,1]$, so this only increases the nonnegative sum. Summing over $\psi_0$ and the other images now leaves $q-|B|$ images not fixed by $\phi_0$. Drop the restrictions requiring images to be distinct. By symmetry and the row-sum bound of each $\bmW_e\in\calP_d$, summing over the image of a leaf contributes at most $d$. Repeating from the leaves towards the roots bounds the sum over these images by $d^{q-|B|}$. Thus each identification contributes at most $O\bigl(d^{1-q}d^{q-|B|}\bigr)=O\bigl(d^{1-|B|}\bigr)$. There are $O(1)$ identifications and sets $S$, and $2^{|S|}=O(1)$. Consequently, we have
\[
\sum_{\psi_0:B\to[n]}|\Cov(U_{\phi_0},U_{\psi_0})|
\le\sum_{S\in\mathcal S}2^{|S|}\sum_{\psi_0:B\to[n]}|c_{\phi_0,\psi_0}(S)|
=O\left(d^{1-|B|}\right) = O(d^{-r}),
\]
uniformly over $\phi_0$.
\end{proof}

\end{proof}

By \Cref{gh:direct-correction-mean} and Chebyshev's inequality, we have
$\Pr[
\left|\widetilde\Delta-\Delta\right|>\sqrt{2\Var[\widetilde{\Delta}]/\zeta}]\le\zeta/2$.
Combining this with \Cref{gh:direct-correction-agreement} and the union bound yields
\begin{align*}
&\Pr\left[
\left|\widehat\Delta-\Delta\right|>\sqrt{\frac{2\Var\left[\widetilde{\Delta}\right]}{\zeta}}\right]\le\Pr[\widehat\Delta\ne\widetilde\Delta]
+\Pr\left[
|\widetilde\Delta-\Delta|>\sqrt{\frac{2\Var\left[\widetilde{\Delta}\right]}{\zeta}}
\right]\le\frac{\zeta}{4}+\frac{\zeta}{2}<\zeta,
\end{align*}
which proves the stated accuracy guarantee by \Cref{gh:direct-correction-variance}.

\paragraph{Rounds and runtime.}
Only vertices with a private outgoing edge require an internal reporting round, followed by the final scalar round. Each such vertex accounts for a distinct private edge, proving the stated round bound. The public hash check, evaluation of the local queries, and normalization and aggregation of the reports run in $\poly(n,d)$ time.
\end{proof}

\subsubsection{Constructing a Public Center}
\label{gh:center-construction}
\label{gh:centers}

We extend the two-query construction in \Cref{overview:degree} to products of incident edge factors. The matrix $\bmF\in\reals^{n\times n^r}$ from \Cref{gh:report-guarantees} records products of the incident factors; it is used to state the error guarantee, not formed by the algorithm. The complete procedure is given in \Cref{gh:center-algorithm}, and we describe its steps below.

\begin{algorithm}[!htb]
\small
\SetAlgoLined
\caption{PC's center subroutine with Gaussian reports.}
\label{gh:center-algorithm}
\KwData{Graph $G=([n],E)$, represented by adjacency row vectors  $\bmA_1,\ldots,\bmA_n\in\{0,1\}^n$, a target integer degree parameter $1\le d\le n-1$, public vertex weights $w_1, \ldots, w_n\in [0,1]$ and incident edge factors $\bmW_1,\ldots,\bmW_{r}$, each marked private or public, privacy parameters $(\veps_0,\delta_0)$ and failure probability $\zeta$.}
\KwResult{A public matrix $\bmP\in\calP_d$.}

 \tcc{Public parameters and sign matrices}
 $[s]\;$Let $s$ be the number of private edge factors and number them first\;
 $[s]\;$Let $m\gets\left\lceil32\cdot3^r\bigl(2d\ln(2n+1)+\ln(4/\zeta)\bigr)\right\rceil$\;
 $[s]\;$Let $\veps_1\gets\veps_0/(2(s+1))$, $\delta_1\gets\delta_0/(2(s+1))$ and $\sigma_1\gets\sqrt{2\ln(2/\delta_1)}/\veps_1$\;
 $[s]\;$Let $C\gets\max\left\{2048(r+1)3^r/\zeta,\left(128(r+1)3^{r/2}\right)^r\right\}$, $L\gets\ln(Cn/\delta_0)$\;
 $[s]\;$Let $\theta_1\gets Cd^{r/2}L^{(r+2s)/2}/(2\veps_0^s)$ and $\theta_2\gets d+2\sigma_1\sqrt{nm}$\;
 $[s]\;$For all $\alpha\in[r]$, sample $\bmS^{(\alpha)}\in \reals^{n\times m}$ with independent uniform entries in $\{-1/\sqrt m,1/\sqrt m\}$\;

 \tcc{First round: private incident factors}
 \For{$i\gets1$ \KwTo $n$}{
  \For{$\alpha\gets1$ \KwTo $s$}{
   $[v_i]\;$Let $\bmB_i^{(\alpha)}\gets\bmA_i\bmS^{(\alpha)}+\calN(0,\sigma_1^2\bmI_m)$ and upload $\bmB_i^{(\alpha)}$ to the server\;
  }
 }
 $[s]\;$For all $s<\alpha\le r$, let $\bmB^{(\alpha)}\gets\bmW_\alpha\bmS^{(\alpha)}$\;
 $[s]\;$Let $\bmB\in\reals^{n\times m}$ be the matrix with entries $B_{i,j}=w_i m^{(r-1)/2}\prod_{\alpha=1}^{r}B^{(\alpha)}_{i,j}$ for all $i\in[n]$ and $j\in[m]$\;
 $[s]\;$\If{$\max_{i\in[n]}\|\bmB_i\|_2>2\theta_1$}{\KwRet{$\bmP=0$}\;}
 \tcc{Second round: private estimates of $\bmA\bmB$}
 $[s]\;$Let $\sigma_2\gets2\theta_1\sigma_1$\;
 \For{$i\gets1$ \KwTo $n$}{
  $[v_i]\;$Let $\bmC_i\gets\bmA_i\bmB+\calN(0,\sigma_2^2\bmI_m)$ and upload $\bmC_i$ to the server\;
 }
 $[s]\;$\If{$\max_{i\in[n]}\|\bmC_i\|_2>2\theta_1\theta_2$}{\KwRet{$\bmP=0$}\;}
\tcc{Fit the public center}
$[s]\;$Compute $\bmP\in\calP_d$ such that
$\|\bmP\bmB-\bmC\|_F^2
 \le\min_{\bmP'\in\calP_d}\|\bmP'\bmB-\bmC\|_F^2+1$\;
\KwRet{$\bmP$.}
\end{algorithm}

Use the $r$ incident factors and weights from \Cref{gh:report-guarantees}; the first $s$ factors are private and equal to $\bmA$, and the others belong to $\calP_d$ and are public. For each factor, use an independent sign matrix $\bmS^{(\alpha)}\in\reals^{n\times m}$ and write $\bmB^{(\alpha)}$ for its noisy private sketch or exact public sketch. We form the combined sketch $\bmB\in\reals^{n\times m}$ by multiplying these sketches coordinate-wise and scaling row $i$ by $w_i m^{(r-1)/2}$. The noise scales $\sigma_1,\sigma_2$ correspond to the two reporting rounds. Before the second query, a public check requires every row of $\bmB$ to have norm at most $2\theta_1$. The parties then release noisy rows of $\bmA\bmB$, forming $\bmC\in\reals^{n\times m}$, and a second public check bounds its row norms by $2\theta_1\theta_2$. These checks bound sensitivity and optimization inputs on every transcript; a failed check returns $\bmP=0$. After both checks pass, the server fits the center by choosing $\bmP\in\calP_d$ whose squared residual $\|\bmP\bmB-\bmC\|_F^2$ is within an additive $1$ of the minimum over $\calP_d$. Now we give the formal proof of \Cref{gh:center}. 

\begin{proof}[Proof of \Cref{gh:center}]
We prove the privacy, accuracy, rounds, and runtime guarantees separately.

\paragraph{Privacy.}
Every row of $\bmS^{(\alpha)}$ has Euclidean norm $1$. Given the preceding transcript, changing one coordinate of a private row changes its first query by a signed row of $\bmS^{(\alpha)}$. Before the second query, the first check bounds the $\ell_2$-norm of every row of $\bmB$ by $2\theta_1$. Thus the two $\ell_2$-sensitivities are $1$ and $2\theta_1$, respectively. By \Cref{lemma:Gaussian mechanism}, each release of $\bmB_i^{(\alpha)}$ or $\bmC_i$ is $(\veps_1,\delta_1)$-DP. Each changed endpoint participates in $s+1$ calls, so \Cref{lemma:basic composition} gives $(\veps_0,\delta_0)$-LEDP. The checks and fitting are postprocessing.

\paragraph{Accuracy.} For accuracy, fix $G\in\calG_{n,d}$ and condition on the preceding centers that determine this call's weights and public edge factors. Write the Gaussian releases as $\bmB^{(\alpha)}=\bmA\bmS^{(\alpha)}+\bmG_1^{(\alpha)}$ for $\alpha\in[s]$ and $\bmC=\bmA\bmB+\bmG_2$, where the noise entries are independent mean-zero Gaussian variables with variances $\sigma_1^2$ and $\sigma_2^2$, respectively, and are independent of the sign matrices $(\bmS^{(\alpha)})_{\alpha \in [r]}$. For analysis, we may generate the Gaussian noises even when an earlier public check returns $\bmP=0$, leaving those samples unobserved. This does not change the algorithm's output. Let $\bmE:=\bmA-\bmP$. We first compare $\|(\bmA-\bmP')\bmF\|_F$ and $\|(\bmA-\bmP')\bmB\|_F$ uniformly over all feasible centers $\bmP'$. The optimization step then controls $\|\bmE\bmB\|_F$, and bounds on the row norms of $\bmB$ and $\bmC$ and on $\|\bmG_2\|_F$ ensure that both public checks pass and give the required error bound. %

\begin{lemma}\label{gh:claim-center-uniform}
With probability at least $1-\zeta/4$, whenever $\max_{i\in[n]}\|\bmB_i\|_2\le2\theta_1$, we have, simultaneously for every $\bmP'\in\calP_d$,
\begin{align*}
\|(\bmA-\bmP')\bmF\|_F^2
\le O\left(\|(\bmA-\bmP')\bmB\|_F^2+nd\theta_1^2\right).
\end{align*}
\end{lemma}

\begin{proof}
We first establish a comparison for a finite set of coefficient row vectors and then extend it to every $\bmx\in\reals^n$ with $\|\bmx\|_1\le2d$. Each row of $\bmA-\bmP'$ satisfies this bound, so summing the resulting inequalities gives the required comparison. Since the bound holds simultaneously for every $\bmP'\in\calP_d$, it also holds for the center $\bmP$ selected by the algorithm.

\textbf{Bounding the second and fourth moments.} For a fixed coefficient row vector $\bma$, let $R_j=\sqrt m(\bma\bmB)_j$ be its rescaled $j$th sketch coordinate. Condition on the sign matrices $\bmS^{(1)},\ldots,\bmS^{(r)}$. Recall that the edges incident to $v$ other than $\{v,u\}$ are $\{v,u_1\},\ldots,\{v,u_{r}\}$. Since the Gaussian noises are independent across factors and have mean zero, expanding the matrix products and following \Cref{gh:feature} gives
\begin{align*}
\E[R_j\mid\bmS^{(1)},\ldots,\bmS^{(r)}]
&=m^{\frac{r}{2}}\sum_{i=1}^n a_iw_i
\prod_{\alpha=1}^r(\bmW_\alpha\bmS^{(\alpha)})_{i,j}=m^{\frac{r}{2}}\sum_{\phi:\{u_1,\dots, u_r\}\to [n]}(\bma\bmF)_\phi\prod_{\alpha=1}^r S^{(\alpha)}_{\phi(u_\alpha),j}.
\end{align*}
The independent uniform signs satisfy $\E\left[S^{(\alpha)}_{t,j}S^{(\alpha)}_{t',j}\right]
=(1/m)\bm1_{t=t'}$. Thus, conditional Jensen's inequality and independence across the sign matrices give
\begin{align*}
\E[R_j^2] &= \E[\E[R_j^2\mid \bmS^{(1)},\ldots,\bmS^{(r)}]]\ge\E\left[\left(\E[R_j\mid\bmS^{(1)},\ldots,\bmS^{(r)}]\right)^2\right]\\
&=m^r\sum_{\phi,\psi:\{u_1,\dots, u_r\}\to [n]}(\bma\bmF)_\phi(\bma\bmF)_\psi
\prod_{\alpha=1}^r
\E\left[S^{(\alpha)}_{\phi(u_\alpha),j}S^{(\alpha)}_{\psi(u_\alpha),j}\right]\\
&=\sum_{\phi,\psi:\{u_1,\dots, u_r\}\to [n]}(\bma\bmF)_\phi(\bma\bmF)_\psi
\prod_{\alpha=1}^r\bm1_{\phi(u_\alpha)=\psi(u_\alpha)}\\
&=\sum_{\phi:\{u_1,\dots, u_r\}\to [n]}(\bma\bmF)_\phi^2
=\|\bma\bmF\|_2^2.
\end{align*}
Also, $R_j$ is linear in each of $r$ independent blocks of signs and standard Gaussian variables. To bound its fourth moment, write $R_j=\sum_t c_tZ_t$ using the variables in the last block. Here the $Z_t$ are independent uniform signs in $\{-1,1\}$ or standard Gaussian variables, with their original scales absorbed into the coefficients $c_t$. Both types are centered and have variance one, and their fourth moments are one and three, respectively. The coefficients $c_t$ are linear in each of the other $r-1$ blocks and independent of the $Z_t$. 

We prove the fourth-moment bound by induction on the number $r$ of independent blocks. For $r=0$, the empty product gives $R_j=\bma\bmF$ deterministically, so $\E[R_j^4]=(\E[R_j^2])^2$, establishing the base case. Now let $r\ge1$ and assume that every expression $X$ linear in each of $r-1$ independent blocks of these variables satisfies $\E[X^4]\le3^{r-1}(\E[X^2])^2$. In particular, each coefficient $c_t$ satisfies $\E[c_t^4]\le3^{r-1}(\E[c_t^2])^2$. Expanding $R_j^4$ and using the independence and centering of the $Z_t$, followed by their fourth-moment bounds, Cauchy--Schwarz, and the induction hypothesis, gives
\begin{align*}
\E[R_j^4]
&=\sum_t\E[c_t^4]\E[Z_t^4]
  +6\sum_{t<t'}\E[c_t^2c_{t'}^2]\E[Z_t^2]\E[Z_{t'}^2]\\
&\le3\sum_t\E[c_t^4]
  +6\sum_{t<t'}\E[c_t^2c_{t'}^2]=3\sum_{t,t'}\E[c_t^2c_{t'}^2]\\
&\le3\sum_{t,t'}\sqrt{\E[c_t^4]\E[c_{t'}^4]}=3\left(\sum_t\sqrt{\E[c_t^4]}\right)^2\\
&\le3\left(\sum_t\sqrt{3^{r-1}(\E[c_t^2])^2}\right)^2=3^r\left(\sum_t\E[c_t^2]\right)^2=3^r\left(\E[R_j^2]\right)^2.
\end{align*}

\textbf{A comparison on a finite set.} Put $p_r=1/(4\cdot3^r)$. If $\E[R_j^2]>0$, the Paley--Zygmund inequality gives
\begin{align*}
\Pr\left[R_j^2\ge\frac{\E[R_j^2]}{2}\right]
\ge\frac{(\E[R_j^2])^2}{4\E[R_j^4]}
\ge p_r.
\end{align*}
The columns are independent and identically distributed, so a Chernoff bound shows that at least $p_rm/2$ columns satisfy this inequality, except with probability $e^{-p_rm/8}$. On this event,
\begin{align*}
\|\bma\bmB\|_2^2
=\frac{1}{m}\sum_{j=1}^mR_j^2
\ge\frac{p_r}{4}\E[R_j^2]
\ge\frac{p_r}{4}\|\bma\bmF\|_2^2.
\end{align*}
When $\E[R_j^2]=0$, the last inequality holds deterministically.

Let $\bme_i$ be the $i$th coordinate row vector, and let $\calA_d$ be the set of all sums of $2d$ vectors from $\{0,\pm\bme_1,\ldots,\pm\bme_n\}$. There are at most $(2n+1)^{2d}$ such sums. By the choice of $m$, a union bound gives
\begin{align*}
\Pr\left[\exists\bma\in\calA_d,\|\bma\bmB\|_2^2<\frac{p_r}{4}\|\bma\bmF\|_2^2\right]
\le(2n+1)^{2d}e^{-\frac{p_rm}{8}}
\le\frac{\zeta}{4}.
\end{align*}
Hence, with probability at least $1-\zeta/4$, for every $\bma\in\calA_d$, we have
\begin{equation}
\|\bma\bmF\|_2^2\le\frac{4}{p_r}\|\bma\bmB\|_2^2.
\label{gh:finite-atom-comparison}
\end{equation}

\textbf{Returning to the residual rows.} Suppose Inequality (\ref{gh:finite-atom-comparison}) holds and $\max_{i\in[n]}\|\bmB_i\|_2\le2\theta_1$. For any $\bmx\in\reals^n$ with $\|\bmx\|_1\le2d$, draw $2d$ independent row vectors $\bmY_1,\ldots,\bmY_{2d}$, each taking the value $\operatorname{sgn}(x_i)\bme_i$ with probability $|x_i|/(2d)$ for each $i\in[n]$, and the zero vector with probability $1-\|\bmx\|_1/(2d)$. Their sum $\bma:=\sum_{t=1}^{2d}\bmY_t$ belongs to $\calA_d$ and has mean $\bmx$. In the following calculations, expectations are over these auxiliary vectors, with $\bmB$ fixed. For each $t\in[2d]$, we have $\E[\bmY_t\bmB]=\bmx\bmB/(2d)$ and 
\begin{align*}
\E[\|\bmY_t\bmB\|_2^2]
&=\sum_{i=1}^n\frac{|x_i|}{2d}
  \operatorname{sgn}(x_i)^2\|\bmB_i\|_2^2=\sum_{i=1}^n\frac{|x_i|}{2d}\|\bmB_i\|_2^2 \le \frac{2\theta_1^2\|\bmx\|_1}{d} \le 4\theta_1^2,
\end{align*}
where the last two inequalities use $\|\bmB_i\|_2\le2\theta_1$ for every $i\in[n]$ and $\|\bmx\|_1\le2d$. Expanding the squared norm and using independence for the cross terms gives
\begin{align*}
\E[\|\bma\bmB\|_2^2]
&=\sum_{t=1}^{2d}\E[\|\bmY_t\bmB\|_2^2]
+\sum_{t\ne t'}\E\left[(\bmY _t\bmB)(\bmY_{t'}\bmB)^\top\right]\\
&\le 8d\theta_1^2
+2d(2d-1)\left\|\frac{\bmx\bmB}{2d}\right\|_2^2\le\|\bmx\bmB\|_2^2+8d\theta_1^2.
\end{align*}
Jensen's inequality and Inequality (\ref{gh:finite-atom-comparison}) now imply
\begin{align*}
\|\bmx\bmF\|_2^2=\|(\E[\bma])\bmF\|_2^2
=\|\E[\bma\bmF]\|_2^2\le\E[\|\bma\bmF\|_2^2]\le\frac{4}{p_r}\E[\|\bma\bmB\|_2^2]\le\frac{4}{p_r}\left(\|\bmx\bmB\|_2^2+8d\theta_1^2\right).
\end{align*}
These auxiliary vectors are used only in the proof after $\bmB$ is fixed, so this comparison holds on the same event for every $\bmx$ with $\|\bmx\|_1\le2d$.

For every $\bmP'\in\calP_d$ and $i\in[n]$, we have
\begin{align*}
\|(\bmA-\bmP')_i\|_1
\le\sum_{j=1}^nA_{i,j}+\sum_{j=1}^nP'_{i,j}
\le2d.
\end{align*}
Applying the rowwise comparison with $\bmx=(\bmA-\bmP')_i$ and summing over $i\in[n]$ gives
\begin{align*}
\|(\bmA-\bmP')\bmF\|_F^2
&\le\frac{4}{p_r}\left(\|(\bmA-\bmP')\bmB\|_F^2+8nd\theta_1^2\right)=O\left(\|(\bmA-\bmP')\bmB\|_F^2+nd\theta_1^2\right).
\end{align*}
\end{proof}

\begin{lemma}\label{gh:claim-center-fitting}
Whenever the algorithm reaches the optimization step, its output satisfies
\begin{align*}
\|\bmE\bmB\|_F\le2\|\bmG_2\|_F+1.
\end{align*}
\end{lemma}

\begin{proof}
Since $\bmA\in\calP_d$, it is an admissible competitor in the optimization step. The optimization guarantee and $\bmC=\bmA\bmB+\bmG_2$ give
\begin{align*}
\|\bmP\bmB-\bmC\|_F
\le\sqrt{\|\bmA\bmB-\bmC\|_F^2+1}
=\sqrt{\|\bmG_2\|_F^2+1}
\le\|\bmG_2\|_F+1.
\end{align*}
Using $\bmE=\bmA-\bmP$ and the triangle inequality, we obtain
\begin{align*}
\|\bmE\bmB\|_F
\le\|\bmA\bmB-\bmC\|_F+\|\bmP\bmB-\bmC\|_F
\le2\|\bmG_2\|_F+1.
\end{align*}
\end{proof}

\begin{lemma}\label{gh:claim-center-noise}
With probability at least $1-\zeta/2$, both public norm checks (on the row norms of $\bmB$ and $\bmC$) pass, and $\|\bmG_2\|_F\le2\sigma_2\sqrt{nm}$.
\end{lemma}

\begin{proof}
We first bound the row norms of $\bmB$ to verify the first public check. We then bound $\|\bmG_2\|_F$ for accuracy and use the same bound to verify the check on the row norms of $\bmC$.

\textbf{The first-round check.} We first bound each factor in the product defining $B_{i,j}$ to obtain a bound on $\|\bmB_i\|_2$. Put $t:=\ln(16(r+1)nm/\zeta)$. Every incident-factor row has squared $\ell_2$-norm at most $d$. Hoeffding's inequality and Gaussian tail bounds, followed by a union bound, give
\begin{align*}
|(\bmW_\alpha\bmS^{(\alpha)})_{i,j}|\le\sqrt{\frac{2dt}{m}},\qquad
|(\bmG_1^{(\alpha)})_{i,j}|\le\sigma_1\sqrt{2t},
\end{align*}
simultaneously for all $i\in[n]$, $j\in[m]$, and $\alpha\in[r]$ in the first inequality and $\alpha\in[s]$ in the second, except with probability at most $2(r+s)nm e^{-t}\le\zeta/4$. Whenever both displayed bounds hold, the product defining $\bmB$ and $w_i\le1$ imply, for every $i\in[n]$,
\begin{align*}
\|\bmB_i\|_2
&\le\sqrt m\max_{j\in[m]}|B_{i,j}|=w_i m^{\frac r2}\max_{j\in[m]}
  \prod_{\alpha=1}^r|B_{i,j}^{(\alpha)}|\\
&\le m^{\frac r2}
  \left(\sqrt{\frac{2dt}{m}}\right)^{r-s}
  \left(\sqrt{\frac{2dt}{m}}+\sigma_1\sqrt{2t}\right)^s\\
&=\left(\sqrt{2dt}\right)^{r-s}
  \left(\sqrt{2dt}+\sigma_1\sqrt{2mt}\right)^s\\
&=(2dt)^{\frac r2}
  \left(1+\sigma_1\sqrt{\frac md}\right)^s.
\end{align*}

To compare this row bound with $2\theta_1$, we now bound $t$ and $\sigma_1\sqrt{m/d}$ in terms of $L$ and $\veps_0$. For $t$, recall that $C\ge2048(r+1)3^r/\zeta$ and $L=\ln(Cn/\delta_0)$, so $L\ge1$, $\ln(2n+1)\le L$, and $\ln(4/\zeta)\le L$. The choice of $m$ therefore gives
\begin{align*}
m\le32\cdot3^r(2d+1)L+1\le128\cdot3^r dL.
\end{align*}
Substituting this bound into the definition of $t$ and using $d\le n$ gives
\begin{align*}
t
&\le\ln\left(\frac{2048(r+1)3^r n^2L}{\zeta}\right)
\le\ln C+2\ln n+\ln L
\le4L.
\end{align*}
For the factor $\sigma_1\sqrt{m/d}$, use $s\le r$ and $\ln(4(s+1)/\delta_0)\le L$ together with the preceding bound on $m$ to obtain
\begin{align*}
\sigma_1\sqrt{\frac{m}{d}}
&=\frac{2(s+1)}{\veps_0}
\sqrt{\frac{2m}{d}\ln\left(\frac{4(s+1)}{\delta_0}\right)}
\le\frac{32(r+1)3^{\frac{r}{2}}L}{\veps_0}.
\end{align*}
Since $L\ge1$ and $\veps_0\le1$, substituting these two bounds into the row bound yields
\begin{align*}
\|\bmB_i\|_2
&\le(8dL)^{\frac{r}{2}}
\left(\frac{33(r+1)3^{\frac{r}{2}}L}{\veps_0}\right)^s\\
&\le\left(128(r+1)3^{\frac{r}{2}}\right)^r
\frac{d^{\frac{r}{2}}L^{\frac{r+2s}{2}}}{\veps_0^s}\\
&\le C\frac{d^{\frac{r}{2}}L^{\frac{r+2s}{2}}}{\veps_0^s}
=2\theta_1.
\end{align*}
The second inequality uses $s\le r$ and $33\sqrt8<128$, and the last inequality is exactly the second lower bound in the definition of $C$. Thus the first norm check passes. 

\textbf{The second-round noise and final check.} The scale $\sigma_2=2\theta_1\sigma_1$ is fixed by the public parameters, and $\bmG_2$ has independent $\calN(0,\sigma_2^2)$ entries and is independent of $\bmB$. Viewing its $nm$ entries as one vector, whose Euclidean norm equals $\|\bmG_2\|_F$, we apply \Cref{lemma:gaussian-norm-tail} with $D=nm$ and $t=1$. This gives
\begin{align*}
\Pr\left[\|\bmG_2\|_F>2\sigma_2\sqrt{nm}\right]
\le e^{-\frac{nm}{2}}\le\frac{\zeta}{4},
\end{align*}
where the last inequality follows from $m\ge2\ln(4/\zeta)$. 
On the event $\|\bmG_2\|_F\le 2\sigma_2\sqrt{nm}$ and whenever the first norm check passes, for every $i\in[n]$ we have
\begin{align*}
\|\bmC_i\|_2
&\le\left\|\sum_{j=1}^n A_{i,j}\bmB_j\right\|_2+\|(\bmG_2)_i\|_2\\
&\le\sum_{j=1}^n A_{i,j}\|\bmB_j\|_2+\|\bmG_2\|_F\\
&\le2d\theta_1+2\sigma_2\sqrt{nm}
=2\theta_1\theta_2,
\end{align*}
so the second norm check also passes. A union bound gives total failure probability at most $\zeta/4+\zeta/4=\zeta/2$.
\end{proof}

On the intersection of the events in \Cref{gh:claim-center-uniform,gh:claim-center-noise}, the algorithm reaches the optimization step, so \Cref{gh:claim-center-fitting} applies. Substituting $\bmP'=\bmP$ into the inequality in \Cref{gh:claim-center-uniform} and then using the bounds on $\|\bmE\bmB\|_F$ and $\|\bmG_2\|_F$, we obtain
\begin{align*}
\|\bmE\bmF\|_F
&\le O\left(\|\bmE\bmB\|_F+\theta_1\sqrt{nd}\right)\\
&\le O\left(\|\bmG_2\|_F+1+\theta_1\sqrt{nd}\right)\\
&\le O\left(\sigma_2\sqrt{nm}\right)\\
&\le O\left(\frac{\sqrt n\,d^{\frac{r+1}{2}}\log^{\frac{r+2s+2}{2}}(\frac{n}{\delta_0})}{\veps_0^{s+1}}\right).
\end{align*}
The third inequality uses $m\ge d$, $\theta_1,\sigma_1\ge1$, and $\sigma_2=2\theta_1\sigma_1$. The last inequality follows from the definitions of $\theta_1$, $m$, and $\sigma_1$, with $L=O(\log(n/\delta_0))$. A union bound gives success probability at least $1-\zeta/4-\zeta/2=1-3\zeta/4>1-\zeta$.

\paragraph{Rounds.}
The parties report $(\bmB_i^{(\alpha)})_{\alpha\in[s]}$ in the first round, if $s>0$, and $\bmC_i$ in the second. We can show that the first-round reports of all calls to \Cref{gh:center-algorithm} can be collected in a single round without changing the joint distribution of the reports and outputs. Fix the block roots, edge orders and endpoint choices in \Cref{gh:full-algorithm} in advance. The first-round reports depend only on $\bmA_i$, the fresh sign matrices and the public parameters; they do not depend on the vertex weights or preceding centers. Hence these reports from all calls can be collected in one initial round, using independent randomness across calls. Once a call's vertex weights and public edge factors are available, the server forms $\bmB$ and the parties need at most one further reporting round for $\bmC_i$. Each call uses only its own reports and preceding centers, so this schedule preserves the joint distribution used in the accuracy analysis.

\paragraph{Runtime.}
For fixed $r$ and $\zeta$, the sketch dimension satisfies $m=O(d\log n)$ so generating the random matrices and computing the reports take polynomial time. It remains to bound the running time of the optimization step.

Use the $N:=n(n-1)/2$ entries $P_{i,j}$ with $i<j$ as independent coordinates, setting $P_{j,i}=P_{i,j}$ and $P_{i,i}=0$. When $n=2$, fitting is a univariate convex quadratic minimization over $[0,1]$, solvable directly in $O(m)$ arithmetic operations. Henceforth assume $n\ge3$. The objective $f(\bmP):=\|\bmP\bmB-\bmC\|_F^2$ is convex and differentiable, and both its value and gradient can be computed in polynomial time. The feasible set $\calP_d$ is closed and convex, described by the $O(n^2)$ linear inequalities. Checking these inequalities either certifies feasibility or supplies a violated inequality separating the candidate from $\calP_d$. Both $f$ and its gradient can be evaluated in polynomial time, so the two required oracles have polynomial arithmetic cost.

In these coordinates, $\calP_d$ is contained in the ball of radius $R=\sqrt{dn/2}$ centered at the origin. It contains the ball of radius $r=d/(2n)$ centered at the vector whose coordinates all equal $d/(2n)$: every off-diagonal entry in this ball lies in $[0,d/n]$, and every row sum is at most $d(n-1)/n\le d$. Throughout the enclosing ball, $\|\bmP\|_F\le \sqrt{nd}$. Together with $\|\bmB\|_F\le2\theta_1\sqrt n$ and $\|\mathbf C\|_F\le2\theta_1\theta_2\sqrt n$, this gives
\[
0\le f(\bmP)\le\bigl(\|\bmP\bmB\|_F+\|\mathbf C\|_F\bigr)^2
\le\bigl(\|\bmP\|_F\|\bmB\|_F+\|\mathbf C\|_F\bigr)^2
\le (2\theta_1n\sqrt{d} + 2\theta_1\theta_2\sqrt{n})^2=:U,
\]
where the second and third inequalities follow from the triangle inequality and submultiplicativity of the Frobenius norm, respectively. Thus the difference between the maximum and minimum of $f$ on the enclosing ball is at most $U$. Applying \Cref{lemma:ellipsoid-method} with $\eta=1$ gives a feasible center with additive objective error at most $1$ after
\[
\left\lceil
2N^2\left(
\ln\left(\frac{R}{r}\right)
+\ln\left(1+U
\right)
\right)
\right\rceil+1=\operatorname{poly}\left(n,d,\veps_0^{-1},\log\left(\frac{1}{\delta_0}\right)\right)
\]
iterations, each with polynomial arithmetic cost in $\operatorname{poly}(n,m)$. Therefore, the total runtime has the claimed worst-case bound.
\end{proof}

\subsection{Pure Privacy} 
\label{gh:pure-extension}

We formalize the pure version of PC described in \Cref{overview:pure} and prove \Cref{gh:pure-main-theorem}.

\subsubsection{The Mechanism and the Counting Algorithm}

We first give a pure-private mechanism for releasing a linear query $\bma\bmW$, where $\bma$ is a private binary row and $\bmW$ is public. The following lemma bounds its distance from the corresponding Gaussian release.

\begin{lemma}\label{gh:pure-linear-query}
For any public matrix $\bmW\in\reals^{n\times m}$, public bound $\Delta_{\bmW}\ge\max_{j\in[n]}\|\bmW_j\|_2$, $\veps_1\in(0,1]$, and $L_1\ge\ln(100(n+m))$, the mechanism $\calR_{\bmW,\Delta_{\bmW},\veps_1,L_1}$ defined by \Cref{gh:pure-query-algorithm} is $\veps_1$-DP with respect to a one-bit change in $\bma\in\{0,1\}^n$. For every $\bma\in\{0,1\}^n$, its output satisfies
\begin{align}
d_{\mathrm{TV}}\left(\calD\left(\calR_{\bmW,\Delta_{\bmW},\veps_1,L_1}(\bma)\right),\calD\left(\bma\bmW+\frac{32\Delta_{\bmW}\sqrt{L_1}}{\veps_1}\bmg\right)\right)
\le8(n+m)e^{-32L_1},
\label{gh:pure-linear-query-tv}
\end{align}
where $\bmg\sim\calN(0,\bmI_m)$, $\calD(\cdot)$ denotes the distribution of a random vector, and $d_{\mathrm{TV}}$ denotes total variation distance. The algorithm runs in time $\operatorname{poly}(n,m,1/\veps_1,e^{L_1})$.
\end{lemma}

\begin{algorithm}[!htbp]
\small
\SetAlgoLined
\caption{A pure-DP mechanism approximating a Gaussian linear release.}
\label{gh:pure-query-algorithm}
\KwData{A private row $\bma\in\{0,1\}^n$, a public matrix $\bmW\in\reals^{n\times m}$, a public bound $\Delta_{\bmW}\ge\max_{j\in[n]}\|\bmW_j\|_2$, a privacy parameter $\veps_1\in(0,1]$, and a comparison parameter $L_1\ge\ln(100(n+m))$.}
\KwResult{A private estimate $\calR_{\bmW,\Delta_{\bmW},\veps_1,L_1}(\bma)$ of $\bma\bmW$.}
\If{$\Delta_{\bmW}=0$}{\KwRet{$\bm0$}\;}
\tcc{Normalize the query and bound the log-density changes}
Let $\bmD\gets\frac{\bmW}{\Delta_{\bmW}}$ and $\mathbf q\gets\bma\bmD$\;
Let $\gamma\gets\frac{\veps_1}{4}$ and $\sigma\gets\frac{32\sqrt{L_1}}{\veps_1}$\;
Let $\Omega\gets\{\mathbf p\in\reals^m:\|\mathbf p\|_\infty\le\gamma,\ |\bmD_j\mathbf p^\top|\le\gamma\text{ for every }j\in[n]\}$\;
Define $h(\mathbf z):=\max_{\mathbf p\in\Omega}\left(\mathbf z\mathbf p^\top-\frac{\sigma^2}{2}\|\mathbf p\|_2^2\right)$\;
\tcc{Sample on a public cube with a prescribed multiplicative error}
Let $R\gets n+\gamma\sigma^2$ and $\eta\gets\veps_1e^{-32L_1}$\;
Sample $\mathbf y\in[-R,R]^m$ with density within a factor $e^{\pm\eta}$, pointwise, of the normalized density proportional to $e^{-h(\mathbf y-\mathbf q)}$, using \Cref{gh:inexact-cube-sampling}\;
\KwRet{$\calR_{\bmW,\Delta_{\bmW},\veps_1,L_1}(\bma)=\Delta_{\bmW}\mathbf y$}\;
\end{algorithm}

The proof of \Cref{gh:pure-linear-query} is given in \Cref{sssec:proof-linear-query}. We now use this mechanism to replace the Gaussian releases in \Cref{gh:center-algorithm} and prove \Cref{gh:pure-main-theorem}.

\begin{proof}[Proof of \Cref{gh:pure-main-theorem}]
We obtain the pure version of PC by modifying its Gaussian protocol (\Cref{gh:full-algorithm}) as follows. If $n < k$ or $\ell = 0$, use the exact returns of \Cref{gh:full-algorithm}. For the remaining inputs, construct $\Pi$ as in that algorithm, let $N=2|\Pi|\ell$, and set $C_0:=100Nk3^k$. Run \Cref{gh:full-algorithm} with $\delta=(C_0n)^{-2048}$ and retain its original round schedule. In each call to \Cref{gh:center-algorithm}, replace the Gaussian mechanism in both reporting rounds with \Cref{gh:pure-query-algorithm}, using fresh independent randomness to generate each report. The public norm checks, the optimization step, and the correction procedures remain unchanged. Using the parameters $\veps_1,\delta_1,\theta_1$ in \Cref{gh:center-algorithm}, set $L_1:=\frac{1}{512}\ln(\frac{2}{\delta_1})$. In the first round, release $\bmB_i^{(\alpha)}\gets\calR_{\bmS^{(\alpha)},1,\veps_1,L_1}(\bmA_i)$ for every $i\in[n]$ and $\alpha\in[s]$; in the second round, release $\bmC_i\gets\calR_{\bmB,2\theta_1,\veps_1,L_1}(\bmA_i)$ for every $i\in[n]$.  All other steps remain unchanged.

The branches of exact returns satisfy all claims, so assume $n\ge k\ge2$ and $\ell\ge1$. We prove the privacy, accuracy, rounds, and runtime guarantees separately.

\paragraph{Privacy.}
We only need to verify the condition $L_1\ge\ln(100(n+m))$ required by \Cref{gh:pure-linear-query}.

Since $r\le k-2$, $\zeta=1/(12N)$, $d\le n$,
$\ln(2n+1)\le n$, and $\ln(48N)\le4N$, the choice of $m$ gives
\begin{align}
n+m
&\le n+1+32\cdot3^{k-2}(2n^2+4N)\notag\\
&\le(64N3^{k-2}+1)n^2
\le C_0n^2.\label{gh:pure-dimension-bound}
\end{align}
Here the second inequality uses $n+1\le n^2$ and
$2n^2+4N\le2Nn^2$, which hold for $n,N\ge2$,
and the last uses $C_0=100Nk3^k$.

The allocations $\delta_0=\delta/N$ and
$\delta_1=\delta_0/(2(s+1))$ therefore give
\begin{align*}
L_1
&=\frac{1}{512}\ln\left(\frac{4N(s+1)}{\delta}\right)
\ge\frac{1}{512}\ln\left(\frac{1}{\delta}\right)
=4\ln(C_0n)\\
&\ge\ln(100C_0n^2)
\ge\ln(100(n+m)),
\end{align*}
where we used $\delta=(C_0n)^{-2048}$ and $C_0\ge100$. Thus each replacement report is $\veps_1$-DP by \Cref{gh:pure-linear-query}, and the same composition and postprocessing arguments used in the privacy proofs for \Cref{gh:full-algorithm,gh:center-algorithm} show that the modified algorithm is $\veps$-LEDP.

\paragraph{Accuracy.}
For the releases of $\bmA_i\bmS^{(\alpha)}$ and $\bmA_i\bmB$, the choice of $L_1$ makes the Gaussian standard deviations in \Cref{gh:pure-linear-query-tv} equal to $\sigma_1$ and $\sigma_2$, respectively.
Since Inequality (\ref{gh:pure-dimension-bound}) and $\delta_1\le\delta=(C_0n)^{-2048}$, the total variation distance for each release is at most
\begin{align*}
8(n+m)e^{-32L_1}
&=8(n+m)\left(\frac{\delta_1}{2}\right)^{\frac{1}{16}}\le8C_0n^2(C_0n)^{-128}.
\end{align*}
There are at most $|\Pi|\ell$ center calls, each making at most $n(s+1)$ such releases. Using $s\le k-2$ and $N=2|\Pi|\ell$, their total number is at most
\begin{align*}
|\Pi|\ell n(k-1)
=\frac{Nn(k-1)}{2}
\le C_0n.
\end{align*}

Conditioned on agreeing preceding transcripts, couple corresponding reports using the total variation bound above, and use the same randomness for all unchanged steps. By a union bound, the probability that  the pure and Gaussian versions of PC produce different outputs under this coupling is at most
\begin{align*}
(C_0n)\cdot8C_0n^2(C_0n)^{-128}
=8C_0^{-126}n^{-125}
<\frac{1}{1000}.
\end{align*}
The proof of \Cref{gh:main-theorem} guarantees the stated Gaussian accuracy with probability at least $11/12$. Since
\[
\ln\left(\frac{n}{\delta}\right)
=2049\ln n+2048\ln C_0
=O(\log n)
\]
for fixed $H$, the pure version of PC satisfies the claimed error bound with probability at least $11/12-1/1000>2/3$.
\paragraph{Rounds and Runtime.} The pure version of PC preserves the reporting schedule and therefore still uses at most $\ell+2$ rounds. By \Cref{gh:pure-linear-query}, the runtime of a replacement report is polynomial in $n,m,\frac{1}{\veps_1},e^{L_1}$. Then we have
\begin{align*}
\frac{1}{\veps_1}
&=\frac{2N(s+1)}{\veps}
=O\left(\frac{1}{\veps}\right),\\
e^{L_1}
&=\left(\frac{4N(s+1)}{\delta}\right)^{\frac{1}{512}}
=\bigl(4N(s+1)\bigr)^{\frac{1}{512}}(C_0n)^4
=O(n^4).
\end{align*}
Together with $m\le C_0n^2$ by Inequality (\ref{gh:pure-dimension-bound}), these bounds imply that each call to \Cref{gh:pure-query-algorithm} takes $\operatorname{poly}(n,d,1/\veps)$ time. There are at most $C_0n$ such calls, and the remaining computations are unchanged, so the pure version of PC also runs in worst-case $\operatorname{poly}(n,d,1/\veps)$ time.
\end{proof}

The following corollary gives variants of PC with fewer reporting rounds.

\begin{corollary}\label{cor:pc-star-remainder}
For any fixed connected pattern $H$ with $k\ge2$ vertices and $\ell\ge1$ edges, any $\veps\in(0,1]$, and any public integer $1\le d\le n-1$, variants of PC are $\veps$-LEDP. On every $n$-vertex graph $G$ with maximum degree at most $d$, their estimates satisfy
\[
|\hf_H(G)-f_H(G)|
\le O\left(\frac{\sqrt n\,d^{k-2}\log^{3\ell-2}(2n)}{\veps^\ell}\right)
\]
with probability at least $2/3$. The algorithms have polynomial arithmetic cost for fixed $H$.
\begin{enumerate}[label=(\roman*)]
\item If distinct vertices $x,y$ satisfy that $H-y$ is a connected star centered at $x$, put $b:=|E(H-x)|$. At most $b+2$ rounds suffice.
\item If $x$ is adjacent to every other vertex and $b:=|E(H-x)|\ge2$, at most $b+3$ rounds suffice.
\end{enumerate}
The algorithms runs in time $\operatorname{poly}(n,d,1/\veps)$. 
\end{corollary}

\begin{proof}
We follow the pure version of \Cref{gh:full-algorithm} described in the proof of \Cref{gh:pure-main-theorem}. Since $H$ is connected, each $H_\pi$ is connected; fix $\pi\in\Pi$ and set $K=H'=H_\pi$.

\paragraph{Edge replacements.}
The two structural conditions are preserved by these quotients. In case (i), $x$ is adjacent to every vertex except possibly $y$. If $x$ and $y$ lie in the same part, the quotient is a star; otherwise, $x$ forms a part by itself, and deleting the part containing $y$ leaves a star centered at the part $\{x\}$. In case (ii), $x$ also forms a part by itself, which is adjacent to every other part in the quotient. Write $c$ for the vertex of $K$ corresponding to the part containing $x$, choosing the star center instead when $K$ itself is a star. Thus $|E(K-c)|\le b$.

Keep the factors incident to $c$ private and replace only those on $E(K-c)$, in a fixed public order. Do not eliminate blocks, so $K=H'$ and all weights and the multiplier $\lambda$ remain one. For each replacement, use the endpoint notation $\{v,u\}$ of \Cref{gh:report-guarantees}. In case (i), select $v$ as the vertex of $K$ corresponding to the part containing $y$, so $K-v$ is the star centered at $c$; every replaced edge is incident to $v$. In case (ii), fix either endpoint as $v$ in advance; the edges incident to $c$ keep $K-v$ connected. Hence each saved replacement satisfies the conditions of \Cref{gh:center,gh:stage}.

\paragraph{The final local reports.}
After these replacements, all factors on $E(K-c)$ are public. Restrict the sum in Equation (\ref{gh:weighted-count}) to maps with $\phi(c)=i$. Every private factor in this sum comes from $\bmA_i$, so party $i$ can evaluate it using its own adjacency vector and the public factors $\bmW_e$. Replace the final exact evaluation of $T_K(\bmW,w)$ in \Cref{gh:full-algorithm} by the sum of the following private reports.

To estimate $T_K(\bmW,w)$, each party $i$ releases its restricted sum with Laplace noise, using the row modification and noise scale specified below. The server sums these reports in place of the final exact evaluation in \Cref{gh:full-algorithm}. To control the sensitivity of the final local reports on all input graphs, party $i$ first limits its adjacency row to at most $d$ ones, retaining the first ones in a fixed public coordinate order, and uses the resulting row in every private factor. This leaves the computation unchanged on graphs of maximum degree at most $d$. A one-bit change in $\bmA_i$ changes the retained row in at most two coordinates. For fixed public factors, the restricted sum consequently has sensitivity at most $2|N_K(c)|d^{k_K-2}$. To see this, expand the difference over the factors incident to $c$. Each changed coordinate fixes the images of $c$ and one neighbor. Extend the private star to a spanning tree of $K$ and delete that edge. Summing the other images along the resulting two rooted trees costs at most $d^{k_K-2}$, since the retained row has at most $d$ ones and every public factor belongs to $\calP_d$. The other factors lie in $[0,1]$ and can be omitted for this bound. Sum over the $|N_K(c)|$ private factors and at most two changed coordinates.

Each party adds independent Laplace noise of scale $4|N_K(c)|d^{k_K-2}/\veps_0$. An edge change affects two parties, so this release is $\veps_0$-LEDP by the Laplace mechanism. The sum of these noises has variance $O_H(nd^{2k_K-4}/\veps_0^2)$ and contributes additive error $O_H(\sqrt n\,d^{k_K-2}/\veps_0)$ with probability at least $1-\zeta$.

\paragraph{Privacy and accuracy.}
Each quotient uses $|E(K-c)|$ center calls, the same number of correction calls, and one final local release. Since $c$ has positive degree, $2|E(K-c)|+1\le2\ell_K\le2\ell$, so the allocation $N=2|\Pi|\ell$, $\veps_0=\veps/N$ covers all calls. The same privacy analysis gives $\veps$-LEDP on all input graphs. On $G\in\calG_{n,d}$, the retained rows equal $\bmA_i$, so the sum of the final reports estimates $T_K(\bmW,w)$ with the error bound above. The remaining error analysis and quotient combination follow as in \Cref{gh:counting-analysis}, giving the claimed accuracy with probability at least $2/3$.

\paragraph{Rounds in case (i).}
For a replacement on $\{v,u\}$, attach $v$ to the star $K-v$ through $\{v,u\}$ and root the resulting tree at $v$, as in \Cref{gh:correction-algorithm}. Since $u\ne c$, the vertex $u$ is the only child of $v$, and $c$ is the only child of $u$. All previously replaced factors are incident to $v$ and, by the prescribed edge directions, occur only in the root query (\ref{gh:local-query-v}). Thus the internal queries (\ref{gh:local-query}) use no center. The server computes the leaf reports, after which the reports at $c$ and $u$ can be collected in rounds one and two, respectively. The center computations do not use correction reports, and the two procedures use independent randomness, so collecting these internal reports early preserves the joint distribution of the centers and correction estimates. As in \Cref{gh:centers}, all first-round center reports are collected in round one, and center $j$ is available by round $j+1$. The corresponding correction reports at the root $v$ are released in round $j+2$, and the final reports for $T_K(\bmW,w)$ are released in round $|E(K-c)|+2\le b+2$. All quotient computations run in parallel, so the same round bounds hold for the complete algorithm.

\paragraph{Rounds in case (ii).}
For a replacement on $\{v,u\}$, choose the star centered at $c$ as the spanning tree of $K-v$, attach $v$ through $\{v,u\}$, and root the resulting tree at $v$. The internal reports are computed first at the children of $c$, then at $c$, and then at $u$. By (\ref{gh:local-query})--(\ref{gh:local-query-v}), these reports use preceding centers but not the new center. For replacement $j$, the preceding centers are available by round $j$, so these three groups of internal reports can be collected by rounds $j+1$, $j+2$, and $j+3$, respectively. The final correction reports at the root $v$ are then released by round $j+4$.

For the last replacement, every edge of $K-v$ not incident to $c$ is already public. Its leaf reports are therefore computed by the server, leaving only the reports at $c$, at $u$, and at $v$. This correction finishes by round $|E(K-c)|+3$; every earlier correction satisfies $j+4\le|E(K-c)|+3$. The final reports for $T_K(\bmW,w)$ finish by round $|E(K-c)|+2$. If $|E(K-c)|=1$, case (i) applies; if it is zero, only the final local release is needed. Hence at most $b+3$ rounds suffice. The same analysis therefore applies to the complete algorithm.

\paragraph{Runtime.}
The local sums can be evaluated by enumerating at most $n^k$ maps. Together with the bounds for the pure center and correction procedures, and the fact that the number of calls depends only on $H$, this gives a total arithmetic cost of $\operatorname{poly}(n,d,1/\veps)$ for fixed $H$.
\end{proof}

\subsubsection[Proof of the Linear Query Guarantee]{Proof of the Linear Query Guarantee}\label{sssec:proof-linear-query}

We prove \Cref{gh:pure-linear-query} using the Gaussian
isoperimetric inequality (\Cref{gh:gaussian-enlargement}) and the following sampling lemma, which provides a pointwise multiplicative guarantee.

\begin{lemma}[Sampling with Approximate Evaluations {\cite[Theorem~3.9 and Lemma~3.8]{lowy2026differentially}}]
\label{gh:inexact-cube-sampling}
For any $R>0$, $\Lambda\ge0$, and $\eta\in(0,1)$, let $\psi:[-R,R]^m\to\reals$ be a convex and $\Lambda$-Lipschitz function (with respect to the Euclidean norm), i.e.,
\[
|\psi(\bmx)-\psi(\mathbf y)|\le\Lambda\|\bmx-\mathbf y\|_2
\]
for all $\bmx,\mathbf y\in[-R,R]^m$. Suppose a deterministic procedure returns $\widehat\psi(\mathbf y)$ with $|\widehat\psi(\mathbf y)-\psi(\mathbf y)|\le\eta/8$ for every $\mathbf y\in[-R,R]^m$. Then an algorithm using these approximate values returns a point in $[-R,R]^m$ with density $\widehat\pi$ satisfying
\[
e^{-\eta}\pi(\mathbf y)\le\widehat\pi(\mathbf y)\le e^\eta\pi(\mathbf y)
\]
for every $\mathbf y\in[-R,R]^m$, where $\pi(\mathbf y):=\frac{e^{-\psi(\mathbf y)}}{\int_{[-R,R]^m}e^{-\psi(\bmx)}\,d\bmx}$. Both the number of evaluations of $\widehat\psi$ and the number of additional arithmetic operations are bounded by $\operatorname{poly}(m,1+R\Lambda,1/\eta)$ in the worst case.
\end{lemma}

\begin{proof}
We specialize the grid-walk analysis of \cite[Lemma~3.8]{lowy2026differentially} to the cube and use the conversion from grid points to continuous samples in \cite[Lemma~6.1]{bassily2014differentially}. We give the parameter choices and error bounds explicitly.

First construct the grid. If $\Lambda=0$, then $\psi$ is constant, and uniform sampling from the cube is exact. Otherwise, set
\[
\alpha_0:=\max\left\{\Lambda\sqrt m,\frac{1}{R}\right\},
\qquad
J:=\left\lceil\frac{8R\alpha_0\sqrt m}{\eta}\right\rceil,
\qquad
\alpha:=\frac{\eta J}{8R\sqrt m}.
\]
Since $\alpha\ge\alpha_0\ge\Lambda\sqrt m$, the function $\psi$ is $\alpha$-Lipschitz with respect to $\|\cdot\|_\infty$. Also, $\alpha(2R)\ge2$. Dividing each coordinate interval into $J$ equal intervals partitions $[-R,R]^m$ into $J^m$ smaller cubes, which we call cells. Each cell has side length
\[
a:=\frac{2R}{J}
=\frac{\eta}{4\alpha\sqrt m}.
\]
We use the center of each cell as a grid point of the walk and denote the set of these points by $\mathcal C$. For $\mathbf y\in[-R,R]^m$, let $c(\mathbf y)$ be the center of the cell containing $\mathbf y$. If a point lies on a boundary shared by several cells, assign it to one of them by a fixed rule so that $c(\mathbf y)$ is uniquely defined. Since $\|\mathbf y-c(\mathbf y)\|_\infty\le\frac{a}{2}$, the variation of the true potential within a cell satisfies
\[
|\psi(\mathbf y)-\psi(c(\mathbf y))|
\le\frac{\alpha a}{2}
=\frac{\eta}{8\sqrt m}
\le\frac{\eta}{8}.
\]

Assign each grid point $\mathbf c\in\mathcal C$ the weight $e^{-\widehat\psi(\mathbf c)}$. Since the evaluation procedure is deterministic, revisiting the same point gives the same weight. We use the grid walk described in \cite[Section~B.3.1]{lowy2026differentially}, with these approximate weights. Start at any fixed grid point in $\mathcal C$. At each step, the walk stays at its current point with probability $\frac{1}{2}$. Otherwise, it selects one of the $2m$ coordinate directions uniformly and proposes moving a distance $a$ in that direction. A proposal outside $\mathcal C$ is rejected; a proposal from $\mathbf c$ to $\mathbf c'\in\mathcal C$ is accepted with probability $\min\{1,\exp(\widehat\psi(\mathbf c)-\widehat\psi(\mathbf c'))\}$. The walk has stationary distribution
\[
\nu(\mathbf c):=
\frac{e^{-\widehat\psi(\mathbf c)}}
{\sum_{\mathbf c'\in\mathcal C}e^{-\widehat\psi(\mathbf c')}}.
\]
Thus, if the current point has distribution $\nu$, one step preserves that distribution.

To bound the number of steps, apply \cite[Lemma~3.8]{lowy2026differentially} with cube side length $\tau=2R$, target error $\xi=\frac{\eta}{2}$, and evaluation error $\zeta=\frac{\eta}{8}$. The true potential $\psi$ is convex and $\alpha$-Lipschitz with respect to $\|\cdot\|_\infty$, with $\alpha\tau\ge2$, and the grid spacing is $a=\frac{\xi}{2\alpha\sqrt m}$. The evaluations have deterministic error at most $\zeta<1$; the cited guarantee does not require $\widehat\psi$ itself to be convex. Write $\nu_T(\mathbf c)$ for the probability that the walk is at $\mathbf c$ after $T$ steps. The result gives a predetermined number $T=\operatorname{poly}(m,1+R\alpha,\frac{1}{\eta})$ of steps such that, regardless of the starting point, every grid point $\mathbf c\in\mathcal C$ satisfies
\[
e^{-\frac{\eta}{2}}\nu(\mathbf c)
\le\nu_T(\mathbf c)
\le e^{\frac{\eta}{2}}\nu(\mathbf c).
\]

After the walk selects a grid point, sample uniformly from the cell centered at that point, as in \cite[Lemma~6.1]{bassily2014differentially}. All cells have volume $a^m$, so the resulting density is
\[
\widehat\pi(\mathbf y)=\frac{\nu_T(c(\mathbf y))}{a^m}.
\]
We compare this density with $\pi$ by accounting for the function evaluation error and the variation within a cell. The two bounds give
\begin{align*}
|\widehat\psi(c(\mathbf y))-\psi(\mathbf y)|
&\le|\widehat\psi(c(\mathbf y))-\psi(c(\mathbf y))|
+|\psi(c(\mathbf y))-\psi(\mathbf y)|\\
&\le\frac{\eta}{8}+\frac{\eta}{8}
=\frac{\eta}{4}.
\end{align*}
Write the two normalizing constants as
\[
Z_\psi:=\int_{[-R,R]^m}e^{-\psi(\mathbf y)}\,d\mathbf y,
\qquad
\widehat Z_\psi:=a^m\sum_{\mathbf c\in\mathcal C}e^{-\widehat\psi(\mathbf c)}
=\int_{[-R,R]^m}e^{-\widehat\psi(c(\mathbf y))}\,d\mathbf y.
\]
Exponentiating the preceding error bound and integrating over the cube yields
\[
e^{-\frac{\eta}{4}}e^{-\psi(\mathbf y)}
\le e^{-\widehat\psi(c(\mathbf y))}
\le e^{\frac{\eta}{4}}e^{-\psi(\mathbf y)},
\qquad
e^{-\frac{\eta}{4}}Z_\psi
\le\widehat Z_\psi
\le e^{\frac{\eta}{4}}Z_\psi.
\]
Using the definitions of $\nu$, $\widehat\pi$, and $\pi$, we can therefore factor their density ratio as
\begin{align*}
\frac{\widehat\pi(\mathbf y)}{\pi(\mathbf y)}
&=\frac{\nu_T(c(\mathbf y))}{\nu(c(\mathbf y))}
  \frac{e^{-\widehat\psi(c(\mathbf y))}}{e^{-\psi(\mathbf y)}}
  \frac{Z_\psi}{\widehat Z_\psi}.
\end{align*}
The three factors lie in $[e^{-\frac{\eta}{2}},e^{\frac{\eta}{2}}]$, $[e^{-\frac{\eta}{4}},e^{\frac{\eta}{4}}]$, and $[e^{-\frac{\eta}{4}},e^{\frac{\eta}{4}}]$, respectively. Multiplying these bounds gives
\[
e^{-\eta}\pi(\mathbf y)\le\widehat\pi(\mathbf y)\le e^\eta\pi(\mathbf y).
\]

Finally, the current grid point is represented by the $m$ coordinate indices of its cell, so the walk never enumerates all $J^m$ grid points. Each step requires at most two evaluations of $\widehat\psi$ and $O(m)$ additional arithmetic operations. The grid parameters satisfy
\[
J\le\frac{8\sqrt m}{\eta}\max\{R\Lambda\sqrt m,1\}+1,
\qquad
R\alpha\le\max\{R\Lambda\sqrt m,1\}+\frac{\eta}{8\sqrt m}.
\]
Thus $J$, $R\alpha$, and $T$ are polynomial in $m,1+R\Lambda,\frac{1}{\eta}$. Sampling uniformly in the cell centered at the final grid point takes $O(m)$ further operations. We stop the walk after the predetermined $T$ steps and return the point sampled from that cell, with no restart or additional acceptance test. This proves the claimed worst-case bounds on function evaluations and additional arithmetic operations.
\end{proof}

\begin{proof}[Proof of \Cref{gh:pure-linear-query}]
If $\Delta_{\bmW}=0$, then $\bmW=0$, so both the mechanism and the Gaussian comparison return zero. Suppose $\Delta_{\bmW}>0$. Use the parameters of \Cref{gh:pure-query-algorithm}, including $\mathbf q=\bma\bmD$ for the input $\bma\in\{0,1\}^n$. Since $\|\bmD_j\|_2\le1$, every input satisfies
\[
\|\mathbf q\|_\infty
\le\sum_{j=1}^n a_j\|\bmD_j\|_\infty
\le\sum_{j=1}^n a_j
\le n.
\]
Thus the public cube $[-R,R]^m$, with $R=n+\gamma\sigma^2$, contains every normalized query value with the same noise margin. We prove privacy, total variation distance, and runtime separately.

\paragraph{Privacy.}
Let $\bma,\bma'\in\{0,1\}^n$ differ only in coordinate $j$, and write $\mathbf q'=\bma'\bmD$. Their normalized query values differ by $\mathbf s:=\mathbf q-\mathbf q'=\pm\bmD_j$. The constraints defining $\Omega$ give
\[
|\mathbf s\mathbf p^\top|=|\bmD_j\mathbf p^\top|\le\gamma
\quad\text{for every }\mathbf p\in\Omega.
\]
Consequently, for every $\mathbf z\in\reals^m$,
\begin{align*}
h(\mathbf z+\mathbf s)
&=\max_{\mathbf p\in\Omega}\left(\mathbf z\mathbf p^\top-\frac{\sigma^2}{2}\|\mathbf p\|_2^2+\mathbf s\mathbf p^\top\right)
\le h(\mathbf z)+\gamma.
\end{align*}
The same argument with $-\mathbf s$ gives $|h(\mathbf z+\mathbf s)-h(\mathbf z)|\le\gamma$. Let $u_{\bma}(\mathbf y):=e^{-h(\mathbf y-\mathbf q)}$ on $[-R,R]^m$, and let $Z_{\bma}:=\int_{[-R,R]^m}u_{\bma}(\mathbf y)\,d\mathbf y$ be its normalizing constant. Define $u_{\bma'}$ and $Z_{\bma'}$ analogously using $\mathbf q'$. The preceding bound implies
\[
e^{-\gamma}u_{\bma'}(\mathbf y)
\le u_{\bma}(\mathbf y)
\le e^\gamma u_{\bma'}(\mathbf y).
\]
Integrating over the common cube gives $e^{-\gamma}Z_{\bma'}\le Z_{\bma}\le e^\gamma Z_{\bma'}$. Thus the normalized densities $p_{\bma}:=\frac{u_{\bma}}{Z_{\bma}}$ satisfy
\[
\frac{p_{\bma}(\mathbf y)}{p_{\bma'}(\mathbf y)}
=\frac{u_{\bma}(\mathbf y)}{u_{\bma'}(\mathbf y)}
 \frac{Z_{\bma'}}{Z_{\bma}}
\le e^{2\gamma}.
\]
The sampler returns a density $\widehat p_{\bma}$ within a factor $e^{\pm\eta}$ of $p_{\bma}$. Therefore,
\[
\frac{\widehat p_{\bma}(\mathbf y)}
     {\widehat p_{\bma'}(\mathbf y)}
\le
\frac{e^\eta p_{\bma}(\mathbf y)}
     {e^{-\eta}p_{\bma'}(\mathbf y)}
\le e^{2\gamma+2\eta}
\le e^{\veps_1},
\]
where the last inequality uses $\gamma=\frac{\veps_1}{4}$ and $\eta=\veps_1e^{-32L_1}\le\frac{\veps_1}{4}$. Scaling the output by the public value $\Delta_{\bmW}$ preserves privacy by postprocessing. Then the whole algorithm is $\veps_1$-DP.

\paragraph{Distributional comparison.}
Let $\bmg\sim\calN(0,\bmI_m)$ and $\rho:=2(n+m)e^{-32L_1}$. Let $Z$ and $Z_{\mathrm G}$ be the normalizing constants of the density proportional to $e^{-h(\mathbf z)}$ and the Gaussian distribution $\calN(0,\sigma^2\bmI_m)$, respectively:
\[
Z:=\int_{\reals^m}e^{-h(\mathbf z)}\,d\mathbf z,
\qquad Z_{\mathrm G}:=(2\pi\sigma^2)^{m/2}.
\]
Let $\bar p_{\bma}$ be the density proportional to $e^{-h(\mathbf y-\mathbf q)}$ on $\reals^m$, and let $\varphi_{\bma}$ be the Gaussian density with mean $\mathbf q$ and covariance $\sigma^2\bmI_m$:
\[
\bar p_{\bma}(\mathbf y)
:=\frac{e^{-h(\mathbf y-\mathbf q)}}{Z},
\qquad
\varphi_{\bma}(\mathbf y)
:=\frac{e^{-\frac{\|\mathbf y-\mathbf q\|_2^2}{2\sigma^2}}}{Z_{\mathrm G}}.
\]
Recall that $p_{\bma}$ is $\bar p_{\bma}$ conditioned on $[-R,R]^m$, and $\widehat p_{\bma}$ is the density returned by the sampler before multiplying by $\Delta_{\bmW}$. We extend both densities by zero outside the cube. We bound the distances between $\bar p_{\bma}$ and $\varphi_{\bma}$, between $p_{\bma}$ and $\bar p_{\bma}$, and between $\widehat p_{\bma}$ and $p_{\bma}$ in the following three lemmas, then combine them by the triangle inequality.

\begin{lemma}\label{gh:pure-tv-gaussian}
$d_{\mathrm{TV}}(\bar p_{\bma},\varphi_{\bma})\le\rho$.
\end{lemma}

\begin{proof}

Without the constraint $\mathbf p\in\Omega$, the maximum defining $h(\mathbf z)$ is attained at $\mathbf p=\mathbf z/\sigma^2$ and equals $\|\mathbf z\|_2^2/(2\sigma^2)$. Hence $e^{-h(\mathbf z)}$ agrees with the unnormalized Gaussian density whenever $\mathbf z/\sigma^2\in\Omega$. We first bound the Gaussian probability that this condition fails. For the Gaussian sample $\mathbf z=\sigma\bmg$, the condition $\mathbf z/\sigma^2\in\Omega$ becomes $\bmg\in\sigma\Omega$. Since $\gamma\sigma=8\sqrt{L_1}$ and $\|\bmD_j\|_2\le1$, Gaussian tail bounds and a union bound give
\begin{align}
\Pr[\bmg\notin\sigma\Omega]
&\le\sum_{\alpha=1}^m\Pr[|g_\alpha|>\gamma\sigma]
+\sum_{j=1}^n\Pr[|\bmD_j\bmg^\top|>\gamma\sigma]\notag\\
&\le2(n+m)e^{-\frac{(\gamma\sigma)^2}{2}}
=\rho.\label{ieq:notinOmega}
\end{align}

We next bound how far $\bmg$ lies from $\sigma\Omega$, since this distance determines how much $e^{-h(\sigma\bmg)}$ exceeds the Gaussian density factor $e^{-\|\bmg\|_2^2/2}$ and allows us to bound the increase in the normalizing constant. Define $D(\bmx):=\operatorname{dist}(\bmx,\sigma\Omega)$ for every $\bmx\in\reals^m$, and let $b:=\Phi^{-1}(1-\rho)$. Since $L_1\ge\ln(100(n+m))$, we have
\[
\rho\le2(n+m)e^{-L_1}\le\frac{1}{50}<1-\Phi(1),
\]
and hence $b=\Phi^{-1}(1-\rho)\ge1$. Since $\sigma\Omega$ is nonempty and closed and $\Pr[\bmg\in\sigma\Omega]\ge\Phi(b)$ by Inequality (\ref{ieq:notinOmega}), \Cref{gh:gaussian-enlargement} gives, for every $t\ge0$,
\begin{align}
\Pr[D(\bmg)>t]
&\le1-\Phi(b+t)
=\frac{1}{\sqrt{2\pi}}\int_b^\infty e^{-\frac{(x+t)^2}{2}}\,\mathrm{d}x\notag\\
&\le e^{-bt-\frac{t^2}{2}}
\frac{1}{\sqrt{2\pi}}\int_b^\infty e^{-\frac{x^2}{2}}\,\mathrm{d}x \label{ieq:farfromOmega}
=\rho e^{-bt-\frac{t^2}{2}},
\end{align}
where the last inequality uses $x\ge b$ and the last equation uses $\rho = 1 - \Phi(b)$.

We next compare the normalizing constants $Z$ and $Z_{\mathrm G}$.
By the definition of $h$, we have
\begin{align}
h(\sigma\bmg)
&=\max_{\mathbf p\in\Omega}
\left(\sigma\bmg\mathbf p^\top-\frac{\sigma^2}{2}\|\mathbf p\|_2^2\right)\notag\\
&=\max_{\mathbf p\in\Omega}
\left(\frac{\|\bmg\|_2^2}{2}
-\frac{\|\bmg-\sigma\mathbf p\|_2^2}{2}\right)\notag\\
&=\frac{\|\bmg\|_2^2}{2}
-\frac12\min_{\mathbf p\in\Omega}\|\bmg-\sigma\mathbf p\|_2^2 
=\frac{\|\bmg\|_2^2-D(\bmg)^2}{2},\label{eq:h(sigmag)}
\end{align}
Changing variables $\mathbf z=\sigma\bmx$, with $d\mathbf z=\sigma^m\,d\bmx$, gives
\begin{align*}
\frac{Z}{Z_{\mathrm G}}
&=\frac{1}{(2\pi\sigma^2)^{m/2}}
  \int_{\reals^m}e^{-h(\mathbf z)}\,d\mathbf z\\
&=\frac{1}{(2\pi)^{m/2}}
  \int_{\reals^m}e^{-h(\sigma\bmx)}\,d\bmx\\
&=\int_{\reals^m}
  \frac{e^{-\frac{\|\bmx\|_2^2}{2}}}{(2\pi)^{m/2}}
  e^{\frac{D(\bmx)^2}{2}}
  \,d\bmx=\E\left[e^{\frac{D(\bmg)^2}{2}}\right],
\end{align*}
where the third equality uses \Cref{eq:h(sigmag)}, and the last uses the density of $\bmg\sim\calN(0,\bmI_m)$.

To bound this expectation, write the integral with random upper limit $D(\bmg)$ using an indicator. Since its integrand is nonnegative, we may interchange expectation and integration:
\begin{align*}
\frac{Z}{Z_G}&=\E\left[e^{\frac{D(\bmg)^2}{2}}\right]
=1+\E\left[\int_0^{D(\bmg)}t e^{\frac{t^2}{2}}\,dt\right]\\
&=1+\E\left[\int_0^\infty
  t e^{\frac{t^2}{2}}\mathbf1_{D(\bmg)>t}\,dt\right]\\
&=1+\int_0^\infty
  t e^{\frac{t^2}{2}}\Pr[D(\bmg)>t]\,dt\\
&\le1+\rho\int_0^\infty
  t e^{\frac{t^2}{2}}e^{-bt-\frac{t^2}{2}}\,dt\\
&=1+\rho\int_0^\infty t e^{-bt}\,dt
=1+\frac{\rho}{b^2}
\le1+\rho,
\end{align*}
where the first inequality uses \Cref{ieq:farfromOmega} and the last uses $b\ge1$.

\Cref{eq:h(sigmag)} also gives $h(\mathbf z)\le\|\mathbf z\|_2^2/(2\sigma^2)$. Therefore,
\[
\frac{\bar p_{\bma}(\mathbf y)}{\varphi_{\bma}(\mathbf y)}
=\frac{Z_{\mathrm G}}{Z}
\exp\left(\frac{\|\mathbf y-\mathbf q\|_2^2}{2\sigma^2}
-h(\mathbf y-\mathbf q)\right)
\ge\frac{1}{1+\rho}.
\]
Consequently,
\[
d_{\mathrm{TV}}(\bar p_{\bma},\varphi_{\bma})
=\int_{\reals^m}[\varphi_{\bma}(\mathbf y)-\bar p_{\bma}(\mathbf y)]_+
\,d\mathbf y
\le\frac{\rho}{1+\rho}\le\rho.
\]

\end{proof}

\begin{lemma}\label{gh:pure-tv-cube}
$d_{\mathrm{TV}}(p_{\bma},\bar p_{\bma})\le2\rho$.
\end{lemma}

\begin{proof}

We first bound the probability that a sample from $\bar p_{\bma}$ lies outside $[-R,R]^m$, using the corresponding Gaussian probability and \Cref{gh:pure-tv-gaussian}. We then show directly that conditioning $\bar p_{\bma}$ on the cube changes the distribution by exactly this probability in total variation distance.

Since $\|\mathbf q\|_\infty\le n$, if $\|\bmg\|_\infty\le\gamma\sigma$, then $\|\mathbf q+\sigma\bmg\|_\infty\le n+\gamma\sigma^2=R$. Gaussian tail bounds and a union bound therefore give
\begin{align*}
\int_{\reals^m\setminus[-R,R]^m}\varphi_{\bma}(\mathbf y)\,d\mathbf y
&=\Pr[\mathbf q+\sigma\bmg\notin[-R,R]^m]\\
&\le\Pr[\|\bmg\|_\infty>\gamma\sigma]
\le\sum_{\alpha=1}^m\Pr[|g_\alpha|>\gamma\sigma]\\
&\le2m e^{-\frac{(\gamma\sigma)^2}{2}}
=2m e^{-32L_1}\le\rho.
\end{align*}
To transfer this bound to $\bar p_{\bma}$, recall that the positive and negative parts of the difference of two probability densities have equal integrals. Thus either part integrates to their total variation distance. Writing $[t]_+:=\max\{t,0\}$, we obtain
\begin{align*}
\int_{\reals^m\setminus[-R,R]^m}\bar p_{\bma}(\mathbf y)\,d\mathbf y
&\le\int_{\reals^m\setminus[-R,R]^m}\varphi_{\bma}(\mathbf y)\,d\mathbf y
+\int_{\reals^m}[\bar p_{\bma}(\mathbf y)-\varphi_{\bma}(\mathbf y)]_+\,d\mathbf y\\
&\le\rho+d_{\mathrm{TV}}(\bar p_{\bma},\varphi_{\bma})
\le2\rho,
\end{align*}
where the last inequality uses \Cref{gh:pure-tv-gaussian}.

It remains to compute the effect of conditioning on the cube. Recall from the privacy proof that
\[
Z_{\bma}:=\int_{[-R,R]^m}e^{-h(\mathbf y-\mathbf q)}\,d\mathbf y.
\]
By the definition of $\bar p_{\bma}$, its probability mass inside the cube is $\frac{Z_{\bma}}{Z}$. In particular, $0<Z_{\bma}\le Z$. On the cube, the conditional density satisfies $p_{\bma}(\mathbf y)=\frac{Z}{Z_{\bma}}\bar p_{\bma}(\mathbf y)\ge\bar p_{\bma}(\mathbf y)$; outside the cube, $p_{\bma}(\mathbf y)=0$. Splitting the integral in the definition of total variation distance over these two regions gives
\begin{align*}
d_{\mathrm{TV}}(p_{\bma},\bar p_{\bma})
&=\frac{1}{2}\int_{\reals^m}|p_{\bma}(\mathbf y)-\bar p_{\bma}(\mathbf y)|\,d\mathbf y\\
&=\frac{1}{2}\int_{[-R,R]^m}(p_{\bma}(\mathbf y)-\bar p_{\bma}(\mathbf y))\,d\mathbf y
+\frac{1}{2}\int_{\reals^m\setminus[-R,R]^m}\bar p_{\bma}(\mathbf y)\,d\mathbf y\\
&=\frac{1}{2}\left(\frac{Z}{Z_{\bma}}-1\right)\frac{Z_{\bma}}{Z}
+\frac{1}{2}\left(1-\frac{Z_{\bma}}{Z}\right)\\
&=1-\frac{Z_{\bma}}{Z}
=\int_{\reals^m\setminus[-R,R]^m}\bar p_{\bma}(\mathbf y)\,d\mathbf y
\le2\rho.
\end{align*}

\end{proof}

\begin{lemma}\label{gh:pure-tv-sampling}
$d_{\mathrm{TV}}(\widehat p_{\bma},p_{\bma})\le2\eta$.
\end{lemma}

\begin{proof}

We convert the sampler's pointwise density guarantee into a bound on total variation distance. By \Cref{gh:inexact-cube-sampling}, for every $\mathbf y\in[-R,R]^m$,
\[
e^{-\eta}p_{\bma}(\mathbf y)
\le\widehat p_{\bma}(\mathbf y)
\le e^\eta p_{\bma}(\mathbf y).
\]
The upper bound implies $[\widehat p_{\bma}(\mathbf y)-p_{\bma}(\mathbf y)]_+\le(e^\eta-1)p_{\bma}(\mathbf y)$. Since both densities vanish outside the cube and each integrates to one, integrating the positive part of their difference gives
\begin{align*}
d_{\mathrm{TV}}(\widehat p_{\bma},p_{\bma})
&=\int_{[-R,R]^m}[\widehat p_{\bma}(\mathbf y)-p_{\bma}(\mathbf y)]_+\,d\mathbf y\\
&\le(e^\eta-1)\int_{[-R,R]^m}p_{\bma}(\mathbf y)\,d\mathbf y
=e^\eta-1.
\end{align*}
Finally, $\eta=\veps_1e^{-32L_1}\in(0,1)$, and convexity of the exponential function gives $e^\eta\le(1-\eta)e^0+\eta e^1=1+(e-1)\eta$. Hence $e^\eta-1\le(e-1)\eta\le2\eta$, as required.

\end{proof}

Combining \Cref{gh:pure-tv-gaussian,gh:pure-tv-cube,gh:pure-tv-sampling} by the triangle inequality,
\begin{align*}
d_{\mathrm{TV}}(\widehat p_{\bma},\varphi_{\bma})
&\le d_{\mathrm{TV}}(\widehat p_{\bma},p_{\bma})
+d_{\mathrm{TV}}(p_{\bma},\bar p_{\bma})
+d_{\mathrm{TV}}(\bar p_{\bma},\varphi_{\bma})\\
&\le2\eta+3\rho
\le4\rho
=8(n+m)e^{-32L_1},
\end{align*}
where $2\eta\le\rho$ follows from $\veps_1\le1$. Scaling both outputs by $\Delta_{\bmW}$ gives \Cref{gh:pure-linear-query-tv}.

\paragraph{Runtime.}
The function $\psi_{\bma}(\mathbf y):=h(\mathbf y-\mathbf q)$ is convex and $\gamma\sqrt m$-Lipschitz. By \Cref{gh:inexact-cube-sampling}, both the number of approximate function evaluations and the number of additional sampling operations are bounded by $\operatorname{poly}(m,1+R\gamma\sqrt m,\frac{1}{\eta})$. It remains to implement each evaluation deterministically with error at most $\frac{\eta}{8}$.

Fix $\mathbf y\in[-R,R]^m$, set $\mathbf z:=\mathbf y-\mathbf q$, and write
\[
g_{\mathbf z}(\mathbf p):=\frac{\sigma^2}{2}\|\mathbf p\|_2^2-\mathbf z\mathbf p^\top,
\qquad
h(\mathbf z)=-\min_{\mathbf p\in\Omega}g_{\mathbf z}(\mathbf p).
\]
The objective is a convex quadratic whose value and gradient each take $O(m)$ arithmetic operations to evaluate. When $m=1$, we have $\Omega=[-\gamma,\gamma]$, so the minimizer is $\Clip_{[-\gamma,\gamma]}(\frac{z}{\sigma^2})$ and $h$ can be evaluated exactly. Henceforth assume $m\ge2$.

The feasible set $\Omega$ is closed and convex, described by $2(n+m)$ linear inequalities. Checking these inequalities in a fixed order either certifies feasibility or supplies a separating hyperplane in $O(nm)$ operations. Since $\|\bmD_j\|_2\le1$, the set contains the ball of radius $\gamma$ centered at the origin and is contained in $E_0:=\{\mathbf p:\|\mathbf p\|_2\le\gamma\sqrt m\}$. Moreover, $\|\mathbf q\|_\infty\le n$ gives $\|\mathbf z\|_2\le\sqrt m(R+n)$. Thus the difference between the maximum and minimum of $g_{\mathbf z}$ on $E_0$ is at most
\[
U:=\sigma^2\gamma^2m+2\gamma m\left(R+n\right)
=\veps_1mn+192mL_1.
\]

Applying \Cref{lemma:ellipsoid-method} with objective tolerance $\frac{\eta}{8}$ gives a feasible point $\widehat{\mathbf p}$ with additive objective error at most $\frac{\eta}{8}$ after at most
\[
\left\lceil
2m^2\left(
\ln\sqrt m
+\ln\left(1+\frac{8U}{\eta}\right)
\right)
\right\rceil+1
\]
iterations, each using $O(nm+m^2)$ arithmetic operations. With a fixed deterministic implementation, $\widehat\psi_{\bma}(\mathbf y):=-g_{\mathbf z}(\widehat{\mathbf p})$ is therefore a deterministic approximation satisfying $|\widehat\psi_{\bma}(\mathbf y)-\psi_{\bma}(\mathbf y)|\le\frac{\eta}{8}$ for every $\mathbf y\in[-R,R]^m$, as required.

Finally, $R\gamma\sqrt m=\sqrt m(\frac{\veps_1n}{4}+64L_1)$ and $\frac{1}{\eta}=\frac{e^{32L_1}}{\veps_1}$. Together with the bound on $U$, these identities show that both the cost of each evaluation and the number of evaluations and additional sampling operations are polynomial in $n,m,\frac{1}{\veps_1},e^{L_1}$. Combining these costs with the $O(nm)$ operations needed to form $\bmD$ and $\mathbf q$ proves the claimed worst-case runtime.

\end{proof}

\section{The Degree-Dependent Lower Bound}
\label{sec:degree-lower}
We now prove a lower bound matching PC's polynomial dependence on $n$ and $d$ and, when $d=\Theta(n)$, LNC's dependence on $n$. We use a reduction from a gap-testing variant of the LDP summation problem. 

\begin{definition}[Summation function]
Let $\SUM_n$ be the following function. For all $x_1,\ldots,x_n \in \{0,1\}$, $\SUM_n(x_1, \ldots, x_n)=\sum_{i=1}^{n}x_i$.
\end{definition}

We use the following lemma as the lower bound for testing the summation problem, which follows from the lower bound for simple hypothesis testing in \cite{joseph2019role}.

\begin{lemma}\label{lem:bit-sum-testing}
There exist constants $c_0,c_1,c_2>0$ such that the following holds for every $\veps\in(0,1]$, $n\in\nats$, $r\in[1,n]$, and
\[
 \delta\in\left[0,
 \frac{c_1\veps^3r^2}{n^3\ln\!\left(n^3/(c_2\veps^2r^2)\right)}\right].
\]
If an $(\veps,\delta)$-LDP algorithm distinguishes $\SUM_n(x)=0$ from $\SUM_n(x)\ge r$ with probability at least $2/3$, then $r\ge c_0\min(n,\sqrt n/\veps)$.
\end{lemma}

\begin{proof}
We first recall the simple hypothesis-testing lower bound that we use. By Theorem~5.3 in \cite{joseph2019role}, let $P_0$ and $P_1$ be two distributions with $\|P_0-P_1\|_{\mathrm{TV}}=\gamma$. For any arbitrary, potentially fully interactive, $(\veps',\delta')$-LDP protocol that distinguishes $P_0$ from $P_1$ with probability at least $2/3$ using $n$ i.i.d.\ samples, if $\delta' \in [0, c_1'(\veps')^3\gamma^2/(n\ln\left(n/(c_2'(\veps')^2\gamma^2)\right))]$ for sufficiently small constants $c_1',c_2'>0$ and $\veps'=O(1)$, then $n\ge c_0'/((\veps')^2\gamma^2)$ for some constant $c_0' > 0$.

Suppose, towards a contradiction, that there exists an algorithm $\calB$ satisfying the assumptions for $r< c_0\min(n,\sqrt{n}/\veps)$, where $c_0 > 0$ is a constant. We now use $\calB$ to construct such a hypothesis-testing protocol. By a constant number $\beta$ of independent repetitions of $\mathcal{B}$ followed by majority vote, we may assume that its success probability on every input satisfying the promise is at least $11/12$. Let $\veps':=\beta\veps$ and $\delta':=\beta\delta$. By \Cref{lemma:basic composition}, the resulting protocol $\calB'$ is $(\veps',\delta')$-LDP. 

Taking $c_0\le1/8$ gives $r\le \min(n/8, \sqrt{n}/(8\veps))$. Consider the two distributions $P_0:=\operatorname{Ber}(0)$ and $P_1:=\operatorname{Ber}(p)$, where $p:=8r/n\le1$. Clearly, $\gamma =\|P_0-P_1\|_{\mathrm{TV}}=p=8r/n$. Let $c_1 := 64c_1'\beta^2$ and $c_2:= 64c_2'\beta^2$. Then we can verify the condition on $\delta'$ by
\begin{align*}
\delta'=\beta\delta \le \frac{c_1\beta\veps^3r^2} {n^3\ln\left(n^3/(c_2\veps^2r^2)\right)} = \frac{64c_1'(\veps')^3r^2}{n^3\ln\left(n^3/(64c_2'(\veps')^2r^2)\right)} =\frac{c_1'(\veps')^3\gamma^2}{n\ln\left(n/(c_2'(\veps')^2\gamma^2)\right)}.
\end{align*}

We can use $\calB'$ to distinguish whether $n$ i.i.d. samples are drawn from $P_0$ or $P_1$ with probability at least $2/3$. If $x_1,\ldots,x_n$ are drawn from $P_0$, then $\SUM_n(x_1, \ldots, x_n)=0$ with probability $1$. Therefore, $\calB'$ outputs the null hypothesis with probability at least $11/12$. If $x_1,\ldots,x_n$ are drawn independently from $P_1$, then $X:=\SUM_n(x_1, \ldots, x_n)\sim\operatorname{Bin}(n,p)$ and $\mathbb{E}[X]=np=8r$. By the Chernoff bound and $r\ge 1$, \begin{align*}
\Pr[X<r]=\Pr\left[X<\frac{1}{8}\mathbb{E}[X]\right]\le \exp\left(-\left(\frac{7}{8}\right)^2\frac{\E[X]}{2}\right)=\exp\left(-\frac{49}{16}r\right) < \frac{1}{12}.
\end{align*}
Hence, $\Pr[X\ge r]\ge11/12$. Conditioned on $X\ge r$, $\calB'$ outputs the alternative hypothesis with probability at least $11/12$. Therefore, under $P_1$,  the hypothesis-testing protocol succeeds with probability at least $(11/12)^2>2/3$.

Taking sufficiently small $c_0 > 0$ such that $c_0'/(64\beta^2c_0^2) > 1$, then we have 
\begin{align*}
n \ge \frac{c_0'}{(\veps')^2\gamma^2} = \frac{c_0'n^2}{64\beta^2\veps^2r^2} > \frac{c_0'n}{64\beta^2c_0^2} > n,
\end{align*}
which yields a contradiction.
\end{proof}

We extend the reduction of \cite{eden2025triangle} from triangle counting on $\calG_{n,n}$ to $H$-counting on $\calG_{n,d}$ for any pattern graph $H$ and parameter $d$.

\begin{theorem}
\label{thm:degree-lower-bound}
For any pattern graph $H$ with $q$ connected components, $k$ vertices and $\ell$ edges $(\ell \ge 1)$, there exist constants $c,C,C_1, C_2>0$ such that for every $n,d\in\nats$ with $1\le d\le n-1$ and $d\ge6(k-2)$, $\veps\in (0, 1]$, $\alpha\in(0,n^{q}d^{k - q - 1}/(3C))$, and $ \delta \in [0, C_1\veps^3\alpha^2/(n^{2q+1}d^{2k-2q-2}\ln(n^{2q+1}d^{2k-2q-2}/(C_2\veps^2\alpha^2)))]$, every (potentially interactive) $(\veps,\delta)$-LEDP algorithm on $\calG_{n,d}$ for $H$-counting with additive error at most $\alpha$ with probability at least $2/3$ must have $\alpha\ge c\cdot n^{q-1}d^{k-q-1}\min(n,\sqrt{n}/\veps) - n^{q-1}d^{k-q-1}/C$.
\end{theorem}

\begin{proof}
We argue by contradiction to \Cref{lem:bit-sum-testing}. Assume that an $(\veps,\delta)$-LEDP algorithm $\calA$ counts $H$ on $\calG_{n,d}$ with additive error at most $\alpha$ with probability at least $2/3$, where $\veps\in(0,1]$ and
\begin{align*}
 \delta&\in\left[0,
 \frac{C_1\veps^3\alpha^2}{n^{2q+1}d^{2k-2q-2}\ln\!\left(n^{2q+1}d^{2k-2q-2}/(C_2\veps^2\alpha^2)\right)}\right],\\*
 \alpha&<c\,n^{q-1}d^{k-q-1}\min(n,\sqrt n/\veps)-n^{q-1}d^{k-q-1}/C.
\end{align*}

Set $C:=12^{k-2}|\Aut(H)|$. We reduce LDP testing of $\SUM_{n'}(x)=0$ versus $\SUM_{n'}(x)\ge r$ to LEDP subgraph counting on $n$-vertex graphs, with $n'$ and $r$ specified below. If $k=2$, then $H$ is a single edge: set $n':=\lfloor n/2\rfloor$, take vertex sets $V_1=\varnothing$ and $|V_2|=2n'$, and make the remaining $n-2n'$ vertices publicly isolated.

For $k\ge3$, set
\begin{align*}
d_0:=3\lfloor d/3\rfloor,\qquad n_0:=d_0\lfloor n/d_0\rfloor,\qquad d':=d_0/3,\qquad n':=n_0/3.
\end{align*}
Then $3\mid d_0$, $d_0\mid n_0$, $d/2\le d_0\le d$, and $n/2\le n_0\le n$. Since $6(k-2)$ is a multiple of $3$, the assumption $d\ge6(k-2)$ also gives $d_0\ge6(k-2)$, hence $d'\ge2(k-2)$. Take disjoint sets $V_1,V_2$ of sizes $n'$ and $2n'$, respectively, and make the remaining $n-n_0$ vertices publicly isolated. In both cases, $n/6\le n'\le n$, and all vertices outside $V_2$ contain only public information and can be simulated by any local party. We identify the vertices in $V_2$ with $[2n']$, and each party $i\in[n']$ simulates the two vertices $2i-1$ and $2i$ in $V_2$.

We first add the edges that encode the private input. For each $i\in[n']$, we add the edge $\{2i-1,2i\}$ if and only if $x_i=1$. All edges added in the remainder of the construction are public and hence known to every party.

When $k=2$, no public edges are needed, and $G$ is a matching together with isolated vertices, so its maximum degree is at most $1\le d$. When $k\ge3$, partition $V_1$ into $n'/d'=n_0/d_0$ disjoint groups of size $d'$, and partition $V_2$ into the same number of disjoint groups of size $2d'$, keeping each pair $\{2i-1,2i\}$ in one group. Pair each $V_1$ group with one $V_2$ group. For each paired group, add all edges within its $V_1$ group and all edges between its $V_1$ and $V_2$ groups. Every vertex in $V_1$ has degree $(d'-1)+2d'=d_0-1\le d$, and every vertex in $V_2$ has degree at most $d'+1\le d$. Thus, $G\in\calG_{n,d}$ in both cases.

Let $G_0$ denote the full $n$-vertex graph obtained from the above construction when $x_i=0$ for all $i\in[n']$, including the publicly isolated vertices. Since $G_0$ contains only public edges, $f_H(G_0)$ can be computed without accessing the private input. We observe the following property.

\begin{claim}\label{claim:FHgap}
For every $x_1,\dots, x_{n'}\in\{0,1\}$, we have
\begin{align*}
f_H(G)-f_H(G_0)
\ge \frac{2n^{q - 1}d^{k - q - 1}}{C}\cdot \SUM_{n'}(x_1, \ldots,x_{n'}).
\end{align*}
\end{claim}

\begin{proof}
For $k=2$, we have $q=1$, $C=2$, and $f_H(G)-f_H(G_0)=\SUM_{n'}(x)$, so the claim holds with equality. Assume henceforth that $k\ge3$.

Fix an edge $e^\star$ of $H$, and consider a private edge $\{u,v\}$ corresponding to an index $i$ with $x_i=1$. We count only copies of $H$ in which $e^\star$ is mapped to $\{u,v\}$ and every other vertex of $H$ is mapped to $V_1$. Let $H_1,\ldots,H_q$ be the connected components of $H$, where $e^\star\in E(H_1)$.

The remaining vertices of $H_1$ are mapped to the $V_1$ group paired with the group containing $\{u,v\}$. For each of the other $q-1$ connected components, we may independently choose any of the $n_0/d_0$ groups in $V_1$ and map all of its vertices into that group.

Fixing these group choices, suppose that $t_j$ vertices of $H$ are mapped into the $j$-th $V_1$ group. Since $\sum_{j=1}^{n_0/d_0}t_j=k-2$ and $d'\ge2(k-2)$, the number of injective choices is at least
\begin{align*}
\prod_{j=1}^{n_0/d_0}(d')^{\underline{t_j}}
&\ge\prod_{j=1}^{n_0/d_0}\left(d'-(k-2)\right)^{t_j}
\ge\prod_{j=1}^{n_0/d_0}\left(\frac{d'}{2}\right)^{t_j}
=\left(\frac{d'}{2}\right)^{k-2}.
\end{align*}
Accounting for the two orientations of $e^\star$ and for automorphisms of $H$, the number of such copies containing $\{u,v\}$ is therefore at least
\begin{align*}
\frac{2}{|\Aut(H)|}\left(\frac{n_0}{d_0}\right)^{q-1}\left(\frac{d'}{2}\right)^{k-2}
&=\frac{2n_0^{q-1}d_0^{k-q-1}}{6^{k-2}|\Aut(H)|}\\
&\ge\frac{2n^{q-1}d^{k-q-1}}{12^{k-2}|\Aut(H)|}
=\frac{2n^{q-1}d^{k-q-1}}{C},
\end{align*}
where we used $n_0\ge n/2$, $d_0\ge d/2$, and $k-q-1\ge0$.

Every copy counted above contains exactly one private edge, since all other vertices are mapped to $V_1$. Hence the copies counted for different private edges are disjoint. All these copies lie in $V_1\cup V_2$ and are absent from $G_0$. If $H$ has isolated vertices, the publicly isolated vertices may contribute additional copies, but adding private edges cannot destroy any copy. Thus, summing over all present private edges proves the claim for the full $n$-vertex graphs.
\end{proof}

Let $r:=\lfloor C\alpha/(n^{q-1}d^{k-q-1})\rfloor+1$. To distinguish $\SUM_{n'}(x)=0$ from $\SUM_{n'}(x)\ge r$, an LDP protocol $\cal B$ simulates $\cal A$ on the graph $G$ above. Let $\hf_H(G)$ be the output of $\cal A$. Protocol $\cal B$ outputs $0$ if $\hf_H(G)\le f_H(G_0)+\alpha$, and $1$ otherwise. If two inputs to $\cal B$ differ only in the $i$-th coordinate, then the corresponding graphs differ only in the edge $\{2i-1,2i\}$. Hence, $\calB$ is $(\veps,\delta)$-LDP by the LEDP guarantee of $\cal A$ and \Cref{lemma:post-processing}.

Now we prove the correctness of $\calB$. If $\SUM_{n'}(x)=0$, then $G=G_0$, so $\cal B$ is correct with probability at least $2/3$. If $\SUM_{n'}(x)\ge r$, then by Claim~\ref{claim:FHgap}, $f_H(G)-f_H(G_0)\ge 2n^{q - 1}d^{k - q - 1}r/C>2\alpha$. Thus, whenever $\cal A$ has error at most $\alpha$, we have $\hf_H(G) > f_H(G_0) + \alpha$, so $\cal B$ is again correct with probability at least $2/3$.

It remains to verify the hypotheses of \Cref{lem:bit-sum-testing}. Choose $c>0$ such that $6cC<\min\{1,c_0\}$. The assumed upper bound on $\alpha$ and the inequality $n'\ge n/6$ give
\begin{align*}
r\le\frac{C\alpha}{n^{q-1}d^{k-q-1}}+1
&<cC\min\left\{n,\frac{\sqrt n}{\veps}\right\}
\le6cC\min\left\{n',\frac{\sqrt{n'}}{\veps}\right\}.
\end{align*}
In particular, $1\le r\le n'$. We may assume $c_2\le1/2$ in \Cref{lem:bit-sum-testing} by decreasing it if necessary. Set $C_1:=c_1C^2$, $C_2:=c_2C^2$, and $t:=C\alpha/(n^{q-1}d^{k-q-1})$, so $0<t<r$. Then
\begin{align*}
\delta
&\le\frac{C_1\veps^3\alpha^2}{n^{2q+1}d^{2k-2q-2}\ln\!\left(n^{2q+1}d^{2k-2q-2}/(C_2\veps^2\alpha^2)\right)}\\
&=\frac{c_1\veps^3t^2}{n^3\ln\!\left(n^3/(c_2\veps^2t^2)\right)}
\le\frac{c_1\veps^3r^2}{(n')^3\ln\!\left((n')^3/(c_2\veps^2r^2)\right)}.
\end{align*}
The logarithms are positive because $t<r\le n'\le n$, $\veps\le1$, and $c_2\le1/2$; the last inequality follows from $t<r$ and $n'\le n$. Thus, \Cref{lem:bit-sum-testing} applies and yields $r\ge c_0\min(n',\sqrt{n'}/\veps)$, contradicting the upper bound on $r$ since $6cC<c_0$.

\end{proof}

\section*{AI methodology}
\addcontentsline{toc}{section}{AI methodology}

The algorithmic ideas, lower-bound constructions, and proofs in
\Cref{sec:two-round,gh:main,sec:degree-lower} were developed through
extensive, iterative interactions between the authors and the AI models
ChatGPT 5.6 Sol and ChatGPT 6 Astra. These interactions contributed to both the development of the arguments and the refinement of their presentation. The authors independently
verified, developed, and simplified the proposed arguments and take full
responsibility for the correctness, exposition, and attribution of the work.

\bibliographystyle{alpha}
\bibliography{references}

@inproceedings{joseph2020exponential,
  author={Joseph, Matthew and Mao, Jieming and Roth, Aaron},
  title={Exponential Separations in Local Differential Privacy},
  booktitle={Proceedings of the 2020 ACM-SIAM Symposium on Discrete Algorithms (SODA)},
  pages={515--527},
  year={2020},
  doi={10.1137/1.9781611975994.31}
}

@inproceedings{dagan2020interaction,
  author={Dagan, Yuval and Feldman, Vitaly},
  title={Interaction Is Necessary for Distributed Learning with Privacy or Communication Constraints},
  booktitle={Proceedings of the 52nd Annual ACM SIGACT Symposium on Theory of Computing},
  pages={450--462},
  year={2020},
  doi={10.1145/3357713.3384315}
}

@inproceedings{acharya2022role,
  author={Acharya, Jayadev and Canonne, Cl{\'e}ment L. and Sun, Ziteng and Tyagi, Himanshu},
  title={The Role of Interactivity in Structured Estimation},
  booktitle={Proceedings of Thirty Fifth Conference on Learning Theory},
  series={Proceedings of Machine Learning Research},
  volume={178},
  pages={1328--1355},
  year={2022},
  url={https://proceedings.mlr.press/v178/acharya22b.html}
}

@article{eden2025triangle,
  title={Triangle counting with local edge differential privacy},
  author={Eden, Talya and Liu, Quanquan C and Raskhodnikova, Sofya and Smith, Adam},
  journal={Random Structures \& Algorithms},
  volume={66},
  number={4},
  pages={e70002},
  year={2025},
  publisher={Wiley Online Library}
}

@book{wainwright2019high,
  title={High-dimensional statistics: A non-asymptotic viewpoint},
  author={Wainwright, Martin J},
  volume={48},
  year={2019},
  publisher={Cambridge university press}
}

@inproceedings{joseph2019role,
  title={The role of interactivity in local differential privacy},
  author={Joseph, Matthew and Mao, Jieming and Neel, Seth and Roth, Aaron},
  booktitle={2019 IEEE 60th Annual Symposium on Foundations of Computer Science (FOCS)},
  pages={94--105},
  year={2019},
  organization={IEEE}
}

@article{kasiviswanathan2011can,
  title={What can we learn privately?},
  author={Kasiviswanathan, Shiva Prasad and Lee, Homin K and Nissim, Kobbi and Raskhodnikova, Sofya and Smith, Adam},
  journal={SIAM Journal on Computing},
  volume={40},
  number={3},
  pages={793--826},
  year={2011},
  publisher={SIAM}
}

@inproceedings{dwork2006calibrating,
  title={Calibrating noise to sensitivity in private data analysis},
  author={Dwork, Cynthia and McSherry, Frank and Nissim, Kobbi and Smith, Adam},
  booktitle={Theory of cryptography conference},
  pages={265--284},
  year={2006},
  organization={Springer}
}

@inproceedings{dwork2006our,
  title={Our data, ourselves: Privacy via distributed noise generation},
  author={Dwork, Cynthia and Kenthapadi, Krishnaram and McSherry, Frank and Mironov, Ilya and Naor, Moni},
  booktitle={Annual international conference on the theory and applications of cryptographic techniques},
  pages={486--503},
  year={2006},
  organization={Springer}
}

@article{warner1965randomized,
  title={Randomized response: A survey technique for eliminating evasive answer bias},
  author={Warner, Stanley L},
  journal={Journal of the American statistical association},
  volume={60},
  number={309},
  pages={63--69},
  year={1965},
  publisher={Taylor \& Francis}
}

@inproceedings{imola2022communication,
  title={Communication-Efficient triangle counting under local differential privacy},
  author={Imola, Jacob and Murakami, Takao and Chaudhuri, Kamalika},
  booktitle={31st USENIX security symposium (USENIX Security 22)},
  pages={537--554},
  year={2022}
}

@inproceedings{imola2021locally,
  title={Locally differentially private analysis of graph statistics},
  author={Imola, Jacob and Murakami, Takao and Chaudhuri, Kamalika},
  booktitle={30th USENIX security symposium (USENIX Security 21)},
  pages={983--1000},
  year={2021}
}

@article{bassily2014differentially,
  title={Differentially private empirical risk minimization: Efficient algorithms and tight error bounds},
  author={Bassily, Raef and Smith, Adam and Thakurta, Abhradeep},
  journal={arXiv preprint arXiv:1405.7085},
  year={2014}
}

@book{ben2021lectures,
  title={Lectures on modern convex optimization 2020},
  author={Ben-Tal, Aharon and Nemirovski, Arkadi},
  journal={SIAM, Philadelphia},
  year={2021}
}

@article{tarjan1972depth,
  title={Depth-first search and linear graph algorithms},
  author={Tarjan, Robert},
  journal={SIAM journal on computing},
  volume={1},
  number={2},
  pages={146--160},
  year={1972},
  publisher={SIAM}
}

@book{lovasz2012large,
  title={Large networks and graph limits},
  author={Lov{\'a}sz, L{\'a}szl{\'o}},
  volume={60},
  year={2012},
  publisher={American Mathematical Soc.}
}

@inproceedings{suppakitpaisarn2025graphlets,
  author = {Suppakitpaisarn, Vorapong and Ponnoprat, Donlapark and Hirankarn, Nicha and Hillebrand, Quentin},
  title = {Counting Graphlets of Size {$k$} under Local Differential Privacy},
  booktitle = {Proceedings of the 28th International Conference on Artificial Intelligence and Statistics},
  series = {Proceedings of Machine Learning Research},
  volume = {258},
  pages = {5005--5013},
  year = {2025},
  publisher = {PMLR},
  url = {https://proceedings.mlr.press/v258/suppakitpaisarn25a.html}
}

@article{hu2026acyclic,
  author = {Hu, Yihua and Wang, Kuncan and Dong, Wei},
  title = {Acyclic Graph Pattern Counting under Local Differential Privacy},
  journal = {Proceedings of the ACM on Management of Data},
  volume = {4},
  number = {3},
  pages = {129:1--129:26},
  year = {2026},
  doi = {10.1145/3802006},
  url = {https://arxiv.org/abs/2603.19671}
}

@inproceedings{hillebrand2025cycle,
  author = {Hillebrand, Quentin and Suppakitpaisarn, Vorapong and Shibuya, Tetsuo},
  title = {Cycle Counting under Local Differential Privacy for Degeneracy-Bounded Graphs},
  booktitle = {42nd International Symposium on Theoretical Aspects of Computer Science (STACS 2025)},
  series = {Leibniz International Proceedings in Informatics},
  volume = {327},
  pages = {49:1--49:22},
  year = {2025},
  publisher = {Schloss Dagstuhl -- Leibniz-Zentrum f{\"u}r Informatik},
  doi = {10.4230/LIPIcs.STACS.2025.49},
  url = {https://drops.dagstuhl.de/entities/document/10.4230/LIPIcs.STACS.2025.49}
}

@misc{raskhodnikova2026localnode,
  author = {Raskhodnikova, Sofya and Smith, Adam and Wagaman, Connor and Zavyalov, Anatoly},
  title = {Local Node Differential Privacy},
  year = {2026},
  eprint = {2602.15802},
  archivePrefix = {arXiv},
  primaryClass = {cs.DS},
  note = {To appear in FOCS 2026},
  url = {https://arxiv.org/abs/2602.15802}
}

@misc{raskhodnikova2026subgraphnode,
  author = {Raskhodnikova, Sofya and Smith, Adam and Wagaman, Connor and Zavyalov, Anatoly},
  title = {Subgraph Counting with Local Node Differential Privacy},
  year = {2026},
  eprint = {2610.22278},
  archivePrefix = {arXiv},
  primaryClass = {cs.DS},
  note = {To appear in SODA 2027},
  url = {https://arxiv.org/abs/2610.22278}
}

@inproceedings{nissim2007smooth,
  author = {Nissim, Kobbi and Raskhodnikova, Sofya and Smith, Adam},
  title = {Smooth Sensitivity and Sampling in Private Data Analysis},
  booktitle = {Proceedings of the 39th Annual ACM Symposium on Theory of Computing (STOC)},
  pages = {75--84},
  year = {2007},
  publisher = {ACM},
  doi = {10.1145/1250790.1250803},
  url = {https://doi.org/10.1145/1250790.1250803}
}

@article{karwa2014private,
  author = {Karwa, Vishesh and Raskhodnikova, Sofya and Smith, Adam and Yaroslavtsev, Grigory},
  title = {Private Analysis of Graph Structure},
  journal = {ACM Transactions on Database Systems},
  volume = {39},
  number = {3},
  pages = {22:1--22:33},
  year = {2014},
  doi = {10.1145/2611523},
  url = {https://doi.org/10.1145/2611523}
}

@inproceedings{blocki2013restricted,
  author = {Blocki, Jeremiah and Blum, Avrim and Datta, Anupam and Sheffet, Or},
  title = {Differentially Private Data Analysis of Social Networks via Restricted Sensitivity},
  booktitle = {Proceedings of the 4th Conference on Innovations in Theoretical Computer Science (ITCS)},
  pages = {87--96},
  year = {2013},
  publisher = {ACM},
  doi = {10.1145/2422436.2422449},
  url = {https://arxiv.org/abs/1208.4586}
}

@inproceedings{kasiviswanathan2013graphs,
  author = {Kasiviswanathan, Shiva Prasad and Nissim, Kobbi and Raskhodnikova, Sofya and Smith, Adam},
  title = {Analyzing Graphs with Node Differential Privacy},
  booktitle = {Theory of Cryptography (TCC)},
  series = {Lecture Notes in Computer Science},
  volume = {7785},
  pages = {457--476},
  year = {2013},
  publisher = {Springer},
  doi = {10.1007/978-3-642-36594-2_26},
  url = {https://doi.org/10.1007/978-3-642-36594-2_26}
}

@inproceedings{chen2013recursive,
  author = {Chen, Shixi and Zhou, Shuigeng},
  title = {Recursive Mechanism: Towards Node Differential Privacy and Unrestricted Joins},
  booktitle = {Proceedings of the 2013 ACM SIGMOD International Conference on Management of Data},
  pages = {653--664},
  year = {2013},
  publisher = {ACM},
  doi = {10.1145/2463676.2465304},
  url = {https://doi.org/10.1145/2463676.2465304}
}

@inproceedings{qin2017synthetic,
  author = {Qin, Zhan and Yu, Ting and Yang, Yin and Khalil, Issa and Xiao, Xiaokui and Ren, Kui},
  title = {Generating Synthetic Decentralized Social Graphs with Local Differential Privacy},
  booktitle = {Proceedings of the 2017 ACM SIGSAC Conference on Computer and Communications Security (CCS)},
  pages = {425--438},
  year = {2017},
  publisher = {ACM},
  doi = {10.1145/3133956.3134086},
  url = {https://doi.org/10.1145/3133956.3134086}
}

@inproceedings{sun2019extended,
  author = {Sun, Haipei and Xiao, Xiaokui and Khalil, Issa and Yang, Yin and Qin, Zhan and Wang, Hui and Yu, Ting},
  title = {Analyzing Subgraph Statistics from Extended Local Views with Decentralized Differential Privacy},
  booktitle = {Proceedings of the 2019 ACM SIGSAC Conference on Computer and Communications Security (CCS)},
  pages = {703--717},
  year = {2019},
  publisher = {ACM},
  doi = {10.1145/3319535.3354253},
  url = {https://davidyinyang.weebly.com/uploads/9/8/6/2/9862052/ccs19-ddp.pdf}
}

@inproceedings{imola2022shuffle,
  author = {Imola, Jacob and Murakami, Takao and Chaudhuri, Kamalika},
  title = {Differentially Private Triangle and 4-Cycle Counting in the Shuffle Model},
  booktitle = {Proceedings of the 2022 ACM SIGSAC Conference on Computer and Communications Security (CCS)},
  pages = {1505--1519},
  year = {2022},
  publisher = {ACM},
  doi = {10.1145/3548606.3560659},
  url = {https://arxiv.org/abs/2205.01429}
}

@inproceedings{fang2025shuffle,
  author = {Fang, Juanru and Yi, Ke},
  title = {Counting Subgraphs under Shuffle Differential Privacy},
  booktitle = {Proceedings of the 2025 ACM SIGSAC Conference on Computer and Communications Security (CCS)},
  year = {2025},
  publisher = {ACM},
  doi = {10.1145/3719027.3765047},
  url = {https://juanru-fang.github.io/Subgraph.pdf}
}

@article{he2025robust,
  author = {He, Yizhang and Wang, Kai and Zhang, Wenjie and Lin, Xuemin and Zhang, Ying and Ni, Wei},
  title = {Robust Privacy-Preserving Triangle Counting under Edge Local Differential Privacy},
  journal = {Proceedings of the ACM on Management of Data},
  volume = {3},
  number = {3},
  pages = {211:1--211:26},
  year = {2025},
  doi = {10.1145/3725348},
  url = {https://doi.org/10.1145/3725348}
}

@misc{guo2026noisy,
  author = {Guo, Jintao and Zhou, Ying and Luo, Guixun and Li, Chao and Hu, Yan},
  title = {Locally Private Subgraph Counting via Noisy Adjacency Matrix and Differential Privacy on Randomized Data},
  year = {2026},
  eprint = {2507.06508},
  archivePrefix = {arXiv},
  primaryClass = {cs.CR},
  note = {Version 2, revised August 26, 2026},
  url = {https://arxiv.org/abs/2507.06508v2}
}

@article{hillebrand2025hash,
  author = {Hillebrand, Quentin and Suppakitpaisarn, Vorapong and Shibuya, Tetsuo},
  title = {Communication Cost Reduction for Subgraph Counting under Local Differential Privacy via Hash Functions},
  journal = {Transactions on Machine Learning Research},
  year = {2025},
  url = {https://openreview.net/forum?id=N1J236mepp}
}

@inproceedings{HayLMJ09,
  author       = {Michael Hay and
                  Chao Li and
                  Gerome Miklau and
                  David D. Jensen},
  editor       = {Wei Wang and
                  Hillol Kargupta and
                  Sanjay Ranka and
                  Philip S. Yu and
                  Xindong Wu},
  title        = {Accurate Estimation of the Degree Distribution of Private Networks},
  booktitle    = {{ICDM} 2009, The Ninth {IEEE} International Conference on Data Mining,
                  Miami, Florida, USA, 6-9 December 2009},
  pages        = {169--178},
  publisher    = {{IEEE} Computer Society},
  year         = {2009},
  url          = {https://doi.org/10.1109/ICDM.2009.11},
  doi          = {10.1109/ICDM.2009.11},
  bibsource    = {dblp computer science bibliography, https://dblp.org}
}

@article{mundra2025practical,
  author = {Mundra, Pranay and Papamanthou, Charalampos and Shun, Julian and Liu, Quanquan C.},
  title = {Practical and Accurate Local Edge Differentially Private Graph Algorithms},
  journal = {Proceedings of the VLDB Endowment},
  volume = {18},
  number = {11},
  pages = {4199--4213},
  year = {2025},
  doi = {10.14778/3749646.3749687},
  url = {https://www.vldb.org/pvldb/vol18/p4199-mundra.pdf}
}

@inproceedings{chen2026range,
  author = {Chen, Xian and Bai, Ruobing and Peng, Pan},
  title = {Differentially Private Range Subgraph Counting},
  booktitle = {Proceedings of the 43rd International Conference on Machine Learning (ICML)},
  series = {Proceedings of Machine Learning Research},
  volume = {306},
  pages = {13453--13489},
  year = {2026},
  publisher = {PMLR},
  url = {https://proceedings.mlr.press/v306/chen26b.html}
}

@article{milo2002network,
  title={Network motifs: simple building blocks of complex networks},
  author={Milo, Ron and Shen-Orr, Shai and Itzkovitz, Shalev and Kashtan, Nadav and Chklovskii, Dmitri and Alon, Uri},
  journal={Science},
  volume={298},
  number={5594},
  pages={824--827},
  year={2002},
  publisher={American Association for the Advancement of Science}
}

@article{borell1975brunn,
  title={The Brunn-Minkowski inequality in gauss space},
  author={Borell, Christer},
  journal={Inventiones mathematicae},
  volume={30},
  number={2},
  pages={207--216},
  year={1975},
  publisher={Springer-Verlag, Berlin/Heidelberg}
}

@article{lowy2026differentially,
  title={Differentially private bilevel optimization: Efficient algorithms with near-optimal rates},
  author={Lowy, Andrew and Liu, Daogao},
  journal={Advances in Neural Information Processing Systems},
  volume={38},
  pages={27168--27209},
  year={2025}
}

@inproceedings{becchetti2008efficient,
  title={Efficient semi-streaming algorithms for local triangle counting in massive graphs},
  author={Becchetti, Luca and Boldi, Paolo and Castillo, Carlos and Gionis, Aristides},
  booktitle={Proceedings of the 14th ACM SIGKDD international conference on Knowledge discovery and data mining},
  pages={16--24},
  year={2008}
}

@article{gonen2011counting,
  title={Counting stars and other small subgraphs in sublinear-time},
  author={Gonen, Mira and Ron, Dana and Shavitt, Yuval},
  journal={SIAM Journal on Discrete Mathematics},
  volume={25},
  number={3},
  pages={1365--1411},
  year={2011},
  publisher={SIAM}
}

@article{prvzulj2007biological,
  title={Biological network comparison using graphlet degree distribution},
  author={Pr{\v{z}}ulj, Nata{\v{s}}a},
  journal={Bioinformatics},
  volume={23},
  number={2},
  pages={e177--e183},
  year={2007},
  publisher={Oxford University Press}
}

@article{gonen2009approximating,
  title={Approximating the number of network motifs},
  author={Gonen, Mira and Shavitt, Yuval},
  journal={Internet Mathematics},
  volume={6},
  number={3},
  pages={349--372},
  year={2009},
  publisher={Taylor \& Francis}
}

@article{fan2021characterizing,
  title={Characterizing cycle structure in complex networks},
  author={Fan, Tianlong and L{\"u}, Linyuan and Shi, Dinghua and Zhou, Tao},
  journal={Communications Physics},
  volume={4},
  number={1},
  pages={272},
  year={2021},
  publisher={Nature Publishing Group UK, London}
}

@article{palla2005uncovering,
  title={Uncovering the overlapping community structure of complex networks in nature and society},
  author={Palla, Gergely and Der{\'e}nyi, Imre and Farkas, Ill{\'e}s and Vicsek, Tam{\'a}s},
  journal={Nature},
  volume={435},
  number={7043},
  pages={814--818},
  year={2005},
  publisher={Nature Publishing Group UK, London}
}

@inproceedings{dhulipala2022differential,
  title={Differential privacy from locally adjustable graph algorithms: k-core decomposition, low out-degree ordering, and densest subgraphs},
  author={Dhulipala, Laxman and Liu, Quanquan C and Raskhodnikova, Sofya and Shi, Jessica and Shun, Julian and Yu, Shangdi},
  booktitle={2022 IEEE 63rd Annual Symposium on Foundations of Computer Science (FOCS)},
  pages={754--765},
  year={2022},
  organization={IEEE}
}

@article{clauset2004finding,
  title={Finding community structure in very large networks},
  author={Clauset, Aaron and Newman, Mark EJ and Moore, Cristopher},
  journal={Physical Review E—Statistical, Nonlinear, and Soft Matter Physics},
  volume={70},
  number={6},
  pages={066111},
  year={2004},
  publisher={APS}
}

@article{achlioptas2003database,
  title={Database-friendly random projections: Johnson-Lindenstrauss with binary coins},
  author={Achlioptas, Dimitris},
  journal={Journal of computer and System Sciences},
  volume={66},
  number={4},
  pages={671--687},
  year={2003},
  publisher={Elsevier}
}

@article{he2024common,
  title={Common neighborhood estimation over bipartite graphs under local differential privacy},
  author={He, Yizhang and Wang, Kai and Zhang, Wenjie and Lin, Xuemin and Zhang, Ying},
  journal={Proceedings of the ACM on Management of Data},
  volume={2},
  number={6},
  pages={1--26},
  year={2024},
  publisher={ACM, New York, NY, USA}
}

@article{bun2018heavy,
  title={Heavy hitters and the structure of local privacy},
  author={Bun, Mark and Nelson, Jelani and Stemmer, Uri},
  journal={ACM Transactions on Algorithms (TALG)},
  volume={15},
  number={4},
  pages={1--40},
  year={2019},
  publisher={ACM, New York, NY, USA}
}

@InProceedings{gowtham2022introducing,
  title = 	 {Differentially Private Matrix Completion through Low-rank Matrix Factorization},
  author =       {Wang, Lingxiao and Zhao, Boxin and Kolar, Mladen},
  booktitle = 	 {Proceedings of The 26th International Conference on Artificial Intelligence and Statistics},
  pages = 	 {5731--5748},
  year = 	 {2023},
  editor = 	 {Ruiz, Francisco and Dy, Jennifer and van de Meent, Jan-Willem},
  volume = 	 {206},
  series = 	 {Proceedings of Machine Learning Research},
  month = 	 {25--27 Apr},
  publisher =    {PMLR},
  url = 	 {https://proceedings.mlr.press/v206/wang23d.html}
}

@article{lin2026purifying,
  title={Purifying approximate differential privacy with randomized post-processing},
  author={Lin, Yingyu and Wang, Erchi and Ma, Yian and Wang, Yu-Xiang},
  journal={Advances in Neural Information Processing Systems},
  volume={38},
  pages={144259--144316},
  year={2026}
}

\addcontentsline{toc}{section}{References}

\end{document}